\documentclass[twocolumn]{openjournal}

\newcommand{\msun}{{h$^{-1}$ M$_{\odot}$\;}}

\newcommand{\mpc}{{h$^{-1}$Mpc \;}}

\usepackage[utf8x]{inputenc}

\usepackage{graphicx}
\usepackage{natbib}
\usepackage{amsmath}
\usepackage{color}

\received{\today}
\revised{\today}
\accepted{\today}

\shorttitle{Galaxy Quenching from Cosmic Web Detachment}
\shortauthors{Miguel A. Aragon-Calvo et al.}

\begin{document}

\title{Galaxy Quenching from Cosmic Web Detachment}

\email{maragon@astro.unam.mx}

\author{Miguel A. Aragon-Calvo}
\affil{Instituto de Astronomia, Universidad Nacional Autonoma de Mexico \\
Apdo. postal 877, C.P. 22800, Ensenada, Mexico}

\author{Mark C. Neyrinck}
\affiliation{Department of Theoretical Physics, \\
University of the Basque Country UPV/EHU, 48080 Bilbao, Spain}
\affiliation{IKERBASQUE Foundation for Science, Bilbao, Spain}

\author{Joseph Silk}
\affiliation{Sorbonne Universites, UPMC Univ Paris 6 et CNRS, UMR 7095 \\
Institut dAstrophysique de Paris, 98 bis bd Arago, F-75014 Paris, France}
\affiliation{AIM-Paris-Saclay CEA/DSM/IRFU, CNRS, Univ Paris 7, F-91191, Gif-sur-Yvette, France}
\affiliation{BIPAC, University of Oxford, 1 Keble Road, Oxford OX1 3RH, UK DC 20005}

\begin{abstract}


We propose the Cosmic Web Detachment (CWD) model, a framework to interpret the star-formation history of galaxies in a cosmological context. The CWD model unifies several starvation mechanisms known to disrupt or stop star formation into one single physical framework. Galaxies begin accreting star-forming gas at early times via a network of primordial filaments, simply related to the pattern of density fluctuations in the initial conditions. But when shell-crossing occurs on intergalactic scales, this pattern is disrupted, and the galaxy \textit{detaches} from its primordial filaments, ending the accretion of cold gas. We argue that CWD encompasses  known external processes halting star formation, such as harassment, strangulation and starvation. On top of these external processes, internal feedback processes such as AGN contribute to stop in star formation as well.

By explicitly pointing out the non-linear nature of CWD events  we introduce a simple formalism to identify CWD events in N-body simulations.  With it we reproduce and explain, in the context of CWD, several observations including downsizing, the cosmic star formation rate history, the galaxy mass-color diagram and the dependence of the fraction of red galaxies with mass and local density.

\end{abstract}

\keywords{Cosmology: large-scale structure of Universe; galaxies: kinematics and dynamics; methods: data analysis, N-body simulations}


\section{Introduction}

Star formation quenching, its underlying mechanism and what triggers it, is one of the most pressing problems in modern galaxy formation. Galaxies in the Universe are either actively forming stars or in a ``quenched" state of practically no star formation. This dichotomy is reflected in the bi-modality in the color-magnitude diagram \citep{Strateva01,Hogg04,Baldry04,Faber07} clearly separating blue star-forming spiral galaxies from red ``quenched" elliptical galaxies (see \citet{Dekel05} for a review). Most galaxies sit in one of the two groups and only a few galaxies are  found in the ``green valley" in between \citep{Schawinski14}. The clear gap between the two galaxy populations can be interpreted as star formation quenching either having occurred several Gyr ago and/or that it is a fast event \citep{Bell04,Blanton06,Wyder07}.

Several mechanisms have been proposed as responsible for the decrease in star formation activity in galaxies. Currently favored candidates include internal feedback processes such as AGN and supernovae feedback \citep{Silk98,DiMatteo05, Best05}. However, the non-causal connection between black hole mass vs. bulge mass challenges the predominant place given to AGN in galaxy formation \citep{Peng07,Jahnke11} and recent  deep x-ray surveys find similar SFR for normal and Type 2 AGN galaxies of similar stellar mass, showing no evidence for AGN-driven quenching. Cosmic environment has been long recognized as playing a major role shaping the properties of galaxies, specially in dense environments \citep{Alam18} where tidal effects are important \citep{Kraljic18}. Dense environments affect galaxies through a series of processes including  ram pressure and gas stripping \citep{Gunn72,Balsara94,Abadi99,McCarthy08}, harassment \citep{Moore96,Moore98}, strangulation \citep{Bekki02,Fujita04,Kawata08}, preprocessing \citep{Kodama01,Treu03,Goto03}. These processes have one key characteristic in common: they are external influences that affect star formation by preventing gas from reaching galaxies or by removing gas reservoirs. There is evidence from simulation for environmentally induced quenching independent of AGN feedback from purely geometric constraints \citep{Aragon14b} and an undefined but clear cosmological origin \citep{Feldmann15,Peng15}. The spatial correlation of galaxy properties, up to scales of several megaparsecs, also points to the dominant role of environment on galaxy evolution \citep{Dressler80,Weinmann06,Kauffmann13}.

Both observations and simulations point to star formation quenching being the result of both environmental (external) processes and feedback (internal) processes \citep{Hogg03,Blanton05,Peng10,Voort11} but their relative importance is not clear yet. External processes are usually characterized by local density, a first-order environmental descriptor with limited power to encode the intricate geometry of the cosmic web crucial to understand the geometry and dynamics of gas accretion, galaxy interactions, etc. Internal processes seem to be driven by halo mass \citep{Peng10}. However, halo mass and density are correlated, making this relation not straightforward to interpret.

\subsection{The need for a cosmic web galaxy formation model}\label{galform}

Current galaxy formation models, in particular semi-analytic implementations, can reproduce a wide range of observations and provide useful insight on the physical processes occurring inside galaxies \citep{Rees77,White78,Lacey91,Cole91,White91,Cole00,Benson01}. These models intrinsically consider galaxies as isolated entities. Interactions with other galaxies and their cosmic environment are only indirectly treated via their mass accretion and merger history  \citep{Hearing13,Mutch13,Aldo16},
which are insensitive to the particular geometry and dynamics of the cosmic web surrounding galaxies.  However, there is extensive evidence of the effect of cosmic environment  on halo/galaxy properties \citep{Trujillo06,Aragon07,Aragon07b,Hahn07,Dekel09,Pichon11,Laigle15,Darvish16} but this key information is not included in current galaxy formation models making them blind to environmental effects.  Semi-analytic models can not, by construction, provide a physical mechanism for quenching beyond the ones included in its prescription. Any quenching process of large-scale origin will only be reflected indirectly by mass accretion and merger histories, severely limiting their interpretative power.

The empirical \textit{age matching} model is relevant to CWD \citep{Hearing13,Hearing14,Watson15}. In age matching, the time for star formation quenching is assumed to be the first of i) the time when a galaxy reaches $10^{12}$\msun, ii) the formation time of the halo defined as the time when the slope in the mas accretion history changes or iii) the time when the halo is accreted into a larger halo, becoming a satellite. The \textit{age} of the halo computed as above is then matched with observed properties of galaxies associated with time evolution such as color.
Age matching is an empirical model which, despite being based on ad-hoc rules, can reproduce several observations \citep{Hearing14}. However, it does not explain the underlying physical process related to the \textit{age} of a galaxy and so its explanatory and predictive power is limited.  As we will see in the following sections the criteria used to define the age of a galaxy can be associated to non-linear interactions between galaxies and their environment, pointing to a common physical mechanism of cosmological origin.


\begin{figure*}
  \centering
   \includegraphics[width=0.99\textwidth,angle=0.0]{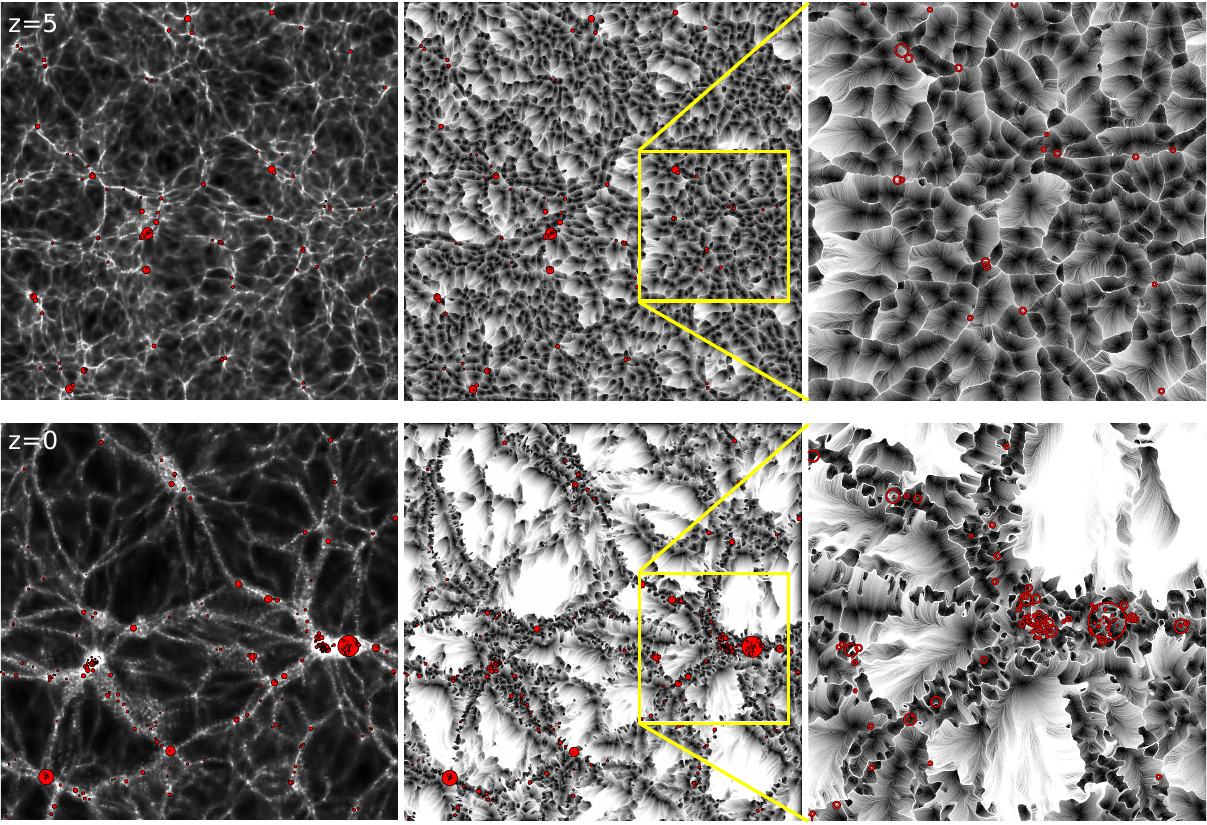}
  \caption{Coherent vs. chaotic velocity field around halos. Left: a 1 h$^{-1}$Mpc thick slice through the density field of a 32 h$^{-1}$Mpc box. Center: the velocity field at scales smaller than 125 h$^{-1}$kpc highlighted using the particle advection visualization technique \citep[see][for a description of the particle advection technique]{Aragon13}. The streams (lines delineating the velocity field) indicate the direction of the velocity field. The velocity field converges at the bright regions. Right: a zoomed region showing details in the velocity field. The red/white circles correspond to halos inside the slice. Top panels show that the velocity field at early times is highly coherent and closely delineates the location of primordial filaments connected to halos. At later times (bottom panels) the velocity field is highly chaotic and halos are not connected to the observed streams. }\label{fig:box_advection}
\end{figure*}

%
\subsection{Star-forming gas accretion via primordial filaments}

Galaxy formation is a complex process that began when, driven by gravity, nodes of a tenuous web of primordial filaments emerged from tiny density fluctuations \citep{Zeldovich70}, \citep[see also][]{Hidding14}. At the nodes of this cosmic network proto-galaxies began to grow feeding via  coherent filamentary streams of cold (T$ <10^5$K) gas \citep{keres05, Dekel09,Danovich12,Harford16}. The coherence of the streams at early times makes gas accretion highly efficient. At $z \sim 2$ most of the gas in star-forming galaxies is accreted via dense narrow filamentary streams that inject cold gas  into the inner regions of galaxies even in the presence of shock heating \citep{Dekel09,Faucher11}. In comparison, far less gas is accreted via the inefficient, isotropic, hot (T$ >10^5$K) accretion mode.

The nature of the cold streams connected to a proto-galaxy is closely linked to its surrounding matter distribution, and in particular to surrounding peaks, as described in the \textit{Cosmic Web theory} \citep{Bond96}. The quadrupolar tidal field configuration associated to a pair of peaks results in the formation of a bridge of matter in between. The Cosmic Web theory is usually invoked to explain the large megaparsec-scale filaments observed in the galaxy distribution. However, the same principle can be applied at smaller scales and earlier times to describe the formation of bridges between proto-galaxies. Primordial filaments form in a configuration directly imprinted by the pattern of fluctuations in the initial conditions, part of a highly coherent velocity field. Figure \ref{fig:box_advection} shows a comparison between the density and velocity fields at $z=5$ and $z=0$. The velocity field was high-pass filtered at 125 kpc in order to highlight dynamics on galactic scales instead of the large-scale flows that dominate the raw velocity field (see \citet{Aragon13} for details). There are important differences between the early and later velocity fields. The velocity field at $z=5$ is laminar and closely follows the filamentary network surrounding halos. All halos can be seen either at the nodes or ridges of coherent filamentary structures in the velocity field.  In contrast, the velocity field at $z=0$ is dominated by chaotic flows and shows no correlation with the  position of the halos.
The different dynamical regimes between primordial and non-linear filaments has not been explicitly noted before in the literature and it is key for a cosmological understanding of gas accretion.

%
\subsection{Primordial vs. large-scale filaments}

Figure \ref{fig:box_advection} shows two different kinds of structures in the cosmic web separated by their evolutionary stage:

\begin{itemize}
\item Primordial filaments, formed immediately after halos begin their collapse, with a marked cellular nature and coherent velocity flows. These are the filamentary streams responsible for gas accretion first described by \citep{Dekel09}.\\
\item Large filaments as those seen in the present-time galaxy distribution. They are formed hierarchically by the collapse of smaller primordial filaments and are larger, more massive and dynamically complex. Such large-scale structures also form a cellular system but at scales of tens of megaparsecs \citep{Joeveer78,Klypin83,Geller89,Icke91,Aragon10c,Einasto11,Aragon13}. 
\end{itemize}

Note that present-time voids may still contain primordial filaments due to their super-Hubble expansion which freezes the development of structure at large scales. Such small-scale filaments are prevalent in high-resolution computer simulations of voids \citep{Gottlober03,Sheth04,Park09,Rieder13} and recently have been observed in deep surveys \citep{Kreckel12,Beygu13,Alpaslan14}.

%
\subsection{Halos as nodes of the primordial filamentary web}

Figure \ref{fig:box_advection} highlights the interplay between halos and their surrounding cosmic web as a function of time and halo mass. At $z=5$ the non-linear mass $M_\ast$ is of the order of $10^7$\msun while at the present time it is close to $10^{13}$\msun (see Fig. \ref{fig:M_star}). The non-linear mass value gives us a rough estimate of the mass-scale at which halos become the dominant component of their surrounding cosmic web \citep{Dekel09}. Halos with masses larger than $M_\ast(z)$ act as nodes of their \textit{local cosmic web} as seen in the coherent streams attached to halos at $z=5$ in Fig. \ref{fig:box_advection}. 
On the other hand,  halos less massive than $M_\ast(z)$ are usually embedded inside larger cosmic web elements as in the case of galaxies embedded in megaparsec-scale filaments. Present-time galactic halos located inside large-scale filaments are part of complex large-scale motions that do not conserve the small-scale coherence seen at early times. 

The picture described above is complicated by the fact that the elements of the cosmic web represent different stages in the large-scale gravitational collapse. While present-time $M_\ast(z=0)$ halos are part of larger collapsed structures such as filaments and clusters, the same halos, when found inside voids, act as nodes of their local filamentary network  \citep{Kreckel12,Beygu13,Alpaslan14}.

%
\subsection{The fate of cold flows and star formation quenching}

The stark difference in the velocity field around halos at early and later times gives us a clue on the fate of the primordial filaments feeding galaxies. The velocity field around halos in present-time over-dense regions has no memory of the primordial filaments originally connected to them. If gas accretion through coherent cold filamentary flows is the most efficient mechanism to inject star-forming gas into galaxies then we should see a clear change in star formation in halos entering non-linear regions. The change in the accretion mode should be fast (within a few Gyr) given the efficient gas-star conversion of cold gas \citep{Dekel09,Voort11} which forces a gas-starved galaxy to stop producing stars within a short timescale \citep{Bauermeister10,Peng15}. Based in our current knowledge of the relation between galaxies, star formation and gas accretion via cold flows we can enumerate the following: 

\begin{itemize}
\item Proto-galaxies are the nodes of a network of primordial filaments.
\item Star-forming cold gas is accreted mainly via primordial coherent filamentary streams.
\item Star formation closely follows cold gas accretion and ends (to a first approximation) when the gas supply is cut off.
\end{itemize}

\noindent These observations point to a quenching process that acts by affecting the accretion of gas through primordial filaments. Any mechanism that is able to separate a galaxy from its web of gas-accreting filaments results in star formation quenching regardless of internal processes.

\begin{figure*}
  \centering
  \includegraphics[width=0.99\textwidth,angle=0.0]{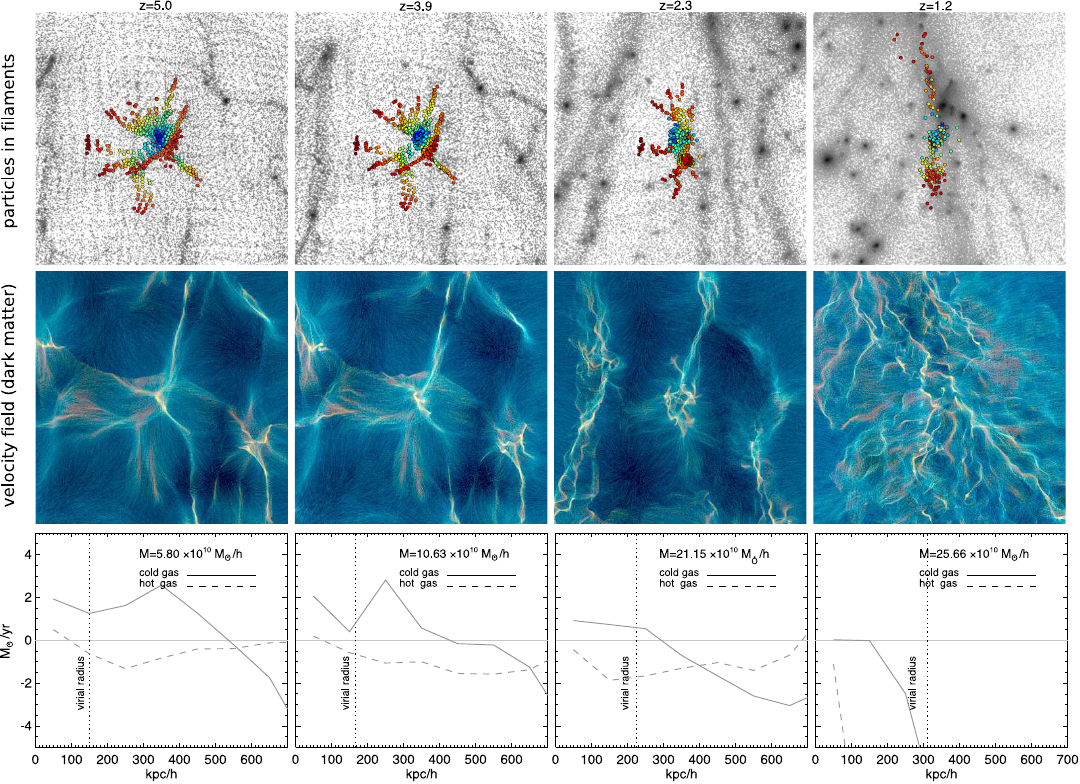}
\caption{{\bf Web Detachment of a $2.3\times10^{11}$\msun galaxy by accretion into a $2.4\times10^{14}$\msun cluster}. The cluster is located at the top outisde the figure.
From top to bottom we show:
{\bf i)} the projected density field centered at the galaxy on a 3 \mpc side box (gray background). Small colored circles correspond to individual particles inside the primordial filamentary web surrounding the galaxy at $z=5$ (see text for details). For reference the particles are colored according to their distance from the central galaxy at $z=5$. The same set of particles (keeping their original colors) is tracked and displayed at different times, showing the disrupting and detachment of the web of filaments as the galaxy is accreted by the cluster.
{\bf ii)} Coherent structures in the dark matter velocity field after removing bulk flows above scales of 250 h$^{-1}$kpc (see Fig. \ref{fig:box_advection} for details). Matter flows from dark-blue regions and accumulates in the light-yellow structures. At early times the velocity field is highly coherent and filamentary, almost laminar streams feed the central galaxy. After the web detachment event at $z\sim3$ the velocity field around the galaxy becomes chaotic and the filamentary streams are lost.
{\bf iii)} Cold gas accretion rate as a function of the distance from the galaxy and divided into ``cold" and ``hot" modes (see text for details). The vertical dotted line shows the virial radius of the galaxy.}
\label{fig:detachment_example}
\end{figure*}

\begin{figure}
  \centering
  \includegraphics[width=0.49\textwidth,angle=0.0]{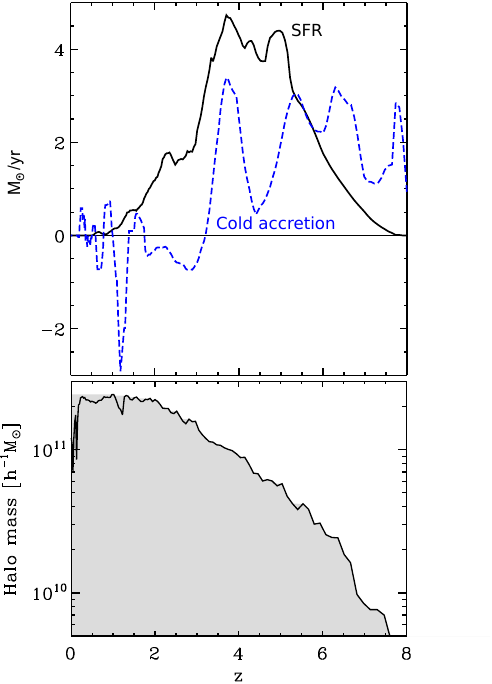}
  \caption{Top: cold gas accretion (measured inside a shell of radius  $r_{vir}-1.5r_{vir}$, dashed blue line) and star formation rate (solid black line) as a function of redshift,  centered in the halo presented in Fig. \ref{fig:detachment_example}.
  Bottom: mass accretion history. The black line is the raw MAH and the solid area the fixed MAH as described in the text.}\label{fig:detachment_example_SFR}
\end{figure}

%
\subsection{Filament detachment by accretion into a cluster}\label{sec:satellite_accretion}

The results presented in this section are based on a zoom simulation extracted from a 64\mpc box. The zoom region is centered in a $2.4\times10^{14}$\msun cluster and has a mass resolution of $2\times10^7$\msun per dark matter particle. The simulation includes gas with cooling and metal enrichment as well as stochastic star formation (see Appendix \ref{sec:simulations} for details on the simulation and Appendix \ref{app:cwd_examples} for other examples).

Figure \ref{fig:detachment_example} shows a satellite galaxy being accreted into a large cluster with present-time masses of $2 \times 10^{11}$ \msun and $2.4\times10^{14}$\msun respectively. This is an extreme example of interaction between a galaxy and its environment resulting in ram-pressure gas stripping \citep{Mori00,Quilis00},  clearly showing the effect of non-linear large-scale dynamics acting on a galaxy and its surrounding network of filaments.  As the galaxy begins to interact with the cluster we see changes in its surrounding matter configuration, mass accretion and star formation.

The top panels in Fig.  \ref{fig:detachment_example} show that at early times ($z=5$) the satellite galaxy is the central node of a network of thin coherent filaments. The filaments were identified by randomly placing test particles around the proto-galaxy, letting them follow the instantaneous velocity field interpolated using a Delaunay tessellation linear interpolation scheme \citep{Schaap00} and applying a high-pass filter at the scale of interests \citep{Aragon13}. The regions where the particles accumulate were used to construct a filament mask which was then used to tag particles inside filaments. The primordial velocity field is composed of coherent laminar streams of gas and dark matter that delineate the location of filaments in the matter distribution. It is worth noting that the (high-pass filtered) velocity field shows the filamentary structures in the cosmic web more clearly than the density field.

Gas accretion was computed in shells of thickness $\Delta R_{shell}$ around the center of the halo $\boldsymbol{r_h}$ with velocity $\boldsymbol{v_h}$ following the approach of \citet{Faucher11b} as follows:

\begin{equation}
\dot{M} = \sum_i m_i \frac{\boldsymbol{v_i} - \boldsymbol{v_h}}{\Delta R_{shell}} \cdot \frac{\boldsymbol{r_i} - \boldsymbol{r_h}}{\left | \boldsymbol{r_i} - \boldsymbol{r_h} \right |}
\end{equation}

\noindent where the sum is over the particles with position $\boldsymbol{r_i}$, mass $m_i$ and velocity $\boldsymbol{v_i}$ inside the shell.  Gas particles were segregated into hot and cold species by their instantaneous temperature at the threshold $T=10^{5} $K.  Figure \ref{fig:detachment_example_SFR} shows a steady increase in the star formation rate from $z=8$ up to its peak at $z \sim 4$. The cold gas accretion rate at $z=5$ is of the order of $\sim 1-3 $ M$_{\odot}$/yr. Around this time the galaxy begins to interact with the proto-cluster and this is reflected in both the cold gas accretion rate and the star formation rate which shows a rapid decline. After its first interaction with the proto-cluster, the galaxy suffers several episodes of gas loss. Around $z \sim 1$ the galaxy begins its accretion into the cluster. The higher density and strong tidal field associated to the cluster induces a mechanical stress in the filamentary web around the galaxy, stretching it until the point when it can no longer remain gravitationally attached (top-right panels Fig. \ref{fig:detachment_example}). After $z \sim 1$, even before the galaxy has been completely accreted by the cluster, there is no recognizable filamentary web around the galaxy. The local velocity field is chaotic and the lack of coherent gas streams connected to the galaxy prevents the accretion of cold gas (lower panels in Fig. \ref{fig:detachment_example}), ending its efficient star formation phase.
From this point the galaxy can not longer accrete cold gas,  depletes its internal reserves of gas and stops forming stars (see Fig. \ref{fig:detachment_example_SFR}).

\begin{figure*}
  \centering
  \includegraphics[width=0.8\textwidth,angle=0.0]{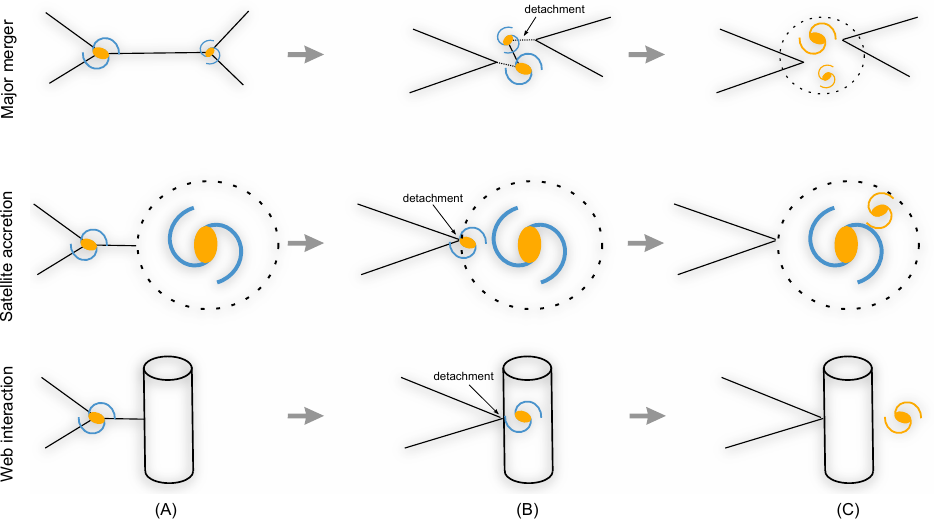}
  \caption{Three toy models of CWD events. From top to bottom: major merger, satellite accretion and cosmic web crossing. These events can be considered different cases of halo accretion.  The basic CWD event occurs as follows:  A) Initially a galaxy accretes cold gas via their web of filaments.  B) The galaxy then suffers a violent detachment event form either a merger, accretion into a larger halo or accretion/crossing of a large-scale structure (here a filament).}\label{fig:toy_model}
\end{figure*}

%
\subsection{Cosmic Web Detachment}\label{sec:detachment_events}

The star formation regulation and subsequent quenching presented in Fig. \ref{fig:detachment_example} is a purely mechanical/gravitational process, requiring no internal feedback from AGN. The rupture in the cold gas accretion channel is sufficient to stop star formation and, in that sense, can be considered a more fundamental process than internal feedback. Subsequent internal feedback from AGN could further prevents gas from entering the galaxy and/or cooling, removing any residual star formation \citep{DiMatteo05,Springel05,Best05}. Figure \ref{fig:detachment_example} highlights the key role of primordial filaments in the star forming history of galaxies and the decisive role of the large-scale matter distribution around a galaxy. The time when a galaxy {\it detaches} from its primordial filamentary web marks a turning point in its star formation history.

The gas stripping event by ram pressure and tidal interactions shown in Fig. \ref{fig:detachment_example} is followed by quenching but it is not the fundamental process responsible for ending star formation in the satellite galaxy. The actual process that induces quenching is the removal of the feeding filaments (via ram pressure) connected to the galaxy. Without them there is no way to effectively accrete cold gas to form stars and the galaxy has to rely on internal reserves. This process which we call \textit{Cosmic Web Detachment} (CWD) is fundamentally a starvation process \citep{Peng15} triggered by non-linear interactions between galaxies their environment.

%
\subsection{Basic Cosmic Web Detachment events}\label{sec:detachment_events}

In this section we describe other non-linear processes that can be described as CWD mechanisms in their ability to trigger star formation quenching. In the rest of this paper we use the terms CWD and web detachment interchangeably.

Figure \ref{fig:toy_model} shows cartoons of basic processes that induce CWD: i) major merger, ii) satellite accretion and iii) cosmic web infall/crossing. As we discuss in the next section, in these processes, matter streams cross on scales larger than the scales of each halo. In all cases presented in Fig. \ref{fig:toy_model} the initially star-forming galaxy is connected to a web of primordial filaments from which it accretes gas. Subsequent non-linear interactions between the galaxy and a nearby structure (other galaxy, a group or a large filament/wall) detaches the galaxy from its web of feeding filaments. Following the CWD event, the galaxy can no longer accrete cold gas and star formation stops. Note that in the case of a major merger the newly merged halo could remain loosely connected to its primordial filaments. However, these filaments do no penetrate to the galaxy's core \citep{Dekel09}. A similar situation occurs in the case of satellite accretion. After web detachment the filaments originally attached to the satellite could remain connected to its parent halo but are not able to inject gas into the satellite galaxy. CWD by interaction with filaments and walls may be an important quenching mechanism for dwarf galaxies \citep{Benitez13}.

\begin{figure}
  \centering
  \includegraphics[width=0.35\textwidth,angle=0.0]{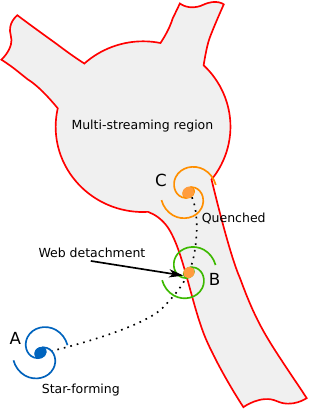}
  \caption{Assigning a detachment time to a galaxy. The red contour encloses regions that have been identified as multi-streaming using our Lagrangian Sheet prescription (see text for details), in this case a cluster and three large filaments connected to it. A galaxy is first detected inside a void (A) at which point it is assumed to be forming stars until it enters a multi-streaming region (B) and from this point it becomes quenched (C). Note that we consider that SF ends after a galaxy enters a multi streaming region regardless of any possible gas accretion from the new environment.}\label{fig:detachment_assignmnet_toy}
\end{figure}

%
\section{A multistreaming formalism for cosmic web Interactions}\label{sec:multistreams}

The CWD model can be used to predict to first order the star formation history of a galaxy, and in particular its quenching time. Identifying CWD events in individual halos is a complex and computationally expensive task that must be done on a case by case basis.  We instead propose to use the dynamical state of the cosmic web to identify the \textit{most likely regions} where CWD events occur. The underlying assumption here is that CWD events are the result of non-linear interactions between galaxies and the cosmic web and that CWD alone can explain the bulk of star formation quenching.

We start by noting that in the three cases presented in Fig.  \ref{fig:toy_model} in order for CWD to occur the trajectories of a galaxy and a nearby structure such as another galaxy, group, filament or wall must cross. As a first approximation we can then associate web detachment events with shell-crossings at some meaningful scale. Consider for instance a major merger event (case A in Fig. \ref{fig:toy_model}). The two merging galaxies correspond each to an initial Lagrangian volume that collapsed and became non-linear and the merger is the union of the two Lagrangian clouds. We can choose a sufficiently large smoothing scale in Lagrangian space where the patch corresponding to each of the two proto-galaxy clouds will  shell-cross at merger. The same principle can be applied to interactions between galaxies and clusters, filaments or walls.

The use of a Lagrangian smoothing to identify shell-crossing events at a scale of interests introduces the need to define such scale:

\begin{equation}
	R_{\rm WD} = \left( \frac{3}{4\pi\rho_m} M_{\rm WD}  \right)^{1/3}.
\label{eqn:rdetach}
\end{equation}

\noindent Where $R_{\rm WD}$ is the scale defining web detachment events, $M_{\rm WD}$ is its equivalent mass and $\rho_m$ is the mean density of the Universe. By smoothing the initial power spectrum at this scale we produce a universe lacking structures less massive than $M_{\rm WD}$ \citep{Aragon10b, Aragon13}, \citep[see also][]{Suhhonenko11}. We then have:

\begin{equation}
	P_{\rm WD}(k) = T^2(k) \; P_{CDM},
\end{equation}

\noindent where the transfer function $T(k)$ is a filter corresponding to a spherical top-hat of radius $R_{\rm WD}$.
In principle we do not know the scale $R_{\rm WD}$, instead of trying to derive a characteristic scale (or range of scales) we use $R_{\rm WD}$ corresponding to the upper halo mass threshold where star formation quenching occurs $M_{\rm WD} \sim 10^{12}$ \msun \citep{Cattaneo06,Birnboim07} as a natural scale for star formation quenching. By doing so we assume that the origin of this mass threshold is not internal processes driving quenching (which could still play an important role) but the dynamical state of the cosmic web around halos of that mass. This gives $R_{\rm WD} \simeq 2\, h^{-1}\, {\rm Mpc}$.

\begin{figure*}
  \centering
  \includegraphics[width=0.32\textwidth,angle=0.0]{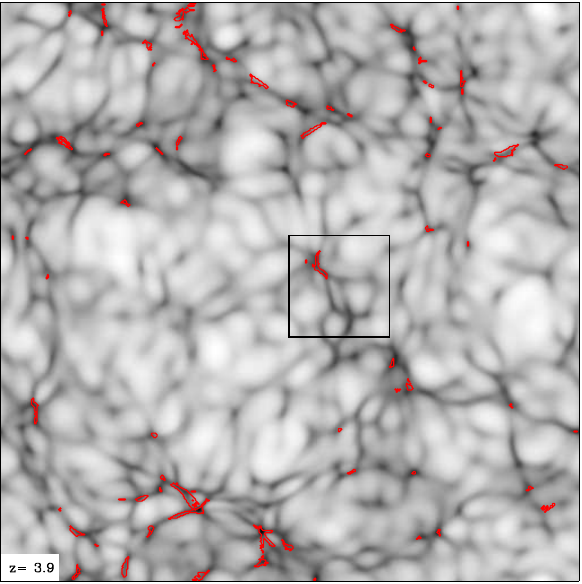}
  \includegraphics[width=0.32\textwidth,angle=0.0]{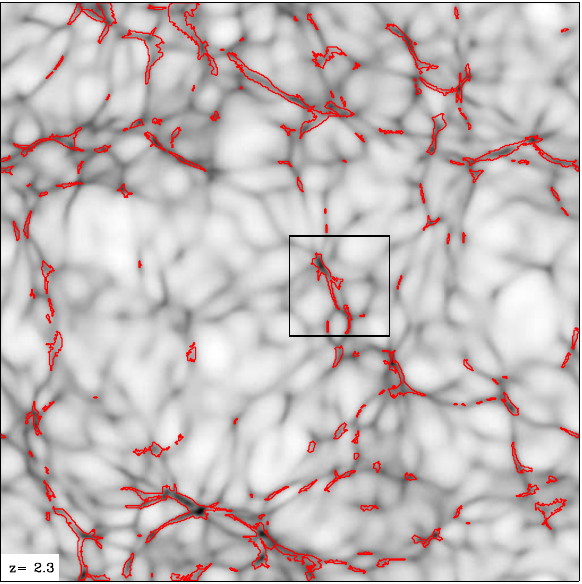}
  \includegraphics[width=0.32\textwidth,angle=0.0]{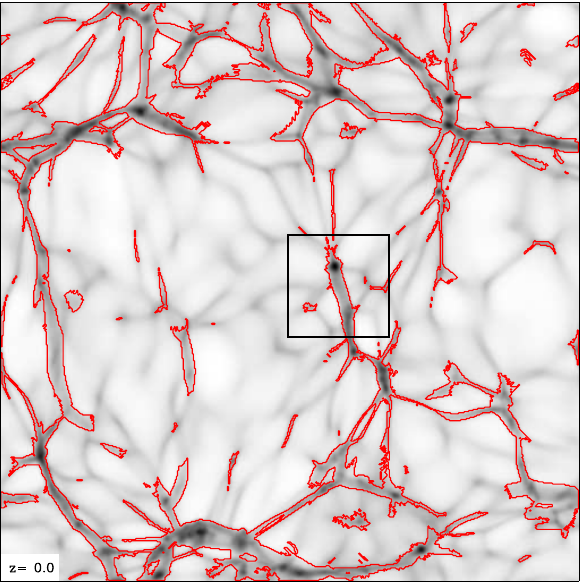}
  \includegraphics[width=0.32\textwidth,angle=0.0]{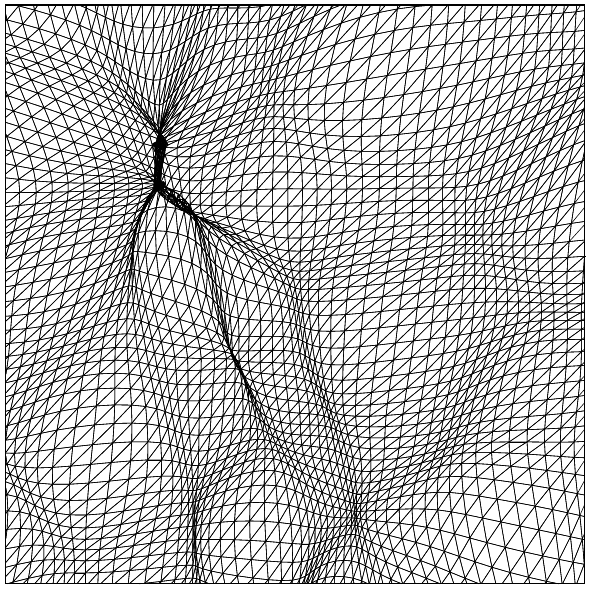}
  \includegraphics[width=0.32\textwidth,angle=0.0]{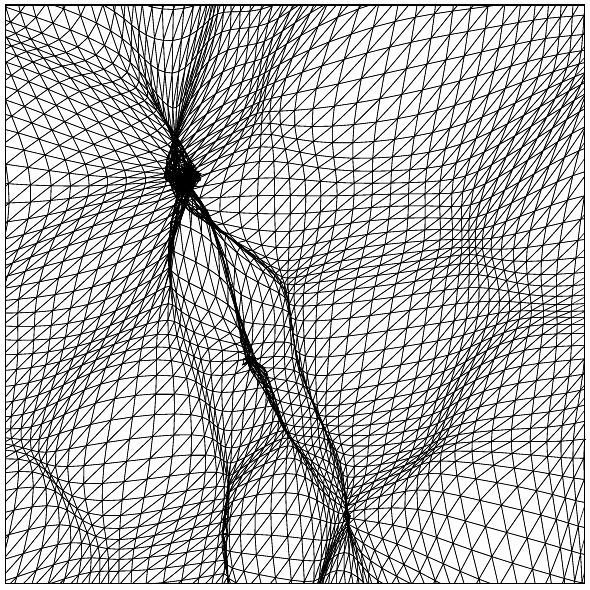}
  \includegraphics[width=0.32\textwidth,angle=0.0]{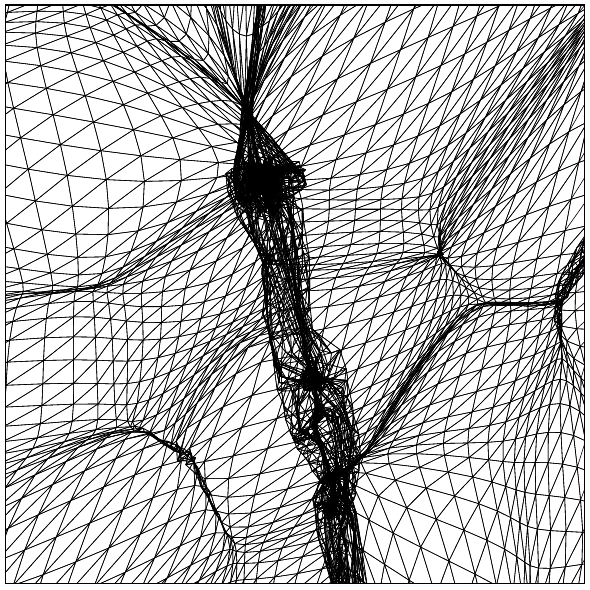}
  \caption{Evolution of shell-crossing regions in a 2\mpc thick slice across the simulation box. Top panels: The gray-scale background represents the projected smoothed density field.  The red contours enclose the regions where shell-crossing has occurred which correspond to CWD regions. Bottom panes: the Lagrangian tessellation used to identify multi-streaming regions inside a zoomed region (black box in top panels). For clarity we show a 2D tessellation at the (Lagrangian) center of the slice shown in the top panels. The actual computation described in the text was performed in 3D.}
\label{fig:multistreams_grid}
\end{figure*}

%
\subsection{Identification of multi-streaming regions}

Following the previous discussion we identify CWD regions in N-body simulations as places where there has been at least one shell-crossing in the Lagrangian smoothed field and assume that halos entering those regions undergo instantaneous CWD. In this approach we assume that star-forming gas is accreted only via primordial filaments and therefore ignore any 
subsequent gas accretion coming from the halo's new environment. 
Figure \ref{fig:detachment_assignmnet_toy} shows a representation of a galaxy being tracked from its star forming phase to the point of web detachment and its subsequent quenched state. In practice we use the most massive progenitor line to track halos across time (see Appendix \ref{sec:M_star}).
In order to identify multi-streaming regions from the particle distribution we use a Lagrangian Sheet approach where the original particle distribution is used to define simplex volume elements \citep{Shandarin12,Abel12}. We modify the original implementation by following Lagrangian volume elements defined by the particle's initial positions and assigning them a constant unitary density. As the simulation evolves adjacent Lagrangian volume elements in high density regions cross each other overlapping, producing \textit{multi-streams} where more than one Lagrangian volume element occupies the same space in Eulerian coordinates \citep{Shandarin11,Shandarin12,Abel12}. The complexity of the multi-streaming regions directly reflect their evolutionary stage as seen also in the {\scshape origami} cosmic web classification scheme \citep{Falck12,Neyrinck12,Neyrinck15,Falck15}


Figure \ref{fig:multistreams_grid} shows the emergence of multi-streaming regions associated with CWD events at three different times. At $z=3.9$  the first multi-streaming regions begin to form around the location of large and dense peaks. At $z=2.3$ several multi-streaming regions have collapsed and megaparsec-scale filamentary features begin to appear. At the nodes of the multi-streaming regions we find the precursors of the large clusters seen at the present time (see Fig. \ref{fig:multistreams_2} for a more detailed time evolution). Note the apparent lack of evolution of the CWD regions after $z\sim1$ in Fig. \ref{fig:multistreams_2}, reflecting the role of the cosmological constant $\Lambda$ in the development of the cosmic web at large scales.
This is also seen in Fig. \ref{fig:multistream_derivative} which shows that the rate of growth of multi-streaming regions begins to slow down around $z \sim 2$. We can see multi-streaming regions even inside voids although they arise at later times compared to denser environments.

By associating multi-streaming regions with CWD processes we define a link between large-scale dynamics and halo gas accretion. Compared to other approaches that describe galaxies as isolated objects having environmental processes entering only indirectly (e.g. via merger trees) our method considers cosmic environment as an integral component in galaxy evolution. This link between galaxies and their environment is the key to the ability of our multi-streaming prescription to successfully reproduce a range of environmental effects.  In the rest of the paper the use the terms multi-streaming and CWD regions interchangeably.

\begin{figure}
  \centering
  \includegraphics[width=0.49\textwidth,angle=0.0]{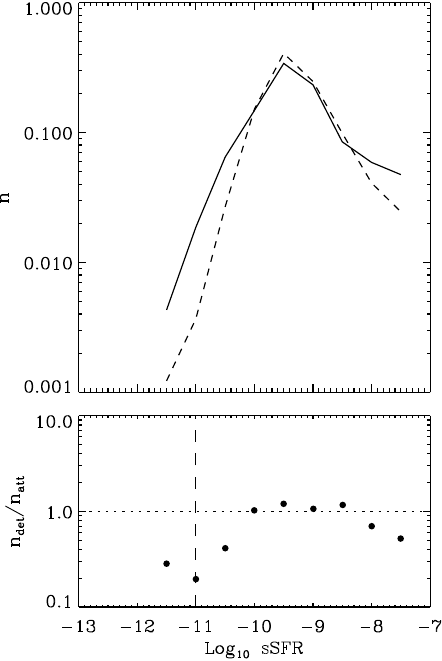}
  \caption{Top: normalized distribution of sSFR of halos inside CWD regions (solid line) and outside CWD regions (dashed line). halos inside CWD regions were restricted to those have entered multi-streaming regions more than 3 Gyr ago. Bottom:  ratio between the two distributions. The vertical line indicates the sSFR threshold used in the literature to divide quenched from main sequence galaxies. }\label{fig:SF_in_CWD_regions}
\end{figure}

%
\subsection{Quenching fraction inside multi-streaming regions}


In Fig. \ref{fig:SF_in_CWD_regions}, we show the ratio of specific star formation rate (sSFR) for galaxies inside and outside multi-stream regions. The sSFR was computed from the 32Mpc box hydrodynamic simulation described in Appendix \ref{sec:simulations}. At sSFR $\sim 10^{-11}$ yr$^{-1}$, there are five times more quenched galaxies inside CWD than outside. Classical quenching mechanisms involving violent interactions (e.g. ram-pressure stripping, harassment, satellite accretion) occur in multi-stream regions more than in more quiescent single-stream regions. The effect of non-linear interaction on sSFR is clear from Fig. \ref{fig:SF_in_CWD_regions} and provides evidence (although circumstantial) of the validity of our approach.

\section{Multi-streaming regions and halo history}

In this section we study the history of halos in order to gain insight on the type of processes occurring when a halo enters a multi-streaming region and how this could affect their star formation. The relation between halo mass accretion history and star formation is, while not completely understood yet, well established \citep{Dekel09,Mutch13,Watson15,Behroozi15,Aldo16}. The analysis we present here shows a direct link between matter accretion and the dynamical state of the local cosmic web and by doing so provides an indirect link to star formation.

\begin{figure}
  \centering
  \includegraphics[width=0.49\textwidth,angle=0.0]{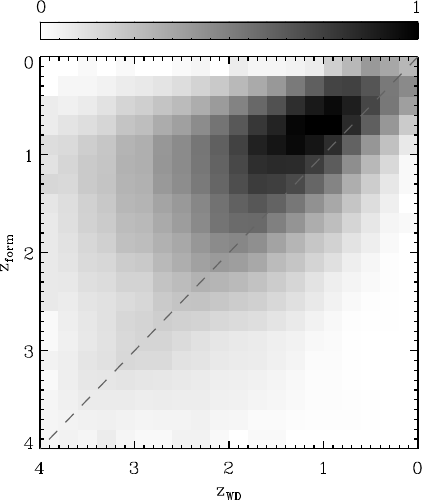}
  \caption{Web detachment redshift $z_{WD}$ vs. halo formation redshift $z_{form}$ (as a two-dimensional histogram) for halos with mass in the range $M >10^{10}$ \msun (see text for details). They grey background indicate the counts inside each two-dimensional bin. The dashed line and color bar (normalized to 1 at maximum value of the 2D histogram) are shown as reference.} \label{fig:t_det_vs_t_for}
\end{figure}

%
\subsection{Formation time and Web Detachment}\label{sec:formation_time}

Figure \ref{fig:t_det_vs_t_for}  shows a comparison between web detachment redshift $z_{\rm WD}$ (i.e. the redshift when the halo became web-detached) and halo formation redshift $z_{\rm form}$ for halos with masses $M >10^{10}$ \msun. We used the common definition of formation time as the time when the mass accretion history (MAH) of a halo changes its slope \citep{Wechsler02}. This was computed for each halo's mass accretion by fitting a two-component model:

\begin{equation}\label{eq:formation_time}
  M = \left\{
    \begin{array}{rl}
      M_0 + m_1(z-z_0) & \text{if } z < z_0,\\
      M_0 + m_2(z-z_0)  & \text{if } z >= z_0.
    \end{array} \right.
\end{equation}

\noindent Where $M_0$ and $z_0$ are the mass accretion rate and redshift at formation time respectively,  and $m_{1,2}$ indicate the difference in the mass accretion rates. Halo mass loss, typical of sub-halos, is corrected by not allowing the halo's MAH to decrease. In this approach a halo being mass-stripped will retain its mass before stripping and the formation time of the sub-halo will usually be the time of mass stripping.

Figure \ref{fig:t_det_vs_t_for} shows a clear correlation between $z_{\rm WD}$ and $z_{\rm form}$ with a small dispersion in redshift in the interval $0 < z < 2$ after which point the dispersion increases. halos with recent web detachment times also have a recent formation time. halos with an early web detachment show a less marked correlation with formation time. This may be the result of having entered multi-streaming environments at early times which could result in complex mass accretion histories that are not easily characterized by a single property such as formation time.

An important aspect of this diagram is that formation time occurs, in general, after the time of web detachment ($z_{\rm WD} > z_{\rm form}$), in agreement with the picture of halos first entering web-detachment regions that later affect their mass/gas accretion rates. This can be explained as galaxies can become web-detached by, for instance, entering a large filament and can in principle continue accreting mass (dark matter and baryons). If the galaxy subsequently enters a large group this will be reflected in its MAH and will result in a later formation time.

\begin{figure}
  \centering
  \includegraphics[width=0.49\textwidth,angle=0.0]{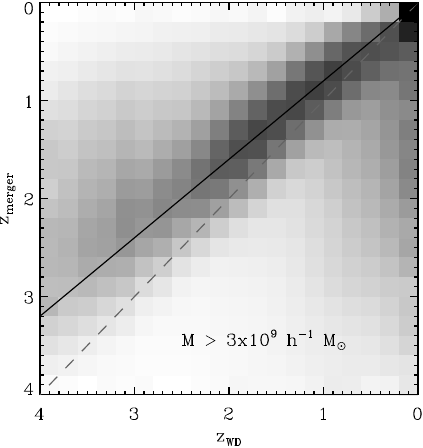}
  \caption{Comparison between web detachment redshift $z_{\scriptsize\textrm{WD}}$ and the redshift of the major merger event $z_{\scriptsize\textrm{merger}}$ (as a two-dimensional histogram) . The sample is constrained to halos that had at least one major merger event with a mass ratio between the two merging halos of $r_{\scriptsize\textrm{merger}} > 1.5$. The thick black line indicating the general trend is shown as reference.} \label{fig:t_det_vs_t_merger}
\end{figure}

%
\subsection{Major merger events}

Major merger events have a significant effect in the mass accretion history of halos. Being highly non-linear large-scale events they provide an efficient CWD mechanism and we should expect a close relation between major mergers and web detachment times. Figure \ref{fig:t_det_vs_t_merger} shows a comparison between the time of web detachment $z_{\scriptsize\textrm{WD}}$ and the time of the first major merger $z_{\scriptsize\textrm{merger}}$. Major mergers were defined as those with a mass ratio of $M_1 / M_2 = 1.5$.

Figure \ref{fig:t_det_vs_t_merger} shows a clear correlation between web detachment events and major mergers. The correlation is even tighter than in the case of formation time (Sec. \ref{sec:formation_time}) and extends to the range $0<z<2$. halos tend to enter web detachment regions earlier than the time of their largest major merger event. This trend is more pronounced at high redshift ($z<2$). While the dispersion between $z_{\scriptsize\textrm{WD}}$ and  $z_{\scriptsize\textrm{merger}}$ increases, the condition  $z_{\scriptsize\textrm{WD}} < z_{\scriptsize\textrm{merger}}$ remains valid at all times. An approximate linear fit gives $z_{\scriptsize\textrm{merger}} \simeq 0.8 \; z_{\scriptsize\textrm{WD}}$. At $z \sim 0$ there is a small population of halos that show no correlation between $z_{\scriptsize\textrm{WD}}$ and $z_{\scriptsize\textrm{merger}}$. The origin of this small population could be purely numerical and will be investigated in future work.

\begin{figure*}
  \centering
  \includegraphics[width=0.99\textwidth,angle=0.0]{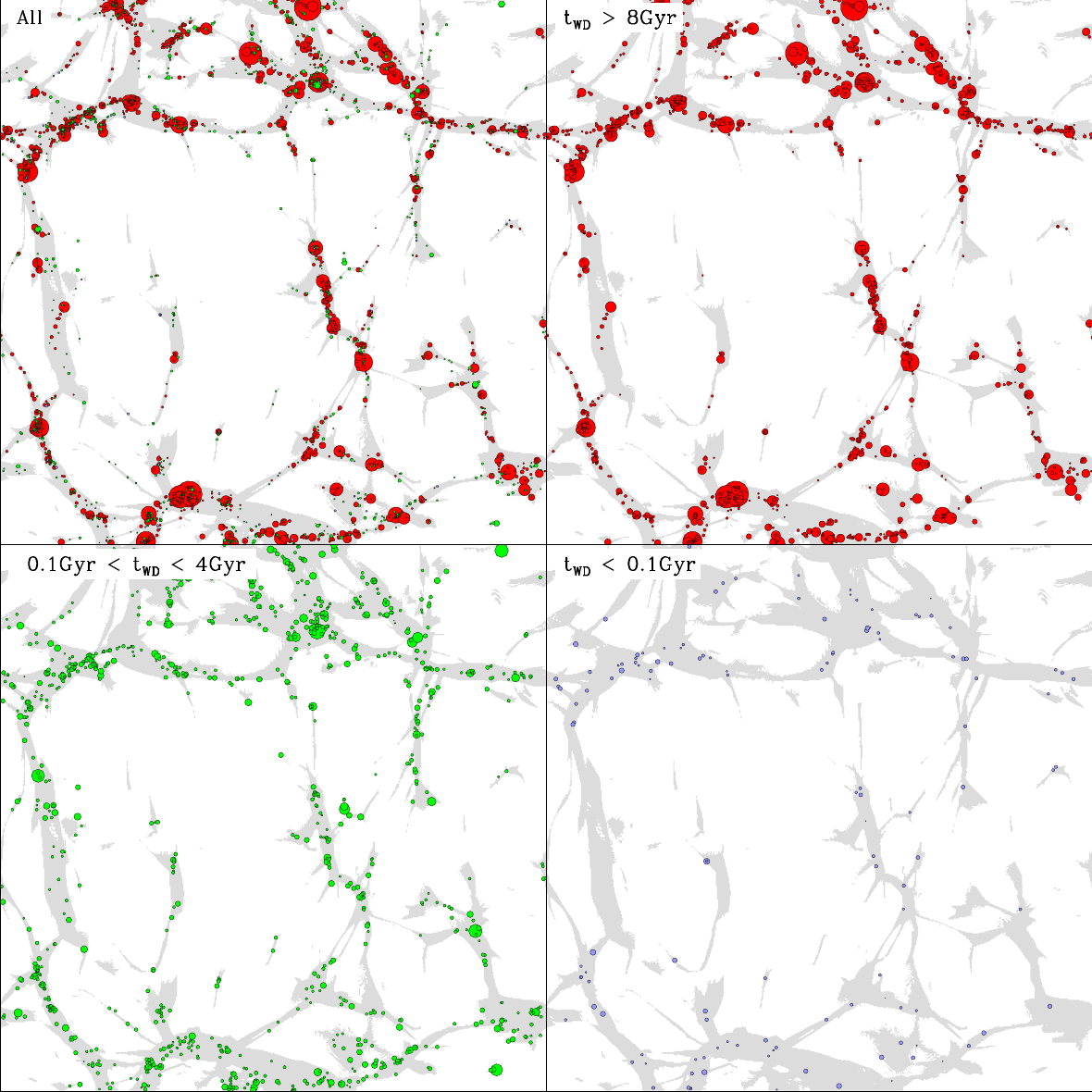}
  \caption{Spatial distribution of halos divided by their web detachment time $t_{WD}$. The circles represent SubFind halos scaled with their viral radius. Grey areas indicate multi-streaming regions. For clarity halos in the bottom panels are scaled with two times their viral radius. From top-left in clockwise order: (All): Distribution of all halos regardless of their detachment time, ($t_{WD}>8$Gyr): early-detachment with no star formation,  ($t_{WD}<4$Gyr): Intermediate detachment time and ($t_{WD}<0.1$Gyr): star-forming halos with very recent detachment. Note the strong correlation between detachment time and halo mass/radius and the different level of clustering as function of $t_{WD}$. }\label{fig:detachment_spatial_distribution}
\end{figure*}

%
\section{Spatial and temporal distribution of CWD events}

We now explore the spatial distribution of halos segregated by CWD times. Figure \ref{fig:detachment_spatial_distribution} shows halos that have been web-detached at early times (more than 8 Gyr ago), halos that have undergone late web detachment (less than 0.1 Gyr ago) and one intermediate case. The CWD model gives a physical underpinning to the traditional Hubble morphological division of galaxies into ``early'' and ``late'' types. Early web-detached halos have quenched their star formation long time ago, while late web-detached halos are either still star forming or in the process of quenching. There is a clear trend between web detachment time and halo radius/mass. halos with early $t_{\rm WD}$ tend to be more massive and more highly clustered indicating that galaxies that entered CWD regions early mark the position of dense peaks in the initial density field. The relation between halo mass and its location inside large dense peaks naturally results in a correlation between web detachment times and spatial location. There seems to be a combination of halos with intermediate and late web detachment times at the interior of the central void in Fig.  \ref{fig:detachment_spatial_distribution}, hinting a relation to the void phenomenon \citep{Peebles01}.

\begin{figure}
  \centering
  \includegraphics[width=0.49\textwidth,angle=0.0]{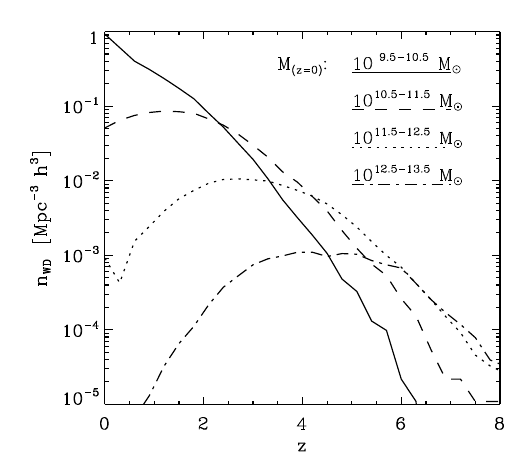}
  \caption{Distribution of web detachment times as function of time divided in four halo mass bins. Massive halos get web-detached at early times while dwarf galaxies web-detach very recently. The distribution of $t_{WD}$ for dwarf galaxies shows that they are the population that is currently undergoing CWD but has not yet reached its peak.}\label{fig:mass_at_z0_vs_detach_distribution}
\end{figure}

%
\subsection{Anti-hierarchical star formation and  CWD}

According to the hierarchical scenario of galaxy formation small galaxies form first (and therefore their population should be composed of old stars) while massive galaxies form later (and as such they should be forming stars at the present time). What we observe is the opposite: massive galaxies are no longer producing stars while low-mass galaxies are actively forming new stars. This apparent anti-hierarchical behavior of the star formation history of galaxies is commonly referred to as ``downsizing" \citep{Cowie96,Heavens04,Thomas05}.

Figure \ref{fig:mass_at_z0_vs_detach_distribution} shows the distribution of web detachment times for several halo mass ranges  (computed at the present time). There is a clear dependence between web detachment time and halo mass. The most relevant aspect of this figure is the shift in the peak from high to low redshift with decreasing halo mass. This indicates that massive halos ($10^{11.5-12.5}$\msun) already became web-detached at early times ($z \sim 2$) while low-mass halos ($10^{9.5-10.5}$\msun) are still attached to their filament network or are just becoming web-detached. The peak in the web detachment time for the lowest mass bin in Fig. \ref{fig:mass_at_z0_vs_detach_distribution} has not been reached at the present time.

The anti-hierarchical trend in Fig.  \ref{fig:mass_at_z0_vs_detach_distribution} of high-mass halos being quenched while low-mass halos being star forming is tantalizingly similar to  the so-called \textit{downsizing}  \citep{Cowie96,Heavens04,Thomas05}. In the context of the CWD model, the key to downsizing is the realization that low-mass star-forming galaxies are found mostly in isolated environments and therefore their surrounding cosmic web has remained unperturbed for most of the history of the Universe. Low-mass isolated galaxies can therefore experience a slow but steady gas accretion (see Fig. 11 in \citet{Aragon13}). On the other hand, massive galaxies form in dense environments where there is a higher rate of web detachment events due to the non-linear dynamics characteristic of such environments. The net effect is the observed anti-hierarchical star formation history.

%
\section{Testing CWD with observations}

In this section we reproduce and, to some extent, interpret several observations in the context of the CWD model. We use a minimum of assumptions for galaxy properties (star formation, color and quenching state) in order to emphasize the effect of CWD and we clearly indicate when we do so.

\begin{figure}
  \centering
  \includegraphics[width=0.49\textwidth,angle=0.0]{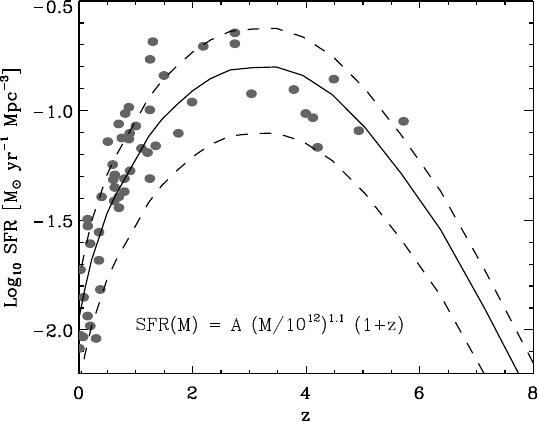}
  \caption{Cosmic star formation rate computed using multi-streams as proxy for web detachment and a simple model for star formation as function of halo mass (eq. \ref{eq:SFR_mass}, also shown in the figure). The three curves from top to bottom correspond to values of the constant $A=150,100,50$ respectively. The gray circles are observed data points taken from \citet{Hopkins06} (error bars omitted for clarity).}\label{fig:madau_plot}
\end{figure}

%
\subsection{The cosmic star formation rate history}

In order to assign star formation rates (SFR) to dark matter halos we start by assuming a simple relation between halo mass before CWD and star formation rate. Based on the fit provided by \citet{Dekel09} we assume an semi-linear dependence between halo mass and gas accretion/star-formation rate as:

\begin{equation}\label{eq:SFR_mass}
  SFR = A \left( \frac{M}{10^{12}}  \right)^{\alpha} \; (1+z)
\end{equation}

\noindent Where the constant $A$ summarizes complex physical processes (which in principle can also depend on halo mass and redshift). At the present time $A$ corresponds to the SFR a halo with a mass of $10^{12}$\msun would experience if it  remained connected to its network of primordial filaments, and the value of the exponent $\alpha = 1.15$ is derived from halo growth rates \citep{Neistein06,Birnboim07}. We assume a SFR given by equation \ref{eq:SFR_mass} before CWD and an instantaneous star formation quenching after the CWD event. This behavior is analogous to a constant star formation rate and a subsequent decline \citep{Blanton06,Reddy12}.

\begin{figure}
  \centering
  \includegraphics[width=0.49\textwidth,angle=0.0]{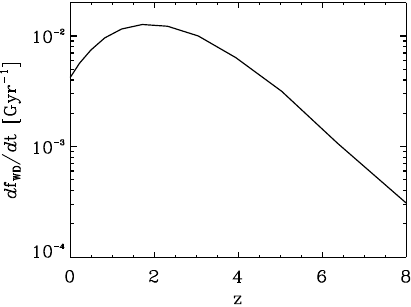}
  \caption{Rate of change of the fraction of volume inside multistreaming regions as a function of time. The peak in the curve indicates the time when more volume in the cosmic web became non-linear at $z \sim 2$.}
  \label{fig:multistream_derivative}
\end{figure}

Figure \ref{fig:madau_plot} shows the cosmic star formation rate history (SFRH) computed by integrating the star formation from all halos which have not yet suffered CWD at a given redshift and assigning star formation rates as a function of halo mass following equation \ref{eq:SFR_mass}. The CWD model can reproduce the general shape of the SFRH curve, in particular its peak at $z \sim 3$ and decline to its present value more than two orders of magnitude lower \citep{Heavens04}. It is interesting to note that the origin in the peak of the SFRH coincides to the peak of CWD events density of massive halos in the mass range $10^{11.5-12.5}$\msun with a peak around $z \sim 2-3$  (Fig. \ref{fig:mass_at_z0_vs_detach_distribution}) indicating the time when the rate of web detachment events was maximum and the (until that point)  increasing star formation rate began to slow down (Fig. \ref{fig:multistream_derivative}).

The peak in the SFR at $z\sim2$ corresponds to just 3.2 Gyr after the Big Bang. At this point the megaparsec-scale structures in the cosmic web are still in full development and will not reach its present-time large-scale features until $z \sim 1$. Figure \ref{fig:multistream_derivative} shows the rate of change of the fraction of volume inside multi-streaming regions as a function of time. The close relation between the rate at which regions in the Universe become non-linear, web-detachment times and the peak in the cosmic star formation rate suggests a causal link between these processes.
Note that the peak in both the density of web detachment events (Fig. \ref{fig:mass_at_z0_vs_detach_distribution}) and the rate of change of volume inside multi-streaming regions (Fig. \ref{fig:multistream_derivative}) is determined only from large-scale dynamics and therefore the resulting peak in the SFH of galaxies can be explained, in the CWD model,  by environmental effects.

\begin{figure}
  \centering
  \includegraphics[width=0.49\textwidth,angle=0.0]{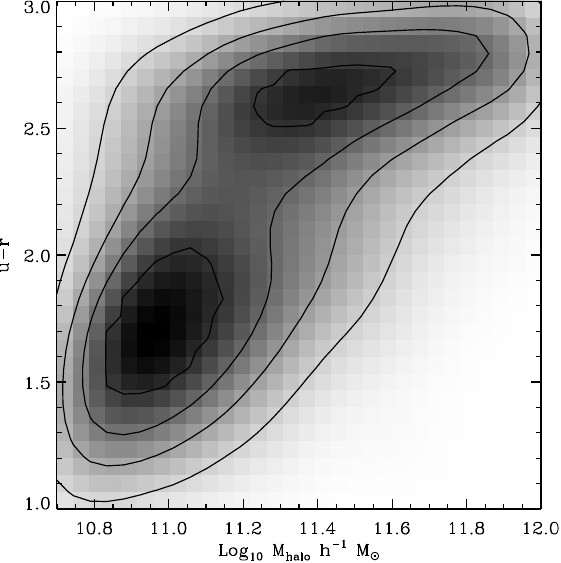}
  \caption{Galaxy color bimodality in the mass-color diagram. The grey background corresponds to the two-dimensional mass-color distribution. We show three contours to highlight the shape of the two regions. The distribution was smoothed for clarity.}\label{fig:color_bimodality}
\end{figure}

%
\subsection{The bimodality in the color distribution of galaxies}\label{sec:sdss}

Figure \ref{fig:color_bimodality} shows the distribution of galaxy colors with halo mass. In order to assign colors to halos we assumed a monotonic relation between a halo's detachment time $t_{\rm WD}$  and the color of its hosted galaxy: halos with an early CWD time host galaxies with an older/redder stellar population compared to galaxies hosted in halos with a later CWD time. 

Galaxy colors were obtained from a template of galaxies selected from the Sloan Digital Sky Survey (see Appendix \ref{app:sdss}). These galaxies were \textit{matched} to halos in our simulation using a similar matching procedure as done by \citet{Hearing13} but using $t_{\rm WD}$ instead of their empirical \textit{age}. Both halos and galaxies were sorted by $t_{\rm WD}$  and $u-r$ color respectively and matched one to one.

Figure \ref{fig:color_bimodality}  shows the familiar bimodality in the color distribution. The color bimodality is an intrinsic property of the SDSS galaxies and not a result of $t_{\rm WD}$. The ability of our CWD implementation to reproduce the mass-color diagram, i.e. the correct location of the blue and red peaks and the extended flattened shape of the red peak into the high mass region, comes from the mapping between a halo mass and its $t_{\rm WD}$ (see Fig. \ref{fig:mass_at_z0_vs_detach_distribution}). 

The region between the blue and red clouds corresponds to galaxies in the green valley. In the CWD framework galaxies in the green valley suffered web detachment around $t_{WD} \sim 8.5$Gyr. However, web detachment is just one of several possible paths as green valley galaxies are most likely a composite population of galaxies in the process of quenching and already quenched galaxies that are undergoing a new star forming phase.

\begin{figure}
  \centering
  \includegraphics[width=0.49\textwidth,angle=0.0]{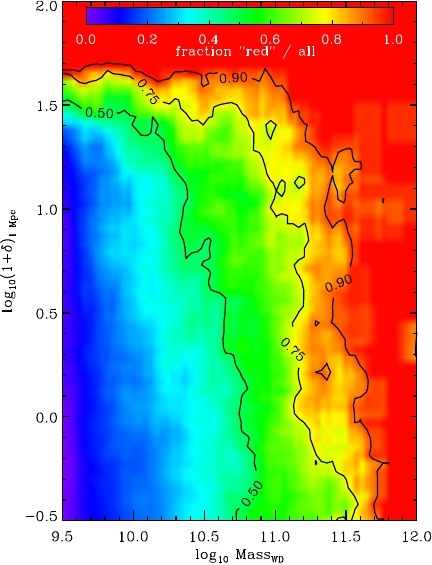}
  \caption{Fraction of red galaxies with respect to the total population as a function of halo mass at the time of detachment and local density (see text for details). }\label{fig:Peng10}
\end{figure}

%
\subsection{Star formation as function of mass and density}

Figure \ref{fig:Peng10} shows the fraction of red galaxies (here defined as galaxies with $t_{WD} > 10$ Gyr) as a function of halo mass and local density. This plot is based on Fig. 6 in \citet{Peng10} where they analyzed a sample of galaxies from the SDSS. The local density was computed using a top-hat window of radius $r=1 h^{-1}$Mpc  centered on each halo and counting the dark matter particles inside the window. It is remarkable that our simple implementation of CWD reproduces to a great degree the relation between fraction of red galaxies, their mass and density. For a given halo mass the fraction of red galaxies in low-density environments is lower than for dense environments and for a given density the fraction increases with halo mass. This is interpreted by \citet{Peng10} as suggesting two separate mechanisms of "environmental quenching" and "mass quenching". However, this relation must be taken with caution since halo mass and density are correlated so the interpretation of two different processes is not straightforward. In the CWD framework the increase in density is associated to an increase in the environment's complexity leading to CWD events and quenching. On the other hand, we expect massive halos to populate dense environments so the increase in the red fraction with mass could also be interpreted as a consequence of the increase in density.

The CWD framework also provides a simple explanation for the lack of quenched dwarf population in voids \citep{Geha12}. Figure \ref{fig:Peng10} shows that the fraction of web-detached halos, compared to the general population, is the lowest for dwarf halos in low density environments. Dwarf galaxies in the low-dense field (i.e. voids and walls) remain connected to their network of filaments. This is seen in our multi-streaming formalism as halos that have not entered yet into non-linear multi-streaming regions. We should therefore expect to find star-forming dwarf galaxies at the present time in voids and walls.

%
\section{Discussion}

%
\subsection{Searching for primordial filaments in voids and walls}
One aspect of CWD worth exploring is its relation with the different elements of the cosmic web. In particular galaxies in voids and walls, given their dynamically young environment, could still be connected to their web of primordial filaments and thus be star forming \citep{Ceccarelli08,Liu15,Moorman16}. The super-Hubble expansion in such environments \citep{Aragon11} can act as a regulator for gas accretion resulting in a slow and steady star formation \citep{Ricciardelli14,Das15}. Present-time voids, and to a lesser extent walls, are ideal environments to search for primordial filaments. Recent deep observations have already found examples of thin filaments connected to void galaxies \citep{Kreckel12,Beygu13,Alpaslan14}.

%
\subsection{Gas accretion in the Milky Way}
The Milky Way has evidence of gas supply from cosmological origin which, given the local planar LSS geometry \citep{Tully87}, we can assume  to be restricted to the plane of the Local Sheet. The Milky Way galaxy is currently forming $\sim 1$ \msun per year. At this rate the gas reserves in the Milky Way would deplete within a timescale shorter than Hubble time \citep{Noeske07,Salim07} but what we observe is a relatively constant  star formation rate and Deuterium abundance in the solar neighborhood in the last several Gyr \citep{Binney00}. The low SFR of the MW, given its mass, could be due to super-Hubble regulation or perhaps interactions with the Local Sheet which may have affected its feeding filaments. Another example of a galaxy with possible cosmological gas accretion is NGC 6946, the ``Fireworks galaxy" featuring a high supernovae rate (9 per century) and a possible detection of filamentary structures \citep{Pisano14}. Based on the CWD model we can predict that most galaxies with a long and steady star formation history which are forming stars at the present time should be found in voids and walls.

%
\subsection{A possible origin for red spirals}
Not all CWD events are the result of violent gravitational interactions. When a galaxy is accreted by a large filament or cluster but without experiencing major mergers this can result in CWD without morphological change.  Good candidates for this scenario are the red spirals seen in dense environments \citep{Masters10} and  SO \citep{Larson80}. This is also consistent with findings that satellites are affected more in color than morphology \citep{Bosch08}.

%
\subsection{SFR fitting and CWD}
Star formation histories of high ­redshift galaxies support the picture of gas­ accretion$\to$star ­formation (web­ attached), star­burst (web­ detaching) and instantaneous or slow decline (web­ detached) \citep{Reddy12, Blanton06}. \citet{Schawinski14} found evidence that galaxies spiral galaxies move through the green valley  when their external gas supply ends but continue to burn their remaining gas into stars. Ellipticals on the other hand require an scenario where both the accreting gas and their reservoirs are consumed in a very short time scale.

%
\subsection{A possible causal link between CWD and AGN?}

The peak in the AGN redshift distribution around $z\sim 2$ \citep{Shaver96,Page01,Hasinger05} is tantalizingly similar to the peak in the density of web detachment for halos in the mass range $\sim 10^{11.5-12.5}$\msun (Fig. \ref{fig:mass_at_z0_vs_detach_distribution}), the peak in the rate of change in the fraction of regions becoming non-linear (Fig. \ref{fig:multistream_derivative}) and the peak in the star formation rate (Fig. \ref{fig:madau_plot}). If we assume a link between AGN and CWD events then the anti-hierarchical behavior seen in the time distribution of CWD events with mass (Fig. \ref{fig:mass_at_z0_vs_detach_distribution}) could explain the difference in the peak in AGN with luminosities cited as AGN downsizing \citep{Hasinger05,Babi07,Enoki14}. While this coincidence alone is not enough to establish a causal link it suggest a relation between these processes.

The link between CWD and AGN activity could be the result of interactions between galaxies and the cosmic web that are capable of perturbing gas trajectories, injecting gas into black holes (in a similar way as in the case of major mergers). However, N-body simulations lacking AGN feedback (like the ones used here) still show a sharp decline in the SF. The observational evidence for AGN quenching  remains controversial; for example, recent deep x-ray surveys find little evidence for AGN-driven quenching \citep{Suh17}. While we can not rule out the contribution of AGN in quenching, we show that it is not necessary to explain the bulk of quenching. This should be studied in more detail in the future.

%
\section{Conclusions}

The CWD model presented here provides a fundamental mechanism to stop star formation in galaxies by separating a galaxy from its main star-forming gas supply. CWD is to first order a purely mechanical/gravitational starvation process. The nature of the particular CWD event determines the fate of the remaining gas inside the galaxy: major mergers will result in most of the gas being consumed in a starburst, often followed by a change in the galaxy's morphology, while less violent CWD events can allow gas reservoirs to be slowly consumed \citep{Bauermeister10}.

Gas strangulation and quenching are two distinct ways of reducing the SFR. Quenching is in principle instantaneous, although in practice it is inefficient. Strangulation cuts the gas supply and allows star formation to continue at a decreasing rate. Evidence has been found for bimodality in the metallicity stellar mass relation between passive and star-forming galaxies that points to both processes in action.

CWD is an starvation process. We have shown that it fits the data on SFR histories, the mass color bimodality and environmental dependence of red galaxy fraction. We therefore propose that it is an effective and natural  implementation of strangulation. Indeed, AGNs are not needed to drive outflows in our approach, even for massive galaxies. Web detachment naturally  regulates the gas supply at all masses. Of course AGNs may well be a consequence both of mass and of gas accretion as well as of mergers, and almost  certainly play a role in quenching massive galaxies, above $\sim 10^{11}$ \msun.

It is important to note that while we do not provide a smoking gun observation for CWD several pieces of evidence point in the same direction, namely that external processes that can induce gas starvation in galaxies are encompassed by one single physical mechanism which we propose to be CWD. More detailed N-body simulations and more sophisticated data analysis techniques are required and will be the subject future studies.

\section{Acknowledgements}

MA would like to thank the many colleagues with whom he discussed, over the course of several years, aspects of the work presented here, in particular Bernard Jones, Alex Szalay and Jaan Einasto. The authors would like to thank the referee for useful comments on the manuscript. 
MA is grateful for support from the Gordon and Betty Moore foundation while at JHU, where this work originated, a Big Data seed grant from the office for Research and Economic Development, UC Riverside and funding from ``Programa de Apoyo a Proyectos de Investigaci\'{o}n e Innovaci\'{o}n Tecnol\'{o}gica'' grant DGAPA-PAPIIT IA104818.
This work has been carried out in part at IAP under  the ILP LABEX (ANR-10-LABX-63) supported by French state funds managed by the ANR within the Investissements d`Avenir programme under reference ANR-11-IDEX-0004-02. The work of MN and JS was supported by ERC Project No. 267117 (DARK) hosted by Universit`e Pierre et Marie Curie (UPMC) - Paris 6, PI J. Silk. JS acknowledges the support of the JHU by NSF grant OIA-1124403. The authors would like to thank Volker Springel for making his code Gadget-3 available.

The SDSS is managed by the Astrophysical Research Consortium for the Participating Institutions. The Participating Institutions are the American Museum of Natural History, Astrophysical Institute Potsdam, University of Basel, University of Cambridge, Case Western Reserve University, University of Chicago, Drexel University, Fermilab, the Institute for Advanced Study, the Japan Participation Group, Johns Hopkins University, the Joint Institute for Nuclear Astrophysics, the Kavli Institute for Particle Astrophysics and Cosmology, the Korean Scientist Group, the Chinese Academy of Sciences (LAMOST), Los Alamos National Laboratory, the Max-Planck-Institute for Astronomy (MPIA), the Max- Planck-Institute for Astrophysics (MPA), New Mexico State University, Ohio State University, University of Pittsburgh, University of Portsmouth, Princeton University, the United States Naval Observatory, and the University of Washington.

\bibliography{refs}

\begin{thebibliography}{141}%
\makeatletter
\providecommand \@ifxundefined [1]{%
 \@ifx{#1\undefined}
}%
\providecommand \@ifnum [1]{%
 \ifnum #1\expandafter \@firstoftwo
 \else \expandafter \@secondoftwo
 \fi
}%
\providecommand \@ifx [1]{%
 \ifx #1\expandafter \@firstoftwo
 \else \expandafter \@secondoftwo
 \fi
}%
\providecommand \natexlab [1]{#1}%
\providecommand \enquote  [1]{``#1''}%
\providecommand \bibnamefont  [1]{#1}%
\providecommand \bibfnamefont [1]{#1}%
\providecommand \citenamefont [1]{#1}%
\providecommand \href@noop [0]{\@secondoftwo}%
\providecommand \href [0]{\begingroup \@sanitize@url \@href}%
\providecommand \@href[1]{\@@startlink{#1}\@@href}%
\providecommand \@@href[1]{\endgroup#1\@@endlink}%
\providecommand \@sanitize@url [0]{\catcode `\\12\catcode `\$12\catcode
  `\&12\catcode `\#12\catcode `\^12\catcode `\_12\catcode `\%12\relax}%
\providecommand \@@startlink[1]{}%
\providecommand \@@endlink[0]{}%
\providecommand \url  [0]{\begingroup\@sanitize@url \@url }%
\providecommand \@url [1]{\endgroup\@href {#1}{\urlprefix }}%
\providecommand \urlprefix  [0]{URL }%
\providecommand \Eprint [0]{\href }%
\providecommand \doibase [0]{http://dx.doi.org/}%
\providecommand \selectlanguage [0]{\@gobble}%
\providecommand \bibinfo  [0]{\@secondoftwo}%
\providecommand \bibfield  [0]{\@secondoftwo}%
\providecommand \translation [1]{[#1]}%
\providecommand \BibitemOpen [0]{}%
\providecommand \bibitemStop [0]{}%
\providecommand \bibitemNoStop [0]{.\EOS\space}%
\providecommand \EOS [0]{\spacefactor3000\relax}%
\providecommand \BibitemShut  [1]{\csname bibitem#1\endcsname}%
\let\auto@bib@innerbib\@empty
\bibitem [{\citenamefont {{Strateva}}\ \emph {et~al.}(2001)\citenamefont
  {{Strateva}}, \citenamefont {{Ivezi{\'c}}}, \citenamefont {{Knapp}},
  \citenamefont {{Narayanan}}, \citenamefont {{Strauss}}, \citenamefont
  {{Gunn}}, \citenamefont {{Lupton}}, \citenamefont {{Schlegel}}, \citenamefont
  {{Bahcall}}, \citenamefont {{Brinkmann}}, \citenamefont {{Brunner}},
  \citenamefont {{Budav{\'a}ri}}, \citenamefont {{Csabai}}, \citenamefont
  {{Castander}}, \citenamefont {{Doi}}, \citenamefont {{Fukugita}},
  \citenamefont {{Gy{\H o}ry}}, \citenamefont {{Hamabe}}, \citenamefont
  {{Hennessy}}, \citenamefont {{Ichikawa}}, \citenamefont {{Kunszt}},
  \citenamefont {{Lamb}}, \citenamefont {{McKay}}, \citenamefont {{Okamura}},
  \citenamefont {{Racusin}}, \citenamefont {{Sekiguchi}}, \citenamefont
  {{Schneider}}, \citenamefont {{Shimasaku}},\ and\ \citenamefont
  {{York}}}]{Strateva01}%
  \BibitemOpen
  \bibfield  {author} {\bibinfo {author} {\bibfnamefont {I.}~\bibnamefont
  {{Strateva}}}, \bibinfo {author} {\bibfnamefont {{\v Z}.}~\bibnamefont
  {{Ivezi{\'c}}}}, \bibinfo {author} {\bibfnamefont {G.~R.}\ \bibnamefont
  {{Knapp}}}, \bibinfo {author} {\bibfnamefont {V.~K.}\ \bibnamefont
  {{Narayanan}}}, \bibinfo {author} {\bibfnamefont {M.~A.}\ \bibnamefont
  {{Strauss}}}, \bibinfo {author} {\bibfnamefont {J.~E.}\ \bibnamefont
  {{Gunn}}}, \bibinfo {author} {\bibfnamefont {R.~H.}\ \bibnamefont
  {{Lupton}}}, \bibinfo {author} {\bibfnamefont {D.}~\bibnamefont
  {{Schlegel}}}, \bibinfo {author} {\bibfnamefont {N.~A.}\ \bibnamefont
  {{Bahcall}}}, \bibinfo {author} {\bibfnamefont {J.}~\bibnamefont
  {{Brinkmann}}}, \bibinfo {author} {\bibfnamefont {R.~J.}\ \bibnamefont
  {{Brunner}}}, \bibinfo {author} {\bibfnamefont {T.}~\bibnamefont
  {{Budav{\'a}ri}}}, \bibinfo {author} {\bibfnamefont {I.}~\bibnamefont
  {{Csabai}}}, \bibinfo {author} {\bibfnamefont {F.~J.}\ \bibnamefont
  {{Castander}}}, \bibinfo {author} {\bibfnamefont {M.}~\bibnamefont {{Doi}}},
  \bibinfo {author} {\bibfnamefont {M.}~\bibnamefont {{Fukugita}}}, \bibinfo
  {author} {\bibfnamefont {Z.}~\bibnamefont {{Gy{\H o}ry}}}, \bibinfo {author}
  {\bibfnamefont {M.}~\bibnamefont {{Hamabe}}}, \bibinfo {author}
  {\bibfnamefont {G.}~\bibnamefont {{Hennessy}}}, \bibinfo {author}
  {\bibfnamefont {T.}~\bibnamefont {{Ichikawa}}}, \bibinfo {author}
  {\bibfnamefont {P.~Z.}\ \bibnamefont {{Kunszt}}}, \bibinfo {author}
  {\bibfnamefont {D.~Q.}\ \bibnamefont {{Lamb}}}, \bibinfo {author}
  {\bibfnamefont {T.~A.}\ \bibnamefont {{McKay}}}, \bibinfo {author}
  {\bibfnamefont {S.}~\bibnamefont {{Okamura}}}, \bibinfo {author}
  {\bibfnamefont {J.}~\bibnamefont {{Racusin}}}, \bibinfo {author}
  {\bibfnamefont {M.}~\bibnamefont {{Sekiguchi}}}, \bibinfo {author}
  {\bibfnamefont {D.~P.}\ \bibnamefont {{Schneider}}}, \bibinfo {author}
  {\bibfnamefont {K.}~\bibnamefont {{Shimasaku}}}, \ and\ \bibinfo {author}
  {\bibfnamefont {D.}~\bibnamefont {{York}}},\ }\href {\doibase 10.1086/323301}
  {\bibfield  {journal} {\bibinfo  {journal} {\aj}\ }\textbf {\bibinfo {volume}
  {122}},\ \bibinfo {pages} {1861} (\bibinfo {year} {2001})},\ \Eprint
  {http://arxiv.org/abs/astro-ph/0107201} {astro-ph/0107201} \BibitemShut
  {NoStop}%
\bibitem [{\citenamefont {{Hogg}}\ \emph {et~al.}(2004)\citenamefont {{Hogg}},
  \citenamefont {{Blanton}}, \citenamefont {{Brinchmann}}, \citenamefont
  {{Eisenstein}}, \citenamefont {{Schlegel}}, \citenamefont {{Gunn}},
  \citenamefont {{McKay}}, \citenamefont {{Rix}}, \citenamefont {{Bahcall}},
  \citenamefont {{Brinkmann}},\ and\ \citenamefont {{Meiksin}}}]{Hogg04}%
  \BibitemOpen
  \bibfield  {author} {\bibinfo {author} {\bibfnamefont {D.~W.}\ \bibnamefont
  {{Hogg}}}, \bibinfo {author} {\bibfnamefont {M.~R.}\ \bibnamefont
  {{Blanton}}}, \bibinfo {author} {\bibfnamefont {J.}~\bibnamefont
  {{Brinchmann}}}, \bibinfo {author} {\bibfnamefont {D.~J.}\ \bibnamefont
  {{Eisenstein}}}, \bibinfo {author} {\bibfnamefont {D.~J.}\ \bibnamefont
  {{Schlegel}}}, \bibinfo {author} {\bibfnamefont {J.~E.}\ \bibnamefont
  {{Gunn}}}, \bibinfo {author} {\bibfnamefont {T.~A.}\ \bibnamefont {{McKay}}},
  \bibinfo {author} {\bibfnamefont {H.-W.}\ \bibnamefont {{Rix}}}, \bibinfo
  {author} {\bibfnamefont {N.~A.}\ \bibnamefont {{Bahcall}}}, \bibinfo {author}
  {\bibfnamefont {J.}~\bibnamefont {{Brinkmann}}}, \ and\ \bibinfo {author}
  {\bibfnamefont {A.}~\bibnamefont {{Meiksin}}},\ }\href {\doibase
  10.1086/381749} {\bibfield  {journal} {\bibinfo  {journal} {\apjl}\ }\textbf
  {\bibinfo {volume} {601}},\ \bibinfo {pages} {L29} (\bibinfo {year}
  {2004})},\ \Eprint {http://arxiv.org/abs/astro-ph/0307336} {astro-ph/0307336}
  \BibitemShut {NoStop}%
\bibitem [{\citenamefont {{Baldry}}\ \emph {et~al.}(2004)\citenamefont
  {{Baldry}}, \citenamefont {{Glazebrook}}, \citenamefont {{Brinkmann}},
  \citenamefont {{Ivezi{\'c}}}, \citenamefont {{Lupton}}, \citenamefont
  {{Nichol}},\ and\ \citenamefont {{Szalay}}}]{Baldry04}%
  \BibitemOpen
  \bibfield  {author} {\bibinfo {author} {\bibfnamefont {I.~K.}\ \bibnamefont
  {{Baldry}}}, \bibinfo {author} {\bibfnamefont {K.}~\bibnamefont
  {{Glazebrook}}}, \bibinfo {author} {\bibfnamefont {J.}~\bibnamefont
  {{Brinkmann}}}, \bibinfo {author} {\bibfnamefont {{\v Z}.}~\bibnamefont
  {{Ivezi{\'c}}}}, \bibinfo {author} {\bibfnamefont {R.~H.}\ \bibnamefont
  {{Lupton}}}, \bibinfo {author} {\bibfnamefont {R.~C.}\ \bibnamefont
  {{Nichol}}}, \ and\ \bibinfo {author} {\bibfnamefont {A.~S.}\ \bibnamefont
  {{Szalay}}},\ }\href {\doibase 10.1086/380092} {\bibfield  {journal}
  {\bibinfo  {journal} {\apj}\ }\textbf {\bibinfo {volume} {600}},\ \bibinfo
  {pages} {681} (\bibinfo {year} {2004})},\ \Eprint
  {http://arxiv.org/abs/astro-ph/0309710} {astro-ph/0309710} \BibitemShut
  {NoStop}%
\bibitem [{\citenamefont {{Faber}}\ \emph {et~al.}(2007)\citenamefont
  {{Faber}}, \citenamefont {{Willmer}}, \citenamefont {{Wolf}}, \citenamefont
  {{Koo}}, \citenamefont {{Weiner}}, \citenamefont {{Newman}}, \citenamefont
  {{Im}}, \citenamefont {{Coil}}, \citenamefont {{Conroy}}, \citenamefont
  {{Cooper}}, \citenamefont {{Davis}}, \citenamefont {{Finkbeiner}},
  \citenamefont {{Gerke}}, \citenamefont {{Gebhardt}}, \citenamefont {{Groth}},
  \citenamefont {{Guhathakurta}}, \citenamefont {{Harker}}, \citenamefont
  {{Kaiser}}, \citenamefont {{Kassin}}, \citenamefont {{Kleinheinrich}},
  \citenamefont {{Konidaris}}, \citenamefont {{Kron}}, \citenamefont {{Lin}},
  \citenamefont {{Luppino}}, \citenamefont {{Madgwick}}, \citenamefont
  {{Meisenheimer}}, \citenamefont {{Noeske}}, \citenamefont {{Phillips}},
  \citenamefont {{Sarajedini}}, \citenamefont {{Schiavon}}, \citenamefont
  {{Simard}}, \citenamefont {{Szalay}}, \citenamefont {{Vogt}},\ and\
  \citenamefont {{Yan}}}]{Faber07}%
  \BibitemOpen
  \bibfield  {author} {\bibinfo {author} {\bibfnamefont {S.~M.}\ \bibnamefont
  {{Faber}}}, \bibinfo {author} {\bibfnamefont {C.~N.~A.}\ \bibnamefont
  {{Willmer}}}, \bibinfo {author} {\bibfnamefont {C.}~\bibnamefont {{Wolf}}},
  \bibinfo {author} {\bibfnamefont {D.~C.}\ \bibnamefont {{Koo}}}, \bibinfo
  {author} {\bibfnamefont {B.~J.}\ \bibnamefont {{Weiner}}}, \bibinfo {author}
  {\bibfnamefont {J.~A.}\ \bibnamefont {{Newman}}}, \bibinfo {author}
  {\bibfnamefont {M.}~\bibnamefont {{Im}}}, \bibinfo {author} {\bibfnamefont
  {A.~L.}\ \bibnamefont {{Coil}}}, \bibinfo {author} {\bibfnamefont
  {C.}~\bibnamefont {{Conroy}}}, \bibinfo {author} {\bibfnamefont {M.~C.}\
  \bibnamefont {{Cooper}}}, \bibinfo {author} {\bibfnamefont {M.}~\bibnamefont
  {{Davis}}}, \bibinfo {author} {\bibfnamefont {D.~P.}\ \bibnamefont
  {{Finkbeiner}}}, \bibinfo {author} {\bibfnamefont {B.~F.}\ \bibnamefont
  {{Gerke}}}, \bibinfo {author} {\bibfnamefont {K.}~\bibnamefont {{Gebhardt}}},
  \bibinfo {author} {\bibfnamefont {E.~J.}\ \bibnamefont {{Groth}}}, \bibinfo
  {author} {\bibfnamefont {P.}~\bibnamefont {{Guhathakurta}}}, \bibinfo
  {author} {\bibfnamefont {J.}~\bibnamefont {{Harker}}}, \bibinfo {author}
  {\bibfnamefont {N.}~\bibnamefont {{Kaiser}}}, \bibinfo {author}
  {\bibfnamefont {S.}~\bibnamefont {{Kassin}}}, \bibinfo {author}
  {\bibfnamefont {M.}~\bibnamefont {{Kleinheinrich}}}, \bibinfo {author}
  {\bibfnamefont {N.~P.}\ \bibnamefont {{Konidaris}}}, \bibinfo {author}
  {\bibfnamefont {R.~G.}\ \bibnamefont {{Kron}}}, \bibinfo {author}
  {\bibfnamefont {L.}~\bibnamefont {{Lin}}}, \bibinfo {author} {\bibfnamefont
  {G.}~\bibnamefont {{Luppino}}}, \bibinfo {author} {\bibfnamefont {D.~S.}\
  \bibnamefont {{Madgwick}}}, \bibinfo {author} {\bibfnamefont
  {K.}~\bibnamefont {{Meisenheimer}}}, \bibinfo {author} {\bibfnamefont
  {K.~G.}\ \bibnamefont {{Noeske}}}, \bibinfo {author} {\bibfnamefont {A.~C.}\
  \bibnamefont {{Phillips}}}, \bibinfo {author} {\bibfnamefont {V.~L.}\
  \bibnamefont {{Sarajedini}}}, \bibinfo {author} {\bibfnamefont {R.~P.}\
  \bibnamefont {{Schiavon}}}, \bibinfo {author} {\bibfnamefont
  {L.}~\bibnamefont {{Simard}}}, \bibinfo {author} {\bibfnamefont {A.~S.}\
  \bibnamefont {{Szalay}}}, \bibinfo {author} {\bibfnamefont {N.~P.}\
  \bibnamefont {{Vogt}}}, \ and\ \bibinfo {author} {\bibfnamefont
  {R.}~\bibnamefont {{Yan}}},\ }\href {\doibase 10.1086/519294} {\bibfield
  {journal} {\bibinfo  {journal} {\apj}\ }\textbf {\bibinfo {volume} {665}},\
  \bibinfo {pages} {265} (\bibinfo {year} {2007})},\ \Eprint
  {http://arxiv.org/abs/astro-ph/0506044} {astro-ph/0506044} \BibitemShut
  {NoStop}%
\bibitem [{\citenamefont {{Dekel}}\ and\ \citenamefont
  {{Birnboim}}(2006)}]{Dekel05}%
  \BibitemOpen
  \bibfield  {author} {\bibinfo {author} {\bibfnamefont {A.}~\bibnamefont
  {{Dekel}}}\ and\ \bibinfo {author} {\bibfnamefont {Y.}~\bibnamefont
  {{Birnboim}}},\ }\href {\doibase 10.1111/j.1365-2966.2006.10145.x} {\bibfield
   {journal} {\bibinfo  {journal} {\mnras}\ }\textbf {\bibinfo {volume}
  {368}},\ \bibinfo {pages} {2} (\bibinfo {year} {2006})},\ \Eprint
  {http://arxiv.org/abs/astro-ph/0412300} {astro-ph/0412300} \BibitemShut
  {NoStop}%
\bibitem [{\citenamefont {{Schawinski}}\ \emph {et~al.}(2014)\citenamefont
  {{Schawinski}}, \citenamefont {{Urry}}, \citenamefont {{Simmons}},
  \citenamefont {{Fortson}}, \citenamefont {{Kaviraj}}, \citenamefont {{Keel}},
  \citenamefont {{Lintott}}, \citenamefont {{Masters}}, \citenamefont
  {{Nichol}}, \citenamefont {{Sarzi}}, \citenamefont {{Skibba}}, \citenamefont
  {{Treister}}, \citenamefont {{Willett}}, \citenamefont {{Wong}},\ and\
  \citenamefont {{Yi}}}]{Schawinski14}%
  \BibitemOpen
  \bibfield  {author} {\bibinfo {author} {\bibfnamefont {K.}~\bibnamefont
  {{Schawinski}}}, \bibinfo {author} {\bibfnamefont {C.~M.}\ \bibnamefont
  {{Urry}}}, \bibinfo {author} {\bibfnamefont {B.~D.}\ \bibnamefont
  {{Simmons}}}, \bibinfo {author} {\bibfnamefont {L.}~\bibnamefont
  {{Fortson}}}, \bibinfo {author} {\bibfnamefont {S.}~\bibnamefont
  {{Kaviraj}}}, \bibinfo {author} {\bibfnamefont {W.~C.}\ \bibnamefont
  {{Keel}}}, \bibinfo {author} {\bibfnamefont {C.~J.}\ \bibnamefont
  {{Lintott}}}, \bibinfo {author} {\bibfnamefont {K.~L.}\ \bibnamefont
  {{Masters}}}, \bibinfo {author} {\bibfnamefont {R.~C.}\ \bibnamefont
  {{Nichol}}}, \bibinfo {author} {\bibfnamefont {M.}~\bibnamefont {{Sarzi}}},
  \bibinfo {author} {\bibfnamefont {R.}~\bibnamefont {{Skibba}}}, \bibinfo
  {author} {\bibfnamefont {E.}~\bibnamefont {{Treister}}}, \bibinfo {author}
  {\bibfnamefont {K.~W.}\ \bibnamefont {{Willett}}}, \bibinfo {author}
  {\bibfnamefont {O.~I.}\ \bibnamefont {{Wong}}}, \ and\ \bibinfo {author}
  {\bibfnamefont {S.~K.}\ \bibnamefont {{Yi}}},\ }\href {\doibase
  10.1093/mnras/stu327} {\bibfield  {journal} {\bibinfo  {journal} {\mnras}\
  }\textbf {\bibinfo {volume} {440}},\ \bibinfo {pages} {889} (\bibinfo {year}
  {2014})},\ \Eprint {http://arxiv.org/abs/1402.4814} {arXiv:1402.4814
  [astro-ph.GA]} \BibitemShut {NoStop}%
\bibitem [{\citenamefont {{Bell}}\ \emph {et~al.}(2004)\citenamefont {{Bell}},
  \citenamefont {{Wolf}}, \citenamefont {{Meisenheimer}}, \citenamefont
  {{Rix}}, \citenamefont {{Borch}}, \citenamefont {{Dye}}, \citenamefont
  {{Kleinheinrich}}, \citenamefont {{Wisotzki}},\ and\ \citenamefont
  {{McIntosh}}}]{Bell04}%
  \BibitemOpen
  \bibfield  {author} {\bibinfo {author} {\bibfnamefont {E.~F.}\ \bibnamefont
  {{Bell}}}, \bibinfo {author} {\bibfnamefont {C.}~\bibnamefont {{Wolf}}},
  \bibinfo {author} {\bibfnamefont {K.}~\bibnamefont {{Meisenheimer}}},
  \bibinfo {author} {\bibfnamefont {H.-W.}\ \bibnamefont {{Rix}}}, \bibinfo
  {author} {\bibfnamefont {A.}~\bibnamefont {{Borch}}}, \bibinfo {author}
  {\bibfnamefont {S.}~\bibnamefont {{Dye}}}, \bibinfo {author} {\bibfnamefont
  {M.}~\bibnamefont {{Kleinheinrich}}}, \bibinfo {author} {\bibfnamefont
  {L.}~\bibnamefont {{Wisotzki}}}, \ and\ \bibinfo {author} {\bibfnamefont
  {D.~H.}\ \bibnamefont {{McIntosh}}},\ }\href {\doibase 10.1086/420778}
  {\bibfield  {journal} {\bibinfo  {journal} {\apj}\ }\textbf {\bibinfo
  {volume} {608}},\ \bibinfo {pages} {752} (\bibinfo {year} {2004})},\ \Eprint
  {http://arxiv.org/abs/astro-ph/0303394} {astro-ph/0303394} \BibitemShut
  {NoStop}%
\bibitem [{\citenamefont {{Blanton}}(2006)}]{Blanton06}%
  \BibitemOpen
  \bibfield  {author} {\bibinfo {author} {\bibfnamefont {M.~R.}\ \bibnamefont
  {{Blanton}}},\ }\href {\doibase 10.1086/505628} {\bibfield  {journal}
  {\bibinfo  {journal} {\apj}\ }\textbf {\bibinfo {volume} {648}},\ \bibinfo
  {pages} {268} (\bibinfo {year} {2006})},\ \Eprint
  {http://arxiv.org/abs/astro-ph/0512127} {astro-ph/0512127} \BibitemShut
  {NoStop}%
\bibitem [{\citenamefont {{Wyder}}\ \emph {et~al.}(2007)\citenamefont
  {{Wyder}}, \citenamefont {{Martin}}, \citenamefont {{Schiminovich}},
  \citenamefont {{Seibert}}, \citenamefont {{Budav{\'a}ri}}, \citenamefont
  {{Treyer}}, \citenamefont {{Barlow}}, \citenamefont {{Forster}},
  \citenamefont {{Friedman}}, \citenamefont {{Morrissey}}, \citenamefont
  {{Neff}}, \citenamefont {{Small}}, \citenamefont {{Bianchi}}, \citenamefont
  {{Donas}}, \citenamefont {{Heckman}}, \citenamefont {{Lee}}, \citenamefont
  {{Madore}}, \citenamefont {{Milliard}}, \citenamefont {{Rich}}, \citenamefont
  {{Szalay}}, \citenamefont {{Welsh}},\ and\ \citenamefont {{Yi}}}]{Wyder07}%
  \BibitemOpen
  \bibfield  {author} {\bibinfo {author} {\bibfnamefont {T.~K.}\ \bibnamefont
  {{Wyder}}}, \bibinfo {author} {\bibfnamefont {D.~C.}\ \bibnamefont
  {{Martin}}}, \bibinfo {author} {\bibfnamefont {D.}~\bibnamefont
  {{Schiminovich}}}, \bibinfo {author} {\bibfnamefont {M.}~\bibnamefont
  {{Seibert}}}, \bibinfo {author} {\bibfnamefont {T.}~\bibnamefont
  {{Budav{\'a}ri}}}, \bibinfo {author} {\bibfnamefont {M.~A.}\ \bibnamefont
  {{Treyer}}}, \bibinfo {author} {\bibfnamefont {T.~A.}\ \bibnamefont
  {{Barlow}}}, \bibinfo {author} {\bibfnamefont {K.}~\bibnamefont {{Forster}}},
  \bibinfo {author} {\bibfnamefont {P.~G.}\ \bibnamefont {{Friedman}}},
  \bibinfo {author} {\bibfnamefont {P.}~\bibnamefont {{Morrissey}}}, \bibinfo
  {author} {\bibfnamefont {S.~G.}\ \bibnamefont {{Neff}}}, \bibinfo {author}
  {\bibfnamefont {T.}~\bibnamefont {{Small}}}, \bibinfo {author} {\bibfnamefont
  {L.}~\bibnamefont {{Bianchi}}}, \bibinfo {author} {\bibfnamefont
  {J.}~\bibnamefont {{Donas}}}, \bibinfo {author} {\bibfnamefont {T.~M.}\
  \bibnamefont {{Heckman}}}, \bibinfo {author} {\bibfnamefont {Y.-W.}\
  \bibnamefont {{Lee}}}, \bibinfo {author} {\bibfnamefont {B.~F.}\ \bibnamefont
  {{Madore}}}, \bibinfo {author} {\bibfnamefont {B.}~\bibnamefont
  {{Milliard}}}, \bibinfo {author} {\bibfnamefont {R.~M.}\ \bibnamefont
  {{Rich}}}, \bibinfo {author} {\bibfnamefont {A.~S.}\ \bibnamefont
  {{Szalay}}}, \bibinfo {author} {\bibfnamefont {B.~Y.}\ \bibnamefont
  {{Welsh}}}, \ and\ \bibinfo {author} {\bibfnamefont {S.~K.}\ \bibnamefont
  {{Yi}}},\ }\href {\doibase 10.1086/521402} {\bibfield  {journal} {\bibinfo
  {journal} {\apjs}\ }\textbf {\bibinfo {volume} {173}},\ \bibinfo {pages}
  {293} (\bibinfo {year} {2007})},\ \Eprint {http://arxiv.org/abs/0706.3938}
  {arXiv:0706.3938} \BibitemShut {NoStop}%
\bibitem [{\citenamefont {{Silk}}\ and\ \citenamefont {{Rees}}(1998)}]{Silk98}%
  \BibitemOpen
  \bibfield  {author} {\bibinfo {author} {\bibfnamefont {J.}~\bibnamefont
  {{Silk}}}\ and\ \bibinfo {author} {\bibfnamefont {M.~J.}\ \bibnamefont
  {{Rees}}},\ }\href@noop {} {\bibfield  {journal} {\bibinfo  {journal} {\aap}\
  }\textbf {\bibinfo {volume} {331}},\ \bibinfo {pages} {L1} (\bibinfo {year}
  {1998})},\ \Eprint {http://arxiv.org/abs/astro-ph/9801013} {astro-ph/9801013}
  \BibitemShut {NoStop}%
\bibitem [{\citenamefont {{Di Matteo}}\ \emph {et~al.}(2005)\citenamefont {{Di
  Matteo}}, \citenamefont {{Springel}},\ and\ \citenamefont
  {{Hernquist}}}]{DiMatteo05}%
  \BibitemOpen
  \bibfield  {author} {\bibinfo {author} {\bibfnamefont {T.}~\bibnamefont {{Di
  Matteo}}}, \bibinfo {author} {\bibfnamefont {V.}~\bibnamefont {{Springel}}},
  \ and\ \bibinfo {author} {\bibfnamefont {L.}~\bibnamefont {{Hernquist}}},\
  }\href {\doibase 10.1038/nature03335} {\bibfield  {journal} {\bibinfo
  {journal} {\nat}\ }\textbf {\bibinfo {volume} {433}},\ \bibinfo {pages} {604}
  (\bibinfo {year} {2005})},\ \Eprint {http://arxiv.org/abs/astro-ph/0502199}
  {astro-ph/0502199} \BibitemShut {NoStop}%
\bibitem [{\citenamefont {{Best}}\ \emph {et~al.}(2005)\citenamefont {{Best}},
  \citenamefont {{Kauffmann}}, \citenamefont {{Heckman}}, \citenamefont
  {{Brinchmann}}, \citenamefont {{Charlot}}, \citenamefont {{Ivezi{\'c}}},\
  and\ \citenamefont {{White}}}]{Best05}%
  \BibitemOpen
  \bibfield  {author} {\bibinfo {author} {\bibfnamefont {P.~N.}\ \bibnamefont
  {{Best}}}, \bibinfo {author} {\bibfnamefont {G.}~\bibnamefont {{Kauffmann}}},
  \bibinfo {author} {\bibfnamefont {T.~M.}\ \bibnamefont {{Heckman}}}, \bibinfo
  {author} {\bibfnamefont {J.}~\bibnamefont {{Brinchmann}}}, \bibinfo {author}
  {\bibfnamefont {S.}~\bibnamefont {{Charlot}}}, \bibinfo {author}
  {\bibfnamefont {{\v Z}.}~\bibnamefont {{Ivezi{\'c}}}}, \ and\ \bibinfo
  {author} {\bibfnamefont {S.~D.~M.}\ \bibnamefont {{White}}},\ }\href
  {\doibase 10.1111/j.1365-2966.2005.09192.x} {\bibfield  {journal} {\bibinfo
  {journal} {\mnras}\ }\textbf {\bibinfo {volume} {362}},\ \bibinfo {pages}
  {25} (\bibinfo {year} {2005})},\ \Eprint
  {http://arxiv.org/abs/astro-ph/0506269} {astro-ph/0506269} \BibitemShut
  {NoStop}%
\bibitem [{\citenamefont {{Peng}}(2007)}]{Peng07}%
  \BibitemOpen
  \bibfield  {author} {\bibinfo {author} {\bibfnamefont {C.~Y.}\ \bibnamefont
  {{Peng}}},\ }\href {\doibase 10.1086/522774} {\bibfield  {journal} {\bibinfo
  {journal} {\apj}\ }\textbf {\bibinfo {volume} {671}},\ \bibinfo {pages}
  {1098} (\bibinfo {year} {2007})},\ \Eprint {http://arxiv.org/abs/0704.1860}
  {arXiv:0704.1860} \BibitemShut {NoStop}%
\bibitem [{\citenamefont {{Jahnke}}\ and\ \citenamefont
  {{Macci{\`o}}}(2011)}]{Jahnke11}%
  \BibitemOpen
  \bibfield  {author} {\bibinfo {author} {\bibfnamefont {K.}~\bibnamefont
  {{Jahnke}}}\ and\ \bibinfo {author} {\bibfnamefont {A.~V.}\ \bibnamefont
  {{Macci{\`o}}}},\ }\href {\doibase 10.1088/0004-637X/734/2/92} {\bibfield
  {journal} {\bibinfo  {journal} {\apj}\ }\textbf {\bibinfo {volume} {734}},\
  \bibinfo {eid} {92} (\bibinfo {year} {2011})},\ \Eprint
  {http://arxiv.org/abs/1006.0482} {arXiv:1006.0482} \BibitemShut {NoStop}%
\bibitem [{\citenamefont {{Alam}}\ \emph {et~al.}(2018)\citenamefont {{Alam}},
  \citenamefont {{Zu}}, \citenamefont {{Peacock}},\ and\ \citenamefont
  {{Mandelbaum}}}]{Alam18}%
  \BibitemOpen
  \bibfield  {author} {\bibinfo {author} {\bibfnamefont {S.}~\bibnamefont
  {{Alam}}}, \bibinfo {author} {\bibfnamefont {Y.}~\bibnamefont {{Zu}}},
  \bibinfo {author} {\bibfnamefont {J.~A.}\ \bibnamefont {{Peacock}}}, \ and\
  \bibinfo {author} {\bibfnamefont {R.}~\bibnamefont {{Mandelbaum}}},\
  }\href@noop {} {\bibfield  {journal} {\bibinfo  {journal} {ArXiv e-prints}\ }
  (\bibinfo {year} {2018})},\ \Eprint {http://arxiv.org/abs/1801.04878}
  {arXiv:1801.04878} \BibitemShut {NoStop}%
\bibitem [{\citenamefont {{Kraljic}}\ \emph {et~al.}(2018)\citenamefont
  {{Kraljic}}, \citenamefont {{Arnouts}}, \citenamefont {{Pichon}},
  \citenamefont {{Laigle}}, \citenamefont {{de la Torre}}, \citenamefont
  {{Vibert}}, \citenamefont {{Cadiou}}, \citenamefont {{Dubois}}, \citenamefont
  {{Treyer}}, \citenamefont {{Schimd}}, \citenamefont {{Codis}}, \citenamefont
  {{de Lapparent}}, \citenamefont {{Devriendt}}, \citenamefont {{Hwang}},
  \citenamefont {{Le Borgne}}, \citenamefont {{Malavasi}}, \citenamefont
  {{Milliard}}, \citenamefont {{Musso}}, \citenamefont {{Pogosyan}},
  \citenamefont {{Alpaslan}}, \citenamefont {{Bland-Hawthorn}},\ and\
  \citenamefont {{Wright}}}]{Kraljic18}%
  \BibitemOpen
  \bibfield  {author} {\bibinfo {author} {\bibfnamefont {K.}~\bibnamefont
  {{Kraljic}}}, \bibinfo {author} {\bibfnamefont {S.}~\bibnamefont
  {{Arnouts}}}, \bibinfo {author} {\bibfnamefont {C.}~\bibnamefont {{Pichon}}},
  \bibinfo {author} {\bibfnamefont {C.}~\bibnamefont {{Laigle}}}, \bibinfo
  {author} {\bibfnamefont {S.}~\bibnamefont {{de la Torre}}}, \bibinfo {author}
  {\bibfnamefont {D.}~\bibnamefont {{Vibert}}}, \bibinfo {author}
  {\bibfnamefont {C.}~\bibnamefont {{Cadiou}}}, \bibinfo {author}
  {\bibfnamefont {Y.}~\bibnamefont {{Dubois}}}, \bibinfo {author}
  {\bibfnamefont {M.}~\bibnamefont {{Treyer}}}, \bibinfo {author}
  {\bibfnamefont {C.}~\bibnamefont {{Schimd}}}, \bibinfo {author}
  {\bibfnamefont {S.}~\bibnamefont {{Codis}}}, \bibinfo {author} {\bibfnamefont
  {V.}~\bibnamefont {{de Lapparent}}}, \bibinfo {author} {\bibfnamefont
  {J.}~\bibnamefont {{Devriendt}}}, \bibinfo {author} {\bibfnamefont {H.~S.}\
  \bibnamefont {{Hwang}}}, \bibinfo {author} {\bibfnamefont {D.}~\bibnamefont
  {{Le Borgne}}}, \bibinfo {author} {\bibfnamefont {N.}~\bibnamefont
  {{Malavasi}}}, \bibinfo {author} {\bibfnamefont {B.}~\bibnamefont
  {{Milliard}}}, \bibinfo {author} {\bibfnamefont {M.}~\bibnamefont {{Musso}}},
  \bibinfo {author} {\bibfnamefont {D.}~\bibnamefont {{Pogosyan}}}, \bibinfo
  {author} {\bibfnamefont {M.}~\bibnamefont {{Alpaslan}}}, \bibinfo {author}
  {\bibfnamefont {J.}~\bibnamefont {{Bland-Hawthorn}}}, \ and\ \bibinfo
  {author} {\bibfnamefont {A.~H.}\ \bibnamefont {{Wright}}},\ }\href {\doibase
  10.1093/mnras/stx2638} {\bibfield  {journal} {\bibinfo  {journal} {\mnras}\
  }\textbf {\bibinfo {volume} {474}},\ \bibinfo {pages} {547} (\bibinfo {year}
  {2018})},\ \Eprint {http://arxiv.org/abs/1710.02676} {arXiv:1710.02676}
  \BibitemShut {NoStop}%
\bibitem [{\citenamefont {{Gunn}}\ and\ \citenamefont {{Gott}}(1972)}]{Gunn72}%
  \BibitemOpen
  \bibfield  {author} {\bibinfo {author} {\bibfnamefont {J.~E.}\ \bibnamefont
  {{Gunn}}}\ and\ \bibinfo {author} {\bibfnamefont {J.~R.}\ \bibnamefont
  {{Gott}}, \bibfnamefont {III}},\ }\href {\doibase 10.1086/151605} {\bibfield
  {journal} {\bibinfo  {journal} {\apj}\ }\textbf {\bibinfo {volume} {176}},\
  \bibinfo {pages} {1} (\bibinfo {year} {1972})}\BibitemShut {NoStop}%
\bibitem [{\citenamefont {{Balsara}}\ \emph {et~al.}(1994)\citenamefont
  {{Balsara}}, \citenamefont {{Livio}},\ and\ \citenamefont
  {{O'Dea}}}]{Balsara94}%
  \BibitemOpen
  \bibfield  {author} {\bibinfo {author} {\bibfnamefont {D.}~\bibnamefont
  {{Balsara}}}, \bibinfo {author} {\bibfnamefont {M.}~\bibnamefont {{Livio}}},
  \ and\ \bibinfo {author} {\bibfnamefont {C.~P.}\ \bibnamefont {{O'Dea}}},\
  }\href {\doibase 10.1086/174977} {\bibfield  {journal} {\bibinfo  {journal}
  {\apj}\ }\textbf {\bibinfo {volume} {437}},\ \bibinfo {pages} {83} (\bibinfo
  {year} {1994})}\BibitemShut {NoStop}%
\bibitem [{\citenamefont {{Abadi}}\ \emph {et~al.}(1999)\citenamefont
  {{Abadi}}, \citenamefont {{Moore}},\ and\ \citenamefont {{Bower}}}]{Abadi99}%
  \BibitemOpen
  \bibfield  {author} {\bibinfo {author} {\bibfnamefont {M.~G.}\ \bibnamefont
  {{Abadi}}}, \bibinfo {author} {\bibfnamefont {B.}~\bibnamefont {{Moore}}}, \
  and\ \bibinfo {author} {\bibfnamefont {R.~G.}\ \bibnamefont {{Bower}}},\
  }\href {\doibase 10.1046/j.1365-8711.1999.02715.x} {\bibfield  {journal}
  {\bibinfo  {journal} {\mnras}\ }\textbf {\bibinfo {volume} {308}},\ \bibinfo
  {pages} {947} (\bibinfo {year} {1999})},\ \Eprint
  {http://arxiv.org/abs/astro-ph/9903436} {astro-ph/9903436} \BibitemShut
  {NoStop}%
\bibitem [{\citenamefont {{McCarthy}}\ \emph {et~al.}(2008)\citenamefont
  {{McCarthy}}, \citenamefont {{Frenk}}, \citenamefont {{Font}}, \citenamefont
  {{Lacey}}, \citenamefont {{Bower}}, \citenamefont {{Mitchell}}, \citenamefont
  {{Balogh}},\ and\ \citenamefont {{Theuns}}}]{McCarthy08}%
  \BibitemOpen
  \bibfield  {author} {\bibinfo {author} {\bibfnamefont {I.~G.}\ \bibnamefont
  {{McCarthy}}}, \bibinfo {author} {\bibfnamefont {C.~S.}\ \bibnamefont
  {{Frenk}}}, \bibinfo {author} {\bibfnamefont {A.~S.}\ \bibnamefont {{Font}}},
  \bibinfo {author} {\bibfnamefont {C.~G.}\ \bibnamefont {{Lacey}}}, \bibinfo
  {author} {\bibfnamefont {R.~G.}\ \bibnamefont {{Bower}}}, \bibinfo {author}
  {\bibfnamefont {N.~L.}\ \bibnamefont {{Mitchell}}}, \bibinfo {author}
  {\bibfnamefont {M.~L.}\ \bibnamefont {{Balogh}}}, \ and\ \bibinfo {author}
  {\bibfnamefont {T.}~\bibnamefont {{Theuns}}},\ }\href {\doibase
  10.1111/j.1365-2966.2007.12577.x} {\bibfield  {journal} {\bibinfo  {journal}
  {\mnras}\ }\textbf {\bibinfo {volume} {383}},\ \bibinfo {pages} {593}
  (\bibinfo {year} {2008})},\ \Eprint {http://arxiv.org/abs/0710.0964}
  {arXiv:0710.0964} \BibitemShut {NoStop}%
\bibitem [{\citenamefont {{Moore}}\ \emph {et~al.}(1996)\citenamefont
  {{Moore}}, \citenamefont {{Katz}}, \citenamefont {{Lake}}, \citenamefont
  {{Dressler}},\ and\ \citenamefont {{Oemler}}}]{Moore96}%
  \BibitemOpen
  \bibfield  {author} {\bibinfo {author} {\bibfnamefont {B.}~\bibnamefont
  {{Moore}}}, \bibinfo {author} {\bibfnamefont {N.}~\bibnamefont {{Katz}}},
  \bibinfo {author} {\bibfnamefont {G.}~\bibnamefont {{Lake}}}, \bibinfo
  {author} {\bibfnamefont {A.}~\bibnamefont {{Dressler}}}, \ and\ \bibinfo
  {author} {\bibfnamefont {A.}~\bibnamefont {{Oemler}}},\ }\href {\doibase
  10.1038/379613a0} {\bibfield  {journal} {\bibinfo  {journal} {\nat}\ }\textbf
  {\bibinfo {volume} {379}},\ \bibinfo {pages} {613} (\bibinfo {year}
  {1996})},\ \Eprint {http://arxiv.org/abs/astro-ph/9510034} {astro-ph/9510034}
  \BibitemShut {NoStop}%
\bibitem [{\citenamefont {{Moore}}\ \emph {et~al.}(1998)\citenamefont
  {{Moore}}, \citenamefont {{Lake}},\ and\ \citenamefont {{Katz}}}]{Moore98}%
  \BibitemOpen
  \bibfield  {author} {\bibinfo {author} {\bibfnamefont {B.}~\bibnamefont
  {{Moore}}}, \bibinfo {author} {\bibfnamefont {G.}~\bibnamefont {{Lake}}}, \
  and\ \bibinfo {author} {\bibfnamefont {N.}~\bibnamefont {{Katz}}},\ }\href
  {\doibase 10.1086/305264} {\bibfield  {journal} {\bibinfo  {journal} {\apj}\
  }\textbf {\bibinfo {volume} {495}},\ \bibinfo {pages} {139} (\bibinfo {year}
  {1998})},\ \Eprint {http://arxiv.org/abs/astro-ph/9701211} {astro-ph/9701211}
  \BibitemShut {NoStop}%
\bibitem [{\citenamefont {{Bekki}}\ \emph {et~al.}(2002)\citenamefont
  {{Bekki}}, \citenamefont {{Couch}},\ and\ \citenamefont
  {{Shioya}}}]{Bekki02}%
  \BibitemOpen
  \bibfield  {author} {\bibinfo {author} {\bibfnamefont {K.}~\bibnamefont
  {{Bekki}}}, \bibinfo {author} {\bibfnamefont {W.~J.}\ \bibnamefont
  {{Couch}}}, \ and\ \bibinfo {author} {\bibfnamefont {Y.}~\bibnamefont
  {{Shioya}}},\ }\href {\doibase 10.1086/342221} {\bibfield  {journal}
  {\bibinfo  {journal} {\apj}\ }\textbf {\bibinfo {volume} {577}},\ \bibinfo
  {pages} {651} (\bibinfo {year} {2002})},\ \Eprint
  {http://arxiv.org/abs/astro-ph/0206207} {astro-ph/0206207} \BibitemShut
  {NoStop}%
\bibitem [{\citenamefont {{Fujita}}(2004)}]{Fujita04}%
  \BibitemOpen
  \bibfield  {author} {\bibinfo {author} {\bibfnamefont {Y.}~\bibnamefont
  {{Fujita}}},\ }\href {\doibase 10.1093/pasj/56.1.29} {\bibfield  {journal}
  {\bibinfo  {journal} {\pasj}\ }\textbf {\bibinfo {volume} {56}},\ \bibinfo
  {pages} {29} (\bibinfo {year} {2004})},\ \Eprint
  {http://arxiv.org/abs/astro-ph/0311193} {astro-ph/0311193} \BibitemShut
  {NoStop}%
\bibitem [{\citenamefont {{Kawata}}\ and\ \citenamefont
  {{Mulchaey}}(2008)}]{Kawata08}%
  \BibitemOpen
  \bibfield  {author} {\bibinfo {author} {\bibfnamefont {D.}~\bibnamefont
  {{Kawata}}}\ and\ \bibinfo {author} {\bibfnamefont {J.~S.}\ \bibnamefont
  {{Mulchaey}}},\ }\href {\doibase 10.1086/526544} {\bibfield  {journal}
  {\bibinfo  {journal} {\apjl}\ }\textbf {\bibinfo {volume} {672}},\ \bibinfo
  {eid} {L103} (\bibinfo {year} {2008})},\ \Eprint
  {http://arxiv.org/abs/0707.3814} {arXiv:0707.3814} \BibitemShut {NoStop}%
\bibitem [{\citenamefont {{Kodama}}\ and\ \citenamefont
  {{Bower}}(2001)}]{Kodama01}%
  \BibitemOpen
  \bibfield  {author} {\bibinfo {author} {\bibfnamefont {T.}~\bibnamefont
  {{Kodama}}}\ and\ \bibinfo {author} {\bibfnamefont {R.~G.}\ \bibnamefont
  {{Bower}}},\ }\href {\doibase 10.1046/j.1365-8711.2001.03981.x} {\bibfield
  {journal} {\bibinfo  {journal} {\mnras}\ }\textbf {\bibinfo {volume} {321}},\
  \bibinfo {pages} {18} (\bibinfo {year} {2001})},\ \Eprint
  {http://arxiv.org/abs/astro-ph/0005397} {astro-ph/0005397} \BibitemShut
  {NoStop}%
\bibitem [{\citenamefont {{Treu}}\ \emph {et~al.}(2003)\citenamefont {{Treu}},
  \citenamefont {{Ellis}}, \citenamefont {{Kneib}}, \citenamefont {{Dressler}},
  \citenamefont {{Smail}}, \citenamefont {{Czoske}}, \citenamefont {{Oemler}},\
  and\ \citenamefont {{Natarajan}}}]{Treu03}%
  \BibitemOpen
  \bibfield  {author} {\bibinfo {author} {\bibfnamefont {T.}~\bibnamefont
  {{Treu}}}, \bibinfo {author} {\bibfnamefont {R.~S.}\ \bibnamefont {{Ellis}}},
  \bibinfo {author} {\bibfnamefont {J.-P.}\ \bibnamefont {{Kneib}}}, \bibinfo
  {author} {\bibfnamefont {A.}~\bibnamefont {{Dressler}}}, \bibinfo {author}
  {\bibfnamefont {I.}~\bibnamefont {{Smail}}}, \bibinfo {author} {\bibfnamefont
  {O.}~\bibnamefont {{Czoske}}}, \bibinfo {author} {\bibfnamefont
  {A.}~\bibnamefont {{Oemler}}}, \ and\ \bibinfo {author} {\bibfnamefont
  {P.}~\bibnamefont {{Natarajan}}},\ }\href {\doibase 10.1086/375314}
  {\bibfield  {journal} {\bibinfo  {journal} {\apj}\ }\textbf {\bibinfo
  {volume} {591}},\ \bibinfo {pages} {53} (\bibinfo {year} {2003})},\ \Eprint
  {http://arxiv.org/abs/astro-ph/0303267} {astro-ph/0303267} \BibitemShut
  {NoStop}%
\bibitem [{\citenamefont {{Goto}}\ \emph {et~al.}(2003)\citenamefont {{Goto}},
  \citenamefont {{Okamura}}, \citenamefont {{Yagi}}, \citenamefont {{Sheth}},
  \citenamefont {{Bahcall}}, \citenamefont {{Zabel}}, \citenamefont {{Crouch}},
  \citenamefont {{Sekiguchi}}, \citenamefont {{Annis}}, \citenamefont
  {{Bernardi}}, \citenamefont {{Chong}}, \citenamefont {{G{\'o}mez}},
  \citenamefont {{Hansen}}, \citenamefont {{Kim}}, \citenamefont {{Knudson}},
  \citenamefont {{McKay}},\ and\ \citenamefont {{Miller}}}]{Goto03}%
  \BibitemOpen
  \bibfield  {author} {\bibinfo {author} {\bibfnamefont {T.}~\bibnamefont
  {{Goto}}}, \bibinfo {author} {\bibfnamefont {S.}~\bibnamefont {{Okamura}}},
  \bibinfo {author} {\bibfnamefont {M.}~\bibnamefont {{Yagi}}}, \bibinfo
  {author} {\bibfnamefont {R.~K.}\ \bibnamefont {{Sheth}}}, \bibinfo {author}
  {\bibfnamefont {N.~A.}\ \bibnamefont {{Bahcall}}}, \bibinfo {author}
  {\bibfnamefont {S.~A.}\ \bibnamefont {{Zabel}}}, \bibinfo {author}
  {\bibfnamefont {M.~S.}\ \bibnamefont {{Crouch}}}, \bibinfo {author}
  {\bibfnamefont {M.}~\bibnamefont {{Sekiguchi}}}, \bibinfo {author}
  {\bibfnamefont {J.}~\bibnamefont {{Annis}}}, \bibinfo {author} {\bibfnamefont
  {M.}~\bibnamefont {{Bernardi}}}, \bibinfo {author} {\bibfnamefont {S.-S.}\
  \bibnamefont {{Chong}}}, \bibinfo {author} {\bibfnamefont {P.~L.}\
  \bibnamefont {{G{\'o}mez}}}, \bibinfo {author} {\bibfnamefont
  {S.}~\bibnamefont {{Hansen}}}, \bibinfo {author} {\bibfnamefont {R.~S.~J.}\
  \bibnamefont {{Kim}}}, \bibinfo {author} {\bibfnamefont {A.}~\bibnamefont
  {{Knudson}}}, \bibinfo {author} {\bibfnamefont {T.~A.}\ \bibnamefont
  {{McKay}}}, \ and\ \bibinfo {author} {\bibfnamefont {C.~J.}\ \bibnamefont
  {{Miller}}},\ }\href {\doibase 10.1093/pasj/55.4.739} {\bibfield  {journal}
  {\bibinfo  {journal} {\pasj}\ }\textbf {\bibinfo {volume} {55}},\ \bibinfo
  {pages} {739} (\bibinfo {year} {2003})},\ \Eprint
  {http://arxiv.org/abs/astro-ph/0301302} {astro-ph/0301302} \BibitemShut
  {NoStop}%
\bibitem [{\citenamefont {{Aragon-Calvo}}\ \emph {et~al.}(2014)\citenamefont
  {{Aragon-Calvo}}, \citenamefont {{Neyrinck}},\ and\ \citenamefont
  {{Silk}}}]{Aragon14b}%
  \BibitemOpen
  \bibfield  {author} {\bibinfo {author} {\bibfnamefont {M.~A.}\ \bibnamefont
  {{Aragon-Calvo}}}, \bibinfo {author} {\bibfnamefont {M.~C.}\ \bibnamefont
  {{Neyrinck}}}, \ and\ \bibinfo {author} {\bibfnamefont {J.}~\bibnamefont
  {{Silk}}},\ }\href@noop {} {\bibfield  {journal} {\bibinfo  {journal} {ArXiv
  e-prints}\ } (\bibinfo {year} {2014})},\ \Eprint
  {http://arxiv.org/abs/1412.1119} {arXiv:1412.1119} \BibitemShut {NoStop}%
\bibitem [{\citenamefont {{Feldmann}}\ and\ \citenamefont
  {{Mayer}}(2015)}]{Feldmann15}%
  \BibitemOpen
  \bibfield  {author} {\bibinfo {author} {\bibfnamefont {R.}~\bibnamefont
  {{Feldmann}}}\ and\ \bibinfo {author} {\bibfnamefont {L.}~\bibnamefont
  {{Mayer}}},\ }\href {\doibase 10.1093/mnras/stu2207} {\bibfield  {journal}
  {\bibinfo  {journal} {\mnras}\ }\textbf {\bibinfo {volume} {446}},\ \bibinfo
  {pages} {1939} (\bibinfo {year} {2015})},\ \Eprint
  {http://arxiv.org/abs/1404.3212} {arXiv:1404.3212} \BibitemShut {NoStop}%
\bibitem [{\citenamefont {{Peng}}\ \emph {et~al.}(2015)\citenamefont {{Peng}},
  \citenamefont {{Maiolino}},\ and\ \citenamefont {{Cochrane}}}]{Peng15}%
  \BibitemOpen
  \bibfield  {author} {\bibinfo {author} {\bibfnamefont {Y.}~\bibnamefont
  {{Peng}}}, \bibinfo {author} {\bibfnamefont {R.}~\bibnamefont {{Maiolino}}},
  \ and\ \bibinfo {author} {\bibfnamefont {R.}~\bibnamefont {{Cochrane}}},\
  }\href {\doibase 10.1038/nature14439} {\bibfield  {journal} {\bibinfo
  {journal} {\nat}\ }\textbf {\bibinfo {volume} {521}},\ \bibinfo {pages} {192}
  (\bibinfo {year} {2015})},\ \Eprint {http://arxiv.org/abs/1505.03143}
  {arXiv:1505.03143} \BibitemShut {NoStop}%
\bibitem [{\citenamefont {{Dressler}}(1980)}]{Dressler80}%
  \BibitemOpen
  \bibfield  {author} {\bibinfo {author} {\bibfnamefont {A.}~\bibnamefont
  {{Dressler}}},\ }\href {\doibase 10.1086/157753} {\bibfield  {journal}
  {\bibinfo  {journal} {\apj}\ }\textbf {\bibinfo {volume} {236}},\ \bibinfo
  {pages} {351} (\bibinfo {year} {1980})}\BibitemShut {NoStop}%
\bibitem [{\citenamefont {{Weinmann}}\ \emph {et~al.}(2006)\citenamefont
  {{Weinmann}}, \citenamefont {{van den Bosch}}, \citenamefont {{Yang}},\ and\
  \citenamefont {{Mo}}}]{Weinmann06}%
  \BibitemOpen
  \bibfield  {author} {\bibinfo {author} {\bibfnamefont {S.~M.}\ \bibnamefont
  {{Weinmann}}}, \bibinfo {author} {\bibfnamefont {F.~C.}\ \bibnamefont {{van
  den Bosch}}}, \bibinfo {author} {\bibfnamefont {X.}~\bibnamefont {{Yang}}}, \
  and\ \bibinfo {author} {\bibfnamefont {H.~J.}\ \bibnamefont {{Mo}}},\ }\href
  {\doibase 10.1111/j.1365-2966.2005.09865.x} {\bibfield  {journal} {\bibinfo
  {journal} {\mnras}\ }\textbf {\bibinfo {volume} {366}},\ \bibinfo {pages} {2}
  (\bibinfo {year} {2006})},\ \Eprint {http://arxiv.org/abs/astro-ph/0509147}
  {astro-ph/0509147} \BibitemShut {NoStop}%
\bibitem [{\citenamefont {{Kauffmann}}\ \emph {et~al.}(2013)\citenamefont
  {{Kauffmann}}, \citenamefont {{Li}}, \citenamefont {{Zhang}},\ and\
  \citenamefont {{Weinmann}}}]{Kauffmann13}%
  \BibitemOpen
  \bibfield  {author} {\bibinfo {author} {\bibfnamefont {G.}~\bibnamefont
  {{Kauffmann}}}, \bibinfo {author} {\bibfnamefont {C.}~\bibnamefont {{Li}}},
  \bibinfo {author} {\bibfnamefont {W.}~\bibnamefont {{Zhang}}}, \ and\
  \bibinfo {author} {\bibfnamefont {S.}~\bibnamefont {{Weinmann}}},\ }\href
  {\doibase 10.1093/mnras/stt007} {\bibfield  {journal} {\bibinfo  {journal}
  {\mnras}\ }\textbf {\bibinfo {volume} {430}},\ \bibinfo {pages} {1447}
  (\bibinfo {year} {2013})},\ \Eprint {http://arxiv.org/abs/1209.3306}
  {arXiv:1209.3306 [astro-ph.CO]} \BibitemShut {NoStop}%
\bibitem [{\citenamefont {{Hogg}}\ \emph {et~al.}(2003)\citenamefont {{Hogg}},
  \citenamefont {{Blanton}}, \citenamefont {{Eisenstein}}, \citenamefont
  {{Gunn}}, \citenamefont {{Schlegel}}, \citenamefont {{Zehavi}}, \citenamefont
  {{Bahcall}}, \citenamefont {{Brinkmann}}, \citenamefont {{Csabai}},
  \citenamefont {{Schneider}}, \citenamefont {{Weinberg}},\ and\ \citenamefont
  {{York}}}]{Hogg03}%
  \BibitemOpen
  \bibfield  {author} {\bibinfo {author} {\bibfnamefont {D.~W.}\ \bibnamefont
  {{Hogg}}}, \bibinfo {author} {\bibfnamefont {M.~R.}\ \bibnamefont
  {{Blanton}}}, \bibinfo {author} {\bibfnamefont {D.~J.}\ \bibnamefont
  {{Eisenstein}}}, \bibinfo {author} {\bibfnamefont {J.~E.}\ \bibnamefont
  {{Gunn}}}, \bibinfo {author} {\bibfnamefont {D.~J.}\ \bibnamefont
  {{Schlegel}}}, \bibinfo {author} {\bibfnamefont {I.}~\bibnamefont
  {{Zehavi}}}, \bibinfo {author} {\bibfnamefont {N.~A.}\ \bibnamefont
  {{Bahcall}}}, \bibinfo {author} {\bibfnamefont {J.}~\bibnamefont
  {{Brinkmann}}}, \bibinfo {author} {\bibfnamefont {I.}~\bibnamefont
  {{Csabai}}}, \bibinfo {author} {\bibfnamefont {D.~P.}\ \bibnamefont
  {{Schneider}}}, \bibinfo {author} {\bibfnamefont {D.~H.}\ \bibnamefont
  {{Weinberg}}}, \ and\ \bibinfo {author} {\bibfnamefont {D.~G.}\ \bibnamefont
  {{York}}},\ }\href {\doibase 10.1086/374238} {\bibfield  {journal} {\bibinfo
  {journal} {\apjl}\ }\textbf {\bibinfo {volume} {585}},\ \bibinfo {pages} {L5}
  (\bibinfo {year} {2003})},\ \Eprint {http://arxiv.org/abs/astro-ph/0212085}
  {astro-ph/0212085} \BibitemShut {NoStop}%
\bibitem [{\citenamefont {{Blanton}}\ \emph {et~al.}(2005)\citenamefont
  {{Blanton}}, \citenamefont {{Eisenstein}}, \citenamefont {{Hogg}},
  \citenamefont {{Schlegel}},\ and\ \citenamefont {{Brinkmann}}}]{Blanton05}%
  \BibitemOpen
  \bibfield  {author} {\bibinfo {author} {\bibfnamefont {M.~R.}\ \bibnamefont
  {{Blanton}}}, \bibinfo {author} {\bibfnamefont {D.}~\bibnamefont
  {{Eisenstein}}}, \bibinfo {author} {\bibfnamefont {D.~W.}\ \bibnamefont
  {{Hogg}}}, \bibinfo {author} {\bibfnamefont {D.~J.}\ \bibnamefont
  {{Schlegel}}}, \ and\ \bibinfo {author} {\bibfnamefont {J.}~\bibnamefont
  {{Brinkmann}}},\ }\href {\doibase 10.1086/422897} {\bibfield  {journal}
  {\bibinfo  {journal} {\apj}\ }\textbf {\bibinfo {volume} {629}},\ \bibinfo
  {pages} {143} (\bibinfo {year} {2005})},\ \Eprint
  {http://arxiv.org/abs/astro-ph/0310453} {astro-ph/0310453} \BibitemShut
  {NoStop}%
\bibitem [{\citenamefont {{Peng}}\ \emph {et~al.}(2010)\citenamefont {{Peng}},
  \citenamefont {{Lilly}}, \citenamefont {{Kova{\v c}}}, \citenamefont
  {{Bolzonella}}, \citenamefont {{Pozzetti}}, \citenamefont {{Renzini}},
  \citenamefont {{Zamorani}}, \citenamefont {{Ilbert}}, \citenamefont
  {{Knobel}}, \citenamefont {{Iovino}}, \citenamefont {{Maier}}, \citenamefont
  {{Cucciati}}, \citenamefont {{Tasca}}, \citenamefont {{Carollo}},
  \citenamefont {{Silverman}}, \citenamefont {{Kampczyk}}, \citenamefont {{de
  Ravel}}, \citenamefont {{Sanders}}, \citenamefont {{Scoville}}, \citenamefont
  {{Contini}}, \citenamefont {{Mainieri}}, \citenamefont {{Scodeggio}},
  \citenamefont {{Kneib}}, \citenamefont {{Le F{\`e}vre}}, \citenamefont
  {{Bardelli}}, \citenamefont {{Bongiorno}}, \citenamefont {{Caputi}},
  \citenamefont {{Coppa}}, \citenamefont {{de la Torre}}, \citenamefont
  {{Franzetti}}, \citenamefont {{Garilli}}, \citenamefont {{Lamareille}},
  \citenamefont {{Le Borgne}}, \citenamefont {{Le Brun}}, \citenamefont
  {{Mignoli}}, \citenamefont {{Perez Montero}}, \citenamefont {{Pello}},
  \citenamefont {{Ricciardelli}}, \citenamefont {{Tanaka}}, \citenamefont
  {{Tresse}}, \citenamefont {{Vergani}}, \citenamefont {{Welikala}},
  \citenamefont {{Zucca}}, \citenamefont {{Oesch}}, \citenamefont {{Abbas}},
  \citenamefont {{Barnes}}, \citenamefont {{Bordoloi}}, \citenamefont
  {{Bottini}}, \citenamefont {{Cappi}}, \citenamefont {{Cassata}},
  \citenamefont {{Cimatti}}, \citenamefont {{Fumana}}, \citenamefont
  {{Hasinger}}, \citenamefont {{Koekemoer}}, \citenamefont {{Leauthaud}},
  \citenamefont {{Maccagni}}, \citenamefont {{Marinoni}}, \citenamefont
  {{McCracken}}, \citenamefont {{Memeo}}, \citenamefont {{Meneux}},
  \citenamefont {{Nair}}, \citenamefont {{Porciani}}, \citenamefont
  {{Presotto}},\ and\ \citenamefont {{Scaramella}}}]{Peng10}%
  \BibitemOpen
  \bibfield  {author} {\bibinfo {author} {\bibfnamefont {Y.-j.}\ \bibnamefont
  {{Peng}}}, \bibinfo {author} {\bibfnamefont {S.~J.}\ \bibnamefont {{Lilly}}},
  \bibinfo {author} {\bibfnamefont {K.}~\bibnamefont {{Kova{\v c}}}}, \bibinfo
  {author} {\bibfnamefont {M.}~\bibnamefont {{Bolzonella}}}, \bibinfo {author}
  {\bibfnamefont {L.}~\bibnamefont {{Pozzetti}}}, \bibinfo {author}
  {\bibfnamefont {A.}~\bibnamefont {{Renzini}}}, \bibinfo {author}
  {\bibfnamefont {G.}~\bibnamefont {{Zamorani}}}, \bibinfo {author}
  {\bibfnamefont {O.}~\bibnamefont {{Ilbert}}}, \bibinfo {author}
  {\bibfnamefont {C.}~\bibnamefont {{Knobel}}}, \bibinfo {author}
  {\bibfnamefont {A.}~\bibnamefont {{Iovino}}}, \bibinfo {author}
  {\bibfnamefont {C.}~\bibnamefont {{Maier}}}, \bibinfo {author} {\bibfnamefont
  {O.}~\bibnamefont {{Cucciati}}}, \bibinfo {author} {\bibfnamefont
  {L.}~\bibnamefont {{Tasca}}}, \bibinfo {author} {\bibfnamefont {C.~M.}\
  \bibnamefont {{Carollo}}}, \bibinfo {author} {\bibfnamefont {J.}~\bibnamefont
  {{Silverman}}}, \bibinfo {author} {\bibfnamefont {P.}~\bibnamefont
  {{Kampczyk}}}, \bibinfo {author} {\bibfnamefont {L.}~\bibnamefont {{de
  Ravel}}}, \bibinfo {author} {\bibfnamefont {D.}~\bibnamefont {{Sanders}}},
  \bibinfo {author} {\bibfnamefont {N.}~\bibnamefont {{Scoville}}}, \bibinfo
  {author} {\bibfnamefont {T.}~\bibnamefont {{Contini}}}, \bibinfo {author}
  {\bibfnamefont {V.}~\bibnamefont {{Mainieri}}}, \bibinfo {author}
  {\bibfnamefont {M.}~\bibnamefont {{Scodeggio}}}, \bibinfo {author}
  {\bibfnamefont {J.-P.}\ \bibnamefont {{Kneib}}}, \bibinfo {author}
  {\bibfnamefont {O.}~\bibnamefont {{Le F{\`e}vre}}}, \bibinfo {author}
  {\bibfnamefont {S.}~\bibnamefont {{Bardelli}}}, \bibinfo {author}
  {\bibfnamefont {A.}~\bibnamefont {{Bongiorno}}}, \bibinfo {author}
  {\bibfnamefont {K.}~\bibnamefont {{Caputi}}}, \bibinfo {author}
  {\bibfnamefont {G.}~\bibnamefont {{Coppa}}}, \bibinfo {author} {\bibfnamefont
  {S.}~\bibnamefont {{de la Torre}}}, \bibinfo {author} {\bibfnamefont
  {P.}~\bibnamefont {{Franzetti}}}, \bibinfo {author} {\bibfnamefont
  {B.}~\bibnamefont {{Garilli}}}, \bibinfo {author} {\bibfnamefont
  {F.}~\bibnamefont {{Lamareille}}}, \bibinfo {author} {\bibfnamefont {J.-F.}\
  \bibnamefont {{Le Borgne}}}, \bibinfo {author} {\bibfnamefont
  {V.}~\bibnamefont {{Le Brun}}}, \bibinfo {author} {\bibfnamefont
  {M.}~\bibnamefont {{Mignoli}}}, \bibinfo {author} {\bibfnamefont
  {E.}~\bibnamefont {{Perez Montero}}}, \bibinfo {author} {\bibfnamefont
  {R.}~\bibnamefont {{Pello}}}, \bibinfo {author} {\bibfnamefont
  {E.}~\bibnamefont {{Ricciardelli}}}, \bibinfo {author} {\bibfnamefont
  {M.}~\bibnamefont {{Tanaka}}}, \bibinfo {author} {\bibfnamefont
  {L.}~\bibnamefont {{Tresse}}}, \bibinfo {author} {\bibfnamefont
  {D.}~\bibnamefont {{Vergani}}}, \bibinfo {author} {\bibfnamefont
  {N.}~\bibnamefont {{Welikala}}}, \bibinfo {author} {\bibfnamefont
  {E.}~\bibnamefont {{Zucca}}}, \bibinfo {author} {\bibfnamefont
  {P.}~\bibnamefont {{Oesch}}}, \bibinfo {author} {\bibfnamefont
  {U.}~\bibnamefont {{Abbas}}}, \bibinfo {author} {\bibfnamefont
  {L.}~\bibnamefont {{Barnes}}}, \bibinfo {author} {\bibfnamefont
  {R.}~\bibnamefont {{Bordoloi}}}, \bibinfo {author} {\bibfnamefont
  {D.}~\bibnamefont {{Bottini}}}, \bibinfo {author} {\bibfnamefont
  {A.}~\bibnamefont {{Cappi}}}, \bibinfo {author} {\bibfnamefont
  {P.}~\bibnamefont {{Cassata}}}, \bibinfo {author} {\bibfnamefont
  {A.}~\bibnamefont {{Cimatti}}}, \bibinfo {author} {\bibfnamefont
  {M.}~\bibnamefont {{Fumana}}}, \bibinfo {author} {\bibfnamefont
  {G.}~\bibnamefont {{Hasinger}}}, \bibinfo {author} {\bibfnamefont
  {A.}~\bibnamefont {{Koekemoer}}}, \bibinfo {author} {\bibfnamefont
  {A.}~\bibnamefont {{Leauthaud}}}, \bibinfo {author} {\bibfnamefont
  {D.}~\bibnamefont {{Maccagni}}}, \bibinfo {author} {\bibfnamefont
  {C.}~\bibnamefont {{Marinoni}}}, \bibinfo {author} {\bibfnamefont
  {H.}~\bibnamefont {{McCracken}}}, \bibinfo {author} {\bibfnamefont
  {P.}~\bibnamefont {{Memeo}}}, \bibinfo {author} {\bibfnamefont
  {B.}~\bibnamefont {{Meneux}}}, \bibinfo {author} {\bibfnamefont
  {P.}~\bibnamefont {{Nair}}}, \bibinfo {author} {\bibfnamefont
  {C.}~\bibnamefont {{Porciani}}}, \bibinfo {author} {\bibfnamefont
  {V.}~\bibnamefont {{Presotto}}}, \ and\ \bibinfo {author} {\bibfnamefont
  {R.}~\bibnamefont {{Scaramella}}},\ }\href {\doibase
  10.1088/0004-637X/721/1/193} {\bibfield  {journal} {\bibinfo  {journal}
  {\apj}\ }\textbf {\bibinfo {volume} {721}},\ \bibinfo {pages} {193} (\bibinfo
  {year} {2010})},\ \Eprint {http://arxiv.org/abs/1003.4747} {arXiv:1003.4747
  [astro-ph.CO]} \BibitemShut {NoStop}%
\bibitem [{\citenamefont {{van de Voort}}\ \emph {et~al.}(2011)\citenamefont
  {{van de Voort}}, \citenamefont {{Schaye}}, \citenamefont {{Booth}},\ and\
  \citenamefont {{Dalla Vecchia}}}]{Voort11}%
  \BibitemOpen
  \bibfield  {author} {\bibinfo {author} {\bibfnamefont {F.}~\bibnamefont {{van
  de Voort}}}, \bibinfo {author} {\bibfnamefont {J.}~\bibnamefont {{Schaye}}},
  \bibinfo {author} {\bibfnamefont {C.~M.}\ \bibnamefont {{Booth}}}, \ and\
  \bibinfo {author} {\bibfnamefont {C.}~\bibnamefont {{Dalla Vecchia}}},\
  }\href {\doibase 10.1111/j.1365-2966.2011.18896.x} {\bibfield  {journal}
  {\bibinfo  {journal} {\mnras}\ }\textbf {\bibinfo {volume} {415}},\ \bibinfo
  {pages} {2782} (\bibinfo {year} {2011})},\ \Eprint
  {http://arxiv.org/abs/1102.3912} {arXiv:1102.3912} \BibitemShut {NoStop}%
\bibitem [{\citenamefont {{Rees}}\ and\ \citenamefont
  {{Ostriker}}(1977)}]{Rees77}%
  \BibitemOpen
  \bibfield  {author} {\bibinfo {author} {\bibfnamefont {M.~J.}\ \bibnamefont
  {{Rees}}}\ and\ \bibinfo {author} {\bibfnamefont {J.~P.}\ \bibnamefont
  {{Ostriker}}},\ }\href {\doibase 10.1093/mnras/179.4.541} {\bibfield
  {journal} {\bibinfo  {journal} {\mnras}\ }\textbf {\bibinfo {volume} {179}},\
  \bibinfo {pages} {541} (\bibinfo {year} {1977})}\BibitemShut {NoStop}%
\bibitem [{\citenamefont {{White}}\ and\ \citenamefont
  {{Rees}}(1978)}]{White78}%
  \BibitemOpen
  \bibfield  {author} {\bibinfo {author} {\bibfnamefont {S.~D.~M.}\
  \bibnamefont {{White}}}\ and\ \bibinfo {author} {\bibfnamefont {M.~J.}\
  \bibnamefont {{Rees}}},\ }\href {\doibase 10.1093/mnras/183.3.341} {\bibfield
   {journal} {\bibinfo  {journal} {\mnras}\ }\textbf {\bibinfo {volume}
  {183}},\ \bibinfo {pages} {341} (\bibinfo {year} {1978})}\BibitemShut
  {NoStop}%
\bibitem [{\citenamefont {{Lacey}}\ and\ \citenamefont
  {{Silk}}(1991)}]{Lacey91}%
  \BibitemOpen
  \bibfield  {author} {\bibinfo {author} {\bibfnamefont {C.}~\bibnamefont
  {{Lacey}}}\ and\ \bibinfo {author} {\bibfnamefont {J.}~\bibnamefont
  {{Silk}}},\ }\href {\doibase 10.1086/170625} {\bibfield  {journal} {\bibinfo
  {journal} {\apj}\ }\textbf {\bibinfo {volume} {381}},\ \bibinfo {pages} {14}
  (\bibinfo {year} {1991})}\BibitemShut {NoStop}%
\bibitem [{\citenamefont {{Cole}}(1991)}]{Cole91}%
  \BibitemOpen
  \bibfield  {author} {\bibinfo {author} {\bibfnamefont {S.}~\bibnamefont
  {{Cole}}},\ }\href {\doibase 10.1086/169600} {\bibfield  {journal} {\bibinfo
  {journal} {\apj}\ }\textbf {\bibinfo {volume} {367}},\ \bibinfo {pages} {45}
  (\bibinfo {year} {1991})}\BibitemShut {NoStop}%
\bibitem [{\citenamefont {{White}}\ and\ \citenamefont
  {{Frenk}}(1991)}]{White91}%
  \BibitemOpen
  \bibfield  {author} {\bibinfo {author} {\bibfnamefont {S.~D.~M.}\
  \bibnamefont {{White}}}\ and\ \bibinfo {author} {\bibfnamefont {C.~S.}\
  \bibnamefont {{Frenk}}},\ }\href {\doibase 10.1086/170483} {\bibfield
  {journal} {\bibinfo  {journal} {\apj}\ }\textbf {\bibinfo {volume} {379}},\
  \bibinfo {pages} {52} (\bibinfo {year} {1991})}\BibitemShut {NoStop}%
\bibitem [{\citenamefont {{Cole}}\ \emph {et~al.}(2000)\citenamefont {{Cole}},
  \citenamefont {{Lacey}}, \citenamefont {{Baugh}},\ and\ \citenamefont
  {{Frenk}}}]{Cole00}%
  \BibitemOpen
  \bibfield  {author} {\bibinfo {author} {\bibfnamefont {S.}~\bibnamefont
  {{Cole}}}, \bibinfo {author} {\bibfnamefont {C.~G.}\ \bibnamefont {{Lacey}}},
  \bibinfo {author} {\bibfnamefont {C.~M.}\ \bibnamefont {{Baugh}}}, \ and\
  \bibinfo {author} {\bibfnamefont {C.~S.}\ \bibnamefont {{Frenk}}},\ }\href
  {\doibase 10.1046/j.1365-8711.2000.03879.x} {\bibfield  {journal} {\bibinfo
  {journal} {\mnras}\ }\textbf {\bibinfo {volume} {319}},\ \bibinfo {pages}
  {168} (\bibinfo {year} {2000})},\ \Eprint
  {http://arxiv.org/abs/astro-ph/0007281} {astro-ph/0007281} \BibitemShut
  {NoStop}%
\bibitem [{\citenamefont {{Benson}}\ \emph {et~al.}(2001)\citenamefont
  {{Benson}}, \citenamefont {{Pearce}}, \citenamefont {{Frenk}}, \citenamefont
  {{Baugh}},\ and\ \citenamefont {{Jenkins}}}]{Benson01}%
  \BibitemOpen
  \bibfield  {author} {\bibinfo {author} {\bibfnamefont {A.~J.}\ \bibnamefont
  {{Benson}}}, \bibinfo {author} {\bibfnamefont {F.~R.}\ \bibnamefont
  {{Pearce}}}, \bibinfo {author} {\bibfnamefont {C.~S.}\ \bibnamefont
  {{Frenk}}}, \bibinfo {author} {\bibfnamefont {C.~M.}\ \bibnamefont
  {{Baugh}}}, \ and\ \bibinfo {author} {\bibfnamefont {A.}~\bibnamefont
  {{Jenkins}}},\ }\href {\doibase 10.1046/j.1365-8711.2001.03966.x} {\bibfield
  {journal} {\bibinfo  {journal} {\mnras}\ }\textbf {\bibinfo {volume} {320}},\
  \bibinfo {pages} {261} (\bibinfo {year} {2001})},\ \Eprint
  {http://arxiv.org/abs/astro-ph/9912220} {astro-ph/9912220} \BibitemShut
  {NoStop}%
\bibitem [{\citenamefont {{Hearin}}\ and\ \citenamefont
  {{Watson}}(2013)}]{Hearing13}%
  \BibitemOpen
  \bibfield  {author} {\bibinfo {author} {\bibfnamefont {A.~P.}\ \bibnamefont
  {{Hearin}}}\ and\ \bibinfo {author} {\bibfnamefont {D.~F.}\ \bibnamefont
  {{Watson}}},\ }\href {\doibase 10.1093/mnras/stt1374} {\bibfield  {journal}
  {\bibinfo  {journal} {\mnras}\ }\textbf {\bibinfo {volume} {435}},\ \bibinfo
  {pages} {1313} (\bibinfo {year} {2013})},\ \Eprint
  {http://arxiv.org/abs/1304.5557} {arXiv:1304.5557} \BibitemShut {NoStop}%
\bibitem [{\citenamefont {{Mutch}}\ \emph {et~al.}(2013)\citenamefont
  {{Mutch}}, \citenamefont {{Croton}},\ and\ \citenamefont
  {{Poole}}}]{Mutch13}%
  \BibitemOpen
  \bibfield  {author} {\bibinfo {author} {\bibfnamefont {S.~J.}\ \bibnamefont
  {{Mutch}}}, \bibinfo {author} {\bibfnamefont {D.~J.}\ \bibnamefont
  {{Croton}}}, \ and\ \bibinfo {author} {\bibfnamefont {G.~B.}\ \bibnamefont
  {{Poole}}},\ }\href {\doibase 10.1093/mnras/stt1453} {\bibfield  {journal}
  {\bibinfo  {journal} {\mnras}\ }\textbf {\bibinfo {volume} {435}},\ \bibinfo
  {pages} {2445} (\bibinfo {year} {2013})},\ \Eprint
  {http://arxiv.org/abs/1304.2774} {arXiv:1304.2774} \BibitemShut {NoStop}%
\bibitem [{\citenamefont {{Rodr{\'{\i}}guez-Puebla}}\ \emph
  {et~al.}(2016)\citenamefont {{Rodr{\'{\i}}guez-Puebla}}, \citenamefont
  {{Primack}}, \citenamefont {{Behroozi}},\ and\ \citenamefont
  {{Faber}}}]{Aldo16}%
  \BibitemOpen
  \bibfield  {author} {\bibinfo {author} {\bibfnamefont {A.}~\bibnamefont
  {{Rodr{\'{\i}}guez-Puebla}}}, \bibinfo {author} {\bibfnamefont {J.~R.}\
  \bibnamefont {{Primack}}}, \bibinfo {author} {\bibfnamefont {P.}~\bibnamefont
  {{Behroozi}}}, \ and\ \bibinfo {author} {\bibfnamefont {S.~M.}\ \bibnamefont
  {{Faber}}},\ }\href {\doibase 10.1093/mnras/stv2513} {\bibfield  {journal}
  {\bibinfo  {journal} {\mnras}\ }\textbf {\bibinfo {volume} {455}},\ \bibinfo
  {pages} {2592} (\bibinfo {year} {2016})},\ \Eprint
  {http://arxiv.org/abs/1508.04842} {arXiv:1508.04842} \BibitemShut {NoStop}%
\bibitem [{\citenamefont {{Trujillo}}\ \emph {et~al.}(2006)\citenamefont
  {{Trujillo}}, \citenamefont {{Carretero}},\ and\ \citenamefont
  {{Patiri}}}]{Trujillo06}%
  \BibitemOpen
  \bibfield  {author} {\bibinfo {author} {\bibfnamefont {I.}~\bibnamefont
  {{Trujillo}}}, \bibinfo {author} {\bibfnamefont {C.}~\bibnamefont
  {{Carretero}}}, \ and\ \bibinfo {author} {\bibfnamefont {S.~G.}\ \bibnamefont
  {{Patiri}}},\ }\href {\doibase 10.1086/503548} {\bibfield  {journal}
  {\bibinfo  {journal} {\apjl}\ }\textbf {\bibinfo {volume} {640}},\ \bibinfo
  {pages} {L111} (\bibinfo {year} {2006})},\ \Eprint
  {http://arxiv.org/abs/astro-ph/0511680} {astro-ph/0511680} \BibitemShut
  {NoStop}%
\bibitem [{\citenamefont {{Arag{\'o}n-Calvo}}\ \emph
  {et~al.}(2007{\natexlab{a}})\citenamefont {{Arag{\'o}n-Calvo}}, \citenamefont
  {{van de Weygaert}}, \citenamefont {{Jones}},\ and\ \citenamefont {{van der
  Hulst}}}]{Aragon07}%
  \BibitemOpen
  \bibfield  {author} {\bibinfo {author} {\bibfnamefont {M.~A.}\ \bibnamefont
  {{Arag{\'o}n-Calvo}}}, \bibinfo {author} {\bibfnamefont {R.}~\bibnamefont
  {{van de Weygaert}}}, \bibinfo {author} {\bibfnamefont {B.~J.~T.}\
  \bibnamefont {{Jones}}}, \ and\ \bibinfo {author} {\bibfnamefont {J.~M.}\
  \bibnamefont {{van der Hulst}}},\ }\href {\doibase 10.1086/511633} {\bibfield
   {journal} {\bibinfo  {journal} {\apjl}\ }\textbf {\bibinfo {volume} {655}},\
  \bibinfo {pages} {L5} (\bibinfo {year} {2007}{\natexlab{a}})},\ \Eprint
  {http://arxiv.org/abs/astro-ph/0610249} {astro-ph/0610249} \BibitemShut
  {NoStop}%
\bibitem [{\citenamefont {{Arag{\'o}n-Calvo}}\ \emph
  {et~al.}(2007{\natexlab{b}})\citenamefont {{Arag{\'o}n-Calvo}}, \citenamefont
  {{Jones}}, \citenamefont {{van de Weygaert}},\ and\ \citenamefont {{van der
  Hulst}}}]{Aragon07b}%
  \BibitemOpen
  \bibfield  {author} {\bibinfo {author} {\bibfnamefont {M.~A.}\ \bibnamefont
  {{Arag{\'o}n-Calvo}}}, \bibinfo {author} {\bibfnamefont {B.~J.~T.}\
  \bibnamefont {{Jones}}}, \bibinfo {author} {\bibfnamefont {R.}~\bibnamefont
  {{van de Weygaert}}}, \ and\ \bibinfo {author} {\bibfnamefont {J.~M.}\
  \bibnamefont {{van der Hulst}}},\ }\href {\doibase
  10.1051/0004-6361:20077880} {\bibfield  {journal} {\bibinfo  {journal}
  {\aap}\ }\textbf {\bibinfo {volume} {474}},\ \bibinfo {pages} {315} (\bibinfo
  {year} {2007}{\natexlab{b}})},\ \Eprint {http://arxiv.org/abs/0705.2072}
  {arXiv:0705.2072} \BibitemShut {NoStop}%
\bibitem [{\citenamefont {{Hahn}}\ \emph {et~al.}(2007)\citenamefont {{Hahn}},
  \citenamefont {{Carollo}}, \citenamefont {{Porciani}},\ and\ \citenamefont
  {{Dekel}}}]{Hahn07}%
  \BibitemOpen
  \bibfield  {author} {\bibinfo {author} {\bibfnamefont {O.}~\bibnamefont
  {{Hahn}}}, \bibinfo {author} {\bibfnamefont {C.~M.}\ \bibnamefont
  {{Carollo}}}, \bibinfo {author} {\bibfnamefont {C.}~\bibnamefont
  {{Porciani}}}, \ and\ \bibinfo {author} {\bibfnamefont {A.}~\bibnamefont
  {{Dekel}}},\ }\href {\doibase 10.1111/j.1365-2966.2007.12249.x} {\bibfield
  {journal} {\bibinfo  {journal} {\mnras}\ }\textbf {\bibinfo {volume} {381}},\
  \bibinfo {pages} {41} (\bibinfo {year} {2007})},\ \Eprint
  {http://arxiv.org/abs/0704.2595} {arXiv:0704.2595} \BibitemShut {NoStop}%
\bibitem [{\citenamefont {{Dekel}}\ \emph {et~al.}(2009)\citenamefont
  {{Dekel}}, \citenamefont {{Birnboim}}, \citenamefont {{Engel}}, \citenamefont
  {{Freundlich}}, \citenamefont {{Goerdt}}, \citenamefont {{Mumcuoglu}},
  \citenamefont {{Neistein}}, \citenamefont {{Pichon}}, \citenamefont
  {{Teyssier}},\ and\ \citenamefont {{Zinger}}}]{Dekel09}%
  \BibitemOpen
  \bibfield  {author} {\bibinfo {author} {\bibfnamefont {A.}~\bibnamefont
  {{Dekel}}}, \bibinfo {author} {\bibfnamefont {Y.}~\bibnamefont {{Birnboim}}},
  \bibinfo {author} {\bibfnamefont {G.}~\bibnamefont {{Engel}}}, \bibinfo
  {author} {\bibfnamefont {J.}~\bibnamefont {{Freundlich}}}, \bibinfo {author}
  {\bibfnamefont {T.}~\bibnamefont {{Goerdt}}}, \bibinfo {author}
  {\bibfnamefont {M.}~\bibnamefont {{Mumcuoglu}}}, \bibinfo {author}
  {\bibfnamefont {E.}~\bibnamefont {{Neistein}}}, \bibinfo {author}
  {\bibfnamefont {C.}~\bibnamefont {{Pichon}}}, \bibinfo {author}
  {\bibfnamefont {R.}~\bibnamefont {{Teyssier}}}, \ and\ \bibinfo {author}
  {\bibfnamefont {E.}~\bibnamefont {{Zinger}}},\ }\href {\doibase
  10.1038/nature07648} {\bibfield  {journal} {\bibinfo  {journal} {\nat}\
  }\textbf {\bibinfo {volume} {457}},\ \bibinfo {pages} {451} (\bibinfo {year}
  {2009})},\ \Eprint {http://arxiv.org/abs/0808.0553} {arXiv:0808.0553}
  \BibitemShut {NoStop}%
\bibitem [{\citenamefont {{Pichon}}\ \emph {et~al.}(2011)\citenamefont
  {{Pichon}}, \citenamefont {{Pogosyan}}, \citenamefont {{Kimm}}, \citenamefont
  {{Slyz}}, \citenamefont {{Devriendt}},\ and\ \citenamefont
  {{Dubois}}}]{Pichon11}%
  \BibitemOpen
  \bibfield  {author} {\bibinfo {author} {\bibfnamefont {C.}~\bibnamefont
  {{Pichon}}}, \bibinfo {author} {\bibfnamefont {D.}~\bibnamefont
  {{Pogosyan}}}, \bibinfo {author} {\bibfnamefont {T.}~\bibnamefont {{Kimm}}},
  \bibinfo {author} {\bibfnamefont {A.}~\bibnamefont {{Slyz}}}, \bibinfo
  {author} {\bibfnamefont {J.}~\bibnamefont {{Devriendt}}}, \ and\ \bibinfo
  {author} {\bibfnamefont {Y.}~\bibnamefont {{Dubois}}},\ }\href {\doibase
  10.1111/j.1365-2966.2011.19640.x} {\bibfield  {journal} {\bibinfo  {journal}
  {\mnras}\ }\textbf {\bibinfo {volume} {418}},\ \bibinfo {pages} {2493}
  (\bibinfo {year} {2011})},\ \Eprint {http://arxiv.org/abs/1105.0210}
  {arXiv:1105.0210} \BibitemShut {NoStop}%
\bibitem [{\citenamefont {{Laigle}}\ \emph {et~al.}(2015)\citenamefont
  {{Laigle}}, \citenamefont {{Pichon}}, \citenamefont {{Codis}}, \citenamefont
  {{Dubois}}, \citenamefont {{Le Borgne}}, \citenamefont {{Pogosyan}},
  \citenamefont {{Devriendt}}, \citenamefont {{Peirani}}, \citenamefont
  {{Prunet}}, \citenamefont {{Rouberol}}, \citenamefont {{Slyz}},\ and\
  \citenamefont {{Sousbie}}}]{Laigle15}%
  \BibitemOpen
  \bibfield  {author} {\bibinfo {author} {\bibfnamefont {C.}~\bibnamefont
  {{Laigle}}}, \bibinfo {author} {\bibfnamefont {C.}~\bibnamefont {{Pichon}}},
  \bibinfo {author} {\bibfnamefont {S.}~\bibnamefont {{Codis}}}, \bibinfo
  {author} {\bibfnamefont {Y.}~\bibnamefont {{Dubois}}}, \bibinfo {author}
  {\bibfnamefont {D.}~\bibnamefont {{Le Borgne}}}, \bibinfo {author}
  {\bibfnamefont {D.}~\bibnamefont {{Pogosyan}}}, \bibinfo {author}
  {\bibfnamefont {J.}~\bibnamefont {{Devriendt}}}, \bibinfo {author}
  {\bibfnamefont {S.}~\bibnamefont {{Peirani}}}, \bibinfo {author}
  {\bibfnamefont {S.}~\bibnamefont {{Prunet}}}, \bibinfo {author}
  {\bibfnamefont {S.}~\bibnamefont {{Rouberol}}}, \bibinfo {author}
  {\bibfnamefont {A.}~\bibnamefont {{Slyz}}}, \ and\ \bibinfo {author}
  {\bibfnamefont {T.}~\bibnamefont {{Sousbie}}},\ }\href {\doibase
  10.1093/mnras/stu2289} {\bibfield  {journal} {\bibinfo  {journal} {\mnras}\
  }\textbf {\bibinfo {volume} {446}},\ \bibinfo {pages} {2744} (\bibinfo {year}
  {2015})},\ \Eprint {http://arxiv.org/abs/1310.3801} {arXiv:1310.3801}
  \BibitemShut {NoStop}%
\bibitem [{\citenamefont {{Darvish}}\ \emph {et~al.}(2016)\citenamefont
  {{Darvish}}, \citenamefont {{Mobasher}}, \citenamefont {{Sobral}},
  \citenamefont {{Rettura}}, \citenamefont {{Scoville}}, \citenamefont
  {{Faisst}},\ and\ \citenamefont {{Capak}}}]{Darvish16}%
  \BibitemOpen
  \bibfield  {author} {\bibinfo {author} {\bibfnamefont {B.}~\bibnamefont
  {{Darvish}}}, \bibinfo {author} {\bibfnamefont {B.}~\bibnamefont
  {{Mobasher}}}, \bibinfo {author} {\bibfnamefont {D.}~\bibnamefont
  {{Sobral}}}, \bibinfo {author} {\bibfnamefont {A.}~\bibnamefont {{Rettura}}},
  \bibinfo {author} {\bibfnamefont {N.}~\bibnamefont {{Scoville}}}, \bibinfo
  {author} {\bibfnamefont {A.}~\bibnamefont {{Faisst}}}, \ and\ \bibinfo
  {author} {\bibfnamefont {P.}~\bibnamefont {{Capak}}},\ }\href {\doibase
  10.3847/0004-637X/825/2/113} {\bibfield  {journal} {\bibinfo  {journal}
  {\apj}\ }\textbf {\bibinfo {volume} {825}},\ \bibinfo {eid} {113} (\bibinfo
  {year} {2016})},\ \Eprint {http://arxiv.org/abs/1605.03182}
  {arXiv:1605.03182} \BibitemShut {NoStop}%
\bibitem [{\citenamefont {{Hearin}}\ \emph {et~al.}(2014)\citenamefont
  {{Hearin}}, \citenamefont {{Watson}}, \citenamefont {{Becker}}, \citenamefont
  {{Reyes}}, \citenamefont {{Berlind}},\ and\ \citenamefont
  {{Zentner}}}]{Hearing14}%
  \BibitemOpen
  \bibfield  {author} {\bibinfo {author} {\bibfnamefont {A.~P.}\ \bibnamefont
  {{Hearin}}}, \bibinfo {author} {\bibfnamefont {D.~F.}\ \bibnamefont
  {{Watson}}}, \bibinfo {author} {\bibfnamefont {M.~R.}\ \bibnamefont
  {{Becker}}}, \bibinfo {author} {\bibfnamefont {R.}~\bibnamefont {{Reyes}}},
  \bibinfo {author} {\bibfnamefont {A.~A.}\ \bibnamefont {{Berlind}}}, \ and\
  \bibinfo {author} {\bibfnamefont {A.~R.}\ \bibnamefont {{Zentner}}},\ }\href
  {\doibase 10.1093/mnras/stu1443} {\bibfield  {journal} {\bibinfo  {journal}
  {\mnras}\ }\textbf {\bibinfo {volume} {444}},\ \bibinfo {pages} {729}
  (\bibinfo {year} {2014})},\ \Eprint {http://arxiv.org/abs/1310.6747}
  {arXiv:1310.6747} \BibitemShut {NoStop}%
\bibitem [{\citenamefont {{Watson}}\ \emph {et~al.}(2015)\citenamefont
  {{Watson}}, \citenamefont {{Hearin}}, \citenamefont {{Berlind}},
  \citenamefont {{Becker}}, \citenamefont {{Behroozi}}, \citenamefont
  {{Skibba}}, \citenamefont {{Reyes}}, \citenamefont {{Zentner}},\ and\
  \citenamefont {{van den Bosch}}}]{Watson15}%
  \BibitemOpen
  \bibfield  {author} {\bibinfo {author} {\bibfnamefont {D.~F.}\ \bibnamefont
  {{Watson}}}, \bibinfo {author} {\bibfnamefont {A.~P.}\ \bibnamefont
  {{Hearin}}}, \bibinfo {author} {\bibfnamefont {A.~A.}\ \bibnamefont
  {{Berlind}}}, \bibinfo {author} {\bibfnamefont {M.~R.}\ \bibnamefont
  {{Becker}}}, \bibinfo {author} {\bibfnamefont {P.~S.}\ \bibnamefont
  {{Behroozi}}}, \bibinfo {author} {\bibfnamefont {R.~A.}\ \bibnamefont
  {{Skibba}}}, \bibinfo {author} {\bibfnamefont {R.}~\bibnamefont {{Reyes}}},
  \bibinfo {author} {\bibfnamefont {A.~R.}\ \bibnamefont {{Zentner}}}, \ and\
  \bibinfo {author} {\bibfnamefont {F.~C.}\ \bibnamefont {{van den Bosch}}},\
  }\href {\doibase 10.1093/mnras/stu2065} {\bibfield  {journal} {\bibinfo
  {journal} {\mnras}\ }\textbf {\bibinfo {volume} {446}},\ \bibinfo {pages}
  {651} (\bibinfo {year} {2015})},\ \Eprint {http://arxiv.org/abs/1403.1578}
  {arXiv:1403.1578} \BibitemShut {NoStop}%
\bibitem [{\citenamefont {{Aragon-Calvo}}\ and\ \citenamefont
  {{Szalay}}(2013)}]{Aragon13}%
  \BibitemOpen
  \bibfield  {author} {\bibinfo {author} {\bibfnamefont {M.~A.}\ \bibnamefont
  {{Aragon-Calvo}}}\ and\ \bibinfo {author} {\bibfnamefont {A.~S.}\
  \bibnamefont {{Szalay}}},\ }\href {\doibase 10.1093/mnras/sts281} {\bibfield
  {journal} {\bibinfo  {journal} {\mnras}\ }\textbf {\bibinfo {volume} {428}},\
  \bibinfo {pages} {3409} (\bibinfo {year} {2013})},\ \Eprint
  {http://arxiv.org/abs/1203.0248} {arXiv:1203.0248 [astro-ph.CO]} \BibitemShut
  {NoStop}%
\bibitem [{\citenamefont {{Zel'dovich}}(1970)}]{Zeldovich70}%
  \BibitemOpen
  \bibfield  {author} {\bibinfo {author} {\bibfnamefont {Y.~B.}\ \bibnamefont
  {{Zel'dovich}}},\ }\href@noop {} {\bibfield  {journal} {\bibinfo  {journal}
  {\aap}\ }\textbf {\bibinfo {volume} {5}},\ \bibinfo {pages} {84} (\bibinfo
  {year} {1970})}\BibitemShut {NoStop}%
\bibitem [{\citenamefont {{Hidding}}\ \emph {et~al.}(2014)\citenamefont
  {{Hidding}}, \citenamefont {{Shandarin}},\ and\ \citenamefont {{van de
  Weygaert}}}]{Hidding14}%
  \BibitemOpen
  \bibfield  {author} {\bibinfo {author} {\bibfnamefont {J.}~\bibnamefont
  {{Hidding}}}, \bibinfo {author} {\bibfnamefont {S.~F.}\ \bibnamefont
  {{Shandarin}}}, \ and\ \bibinfo {author} {\bibfnamefont {R.}~\bibnamefont
  {{van de Weygaert}}},\ }\href {\doibase 10.1093/mnras/stt2142} {\bibfield
  {journal} {\bibinfo  {journal} {\mnras}\ }\textbf {\bibinfo {volume} {437}},\
  \bibinfo {pages} {3442} (\bibinfo {year} {2014})},\ \Eprint
  {http://arxiv.org/abs/1311.7134} {arXiv:1311.7134 [astro-ph.CO]} \BibitemShut
  {NoStop}%
\bibitem [{\citenamefont {{Kere{\v s}}}\ \emph {et~al.}(2005)\citenamefont
  {{Kere{\v s}}}, \citenamefont {{Katz}}, \citenamefont {{Weinberg}},\ and\
  \citenamefont {{Dav{\'e}}}}]{keres05}%
  \BibitemOpen
  \bibfield  {author} {\bibinfo {author} {\bibfnamefont {D.}~\bibnamefont
  {{Kere{\v s}}}}, \bibinfo {author} {\bibfnamefont {N.}~\bibnamefont
  {{Katz}}}, \bibinfo {author} {\bibfnamefont {D.~H.}\ \bibnamefont
  {{Weinberg}}}, \ and\ \bibinfo {author} {\bibfnamefont {R.}~\bibnamefont
  {{Dav{\'e}}}},\ }\href {\doibase 10.1111/j.1365-2966.2005.09451.x} {\bibfield
   {journal} {\bibinfo  {journal} {\mnras}\ }\textbf {\bibinfo {volume}
  {363}},\ \bibinfo {pages} {2} (\bibinfo {year} {2005})},\ \Eprint
  {http://arxiv.org/abs/astro-ph/0407095} {astro-ph/0407095} \BibitemShut
  {NoStop}%
\bibitem [{\citenamefont {{Danovich}}\ \emph {et~al.}(2012)\citenamefont
  {{Danovich}}, \citenamefont {{Dekel}}, \citenamefont {{Hahn}},\ and\
  \citenamefont {{Teyssier}}}]{Danovich12}%
  \BibitemOpen
  \bibfield  {author} {\bibinfo {author} {\bibfnamefont {M.}~\bibnamefont
  {{Danovich}}}, \bibinfo {author} {\bibfnamefont {A.}~\bibnamefont {{Dekel}}},
  \bibinfo {author} {\bibfnamefont {O.}~\bibnamefont {{Hahn}}}, \ and\ \bibinfo
  {author} {\bibfnamefont {R.}~\bibnamefont {{Teyssier}}},\ }\href {\doibase
  10.1111/j.1365-2966.2012.20751.x} {\bibfield  {journal} {\bibinfo  {journal}
  {\mnras}\ }\textbf {\bibinfo {volume} {422}},\ \bibinfo {pages} {1732}
  (\bibinfo {year} {2012})},\ \Eprint {http://arxiv.org/abs/1110.6209}
  {arXiv:1110.6209} \BibitemShut {NoStop}%
\bibitem [{\citenamefont {{Harford}}\ and\ \citenamefont
  {{Hamilton}}(2016)}]{Harford16}%
  \BibitemOpen
  \bibfield  {author} {\bibinfo {author} {\bibfnamefont {G.~A.}\ \bibnamefont
  {{Harford}}}\ and\ \bibinfo {author} {\bibfnamefont {A.~J.~S.}\ \bibnamefont
  {{Hamilton}}},\ }\href@noop {} {\bibfield  {journal} {\bibinfo  {journal}
  {ArXiv e-prints}\ } (\bibinfo {year} {2016})},\ \Eprint
  {http://arxiv.org/abs/1601.01737} {arXiv:1601.01737} \BibitemShut {NoStop}%
\bibitem [{\citenamefont {{Faucher-Gigu{\`e}re}}\ and\ \citenamefont {{Kere{\v
  s}}}(2011)}]{Faucher11}%
  \BibitemOpen
  \bibfield  {author} {\bibinfo {author} {\bibfnamefont {C.-A.}\ \bibnamefont
  {{Faucher-Gigu{\`e}re}}}\ and\ \bibinfo {author} {\bibfnamefont
  {D.}~\bibnamefont {{Kere{\v s}}}},\ }\href {\doibase
  10.1111/j.1745-3933.2011.01018.x} {\bibfield  {journal} {\bibinfo  {journal}
  {\mnras}\ }\textbf {\bibinfo {volume} {412}},\ \bibinfo {pages} {L118}
  (\bibinfo {year} {2011})},\ \Eprint {http://arxiv.org/abs/1011.1693}
  {arXiv:1011.1693 [astro-ph.CO]} \BibitemShut {NoStop}%
\bibitem [{\citenamefont {{Bond}}\ \emph {et~al.}(1996)\citenamefont {{Bond}},
  \citenamefont {{Kofman}},\ and\ \citenamefont {{Pogosyan}}}]{Bond96}%
  \BibitemOpen
  \bibfield  {author} {\bibinfo {author} {\bibfnamefont {J.~R.}\ \bibnamefont
  {{Bond}}}, \bibinfo {author} {\bibfnamefont {L.}~\bibnamefont {{Kofman}}}, \
  and\ \bibinfo {author} {\bibfnamefont {D.}~\bibnamefont {{Pogosyan}}},\
  }\href {\doibase 10.1038/380603a0} {\bibfield  {journal} {\bibinfo  {journal}
  {\nat}\ }\textbf {\bibinfo {volume} {380}},\ \bibinfo {pages} {603} (\bibinfo
  {year} {1996})},\ \Eprint {http://arxiv.org/abs/astro-ph/9512141}
  {astro-ph/9512141} \BibitemShut {NoStop}%
\bibitem [{\citenamefont {{Joeveer}}\ and\ \citenamefont
  {{Einasto}}(1978)}]{Joeveer78}%
  \BibitemOpen
  \bibfield  {author} {\bibinfo {author} {\bibfnamefont {M.}~\bibnamefont
  {{Joeveer}}}\ and\ \bibinfo {author} {\bibfnamefont {J.}~\bibnamefont
  {{Einasto}}},\ }in\ \href@noop {} {\emph {\bibinfo {booktitle} {Large Scale
  Structures in the Universe}}},\ \bibinfo {series} {IAU Symposium},
  Vol.~\bibinfo {volume} {79},\ \bibinfo {editor} {edited by\ \bibinfo {editor}
  {\bibfnamefont {M.~S.}\ \bibnamefont {{Longair}}}\ and\ \bibinfo {editor}
  {\bibfnamefont {J.}~\bibnamefont {{Einasto}}}}\ (\bibinfo {year} {1978})\
  pp.\ \bibinfo {pages} {241--250}\BibitemShut {NoStop}%
\bibitem [{\citenamefont {{Klypin}}\ and\ \citenamefont
  {{Shandarin}}(1983)}]{Klypin83}%
  \BibitemOpen
  \bibfield  {author} {\bibinfo {author} {\bibfnamefont {A.~A.}\ \bibnamefont
  {{Klypin}}}\ and\ \bibinfo {author} {\bibfnamefont {S.~F.}\ \bibnamefont
  {{Shandarin}}},\ }\href {\doibase 10.1093/mnras/204.3.891} {\bibfield
  {journal} {\bibinfo  {journal} {\mnras}\ }\textbf {\bibinfo {volume} {204}},\
  \bibinfo {pages} {891} (\bibinfo {year} {1983})}\BibitemShut {NoStop}%
\bibitem [{\citenamefont {{Geller}}\ and\ \citenamefont
  {{Huchra}}(1989)}]{Geller89}%
  \BibitemOpen
  \bibfield  {author} {\bibinfo {author} {\bibfnamefont {M.~J.}\ \bibnamefont
  {{Geller}}}\ and\ \bibinfo {author} {\bibfnamefont {J.~P.}\ \bibnamefont
  {{Huchra}}},\ }\href {\doibase 10.1126/science.246.4932.897} {\bibfield
  {journal} {\bibinfo  {journal} {Science}\ }\textbf {\bibinfo {volume}
  {246}},\ \bibinfo {pages} {897} (\bibinfo {year} {1989})}\BibitemShut
  {NoStop}%
\bibitem [{\citenamefont {{Icke}}\ and\ \citenamefont {{van de
  Weygaert}}(1991)}]{Icke91}%
  \BibitemOpen
  \bibfield  {author} {\bibinfo {author} {\bibfnamefont {V.}~\bibnamefont
  {{Icke}}}\ and\ \bibinfo {author} {\bibfnamefont {R.}~\bibnamefont {{van de
  Weygaert}}},\ }\href@noop {} {\bibfield  {journal} {\bibinfo  {journal}
  {\qjras}\ }\textbf {\bibinfo {volume} {32}},\ \bibinfo {pages} {85} (\bibinfo
  {year} {1991})}\BibitemShut {NoStop}%
\bibitem [{\citenamefont {{Arag{\'o}n-Calvo}}\ \emph
  {et~al.}(2010)\citenamefont {{Arag{\'o}n-Calvo}}, \citenamefont {{van de
  Weygaert}},\ and\ \citenamefont {{Jones}}}]{Aragon10c}%
  \BibitemOpen
  \bibfield  {author} {\bibinfo {author} {\bibfnamefont {M.~A.}\ \bibnamefont
  {{Arag{\'o}n-Calvo}}}, \bibinfo {author} {\bibfnamefont {R.}~\bibnamefont
  {{van de Weygaert}}}, \ and\ \bibinfo {author} {\bibfnamefont {B.~J.~T.}\
  \bibnamefont {{Jones}}},\ }\href {\doibase 10.1111/j.1365-2966.2010.17263.x}
  {\bibfield  {journal} {\bibinfo  {journal} {\mnras}\ }\textbf {\bibinfo
  {volume} {408}},\ \bibinfo {pages} {2163} (\bibinfo {year} {2010})},\ \Eprint
  {http://arxiv.org/abs/1007.0742} {arXiv:1007.0742 [astro-ph.CO]} \BibitemShut
  {NoStop}%
\bibitem [{\citenamefont {{Einasto}}\ \emph {et~al.}(2011)\citenamefont
  {{Einasto}}, \citenamefont {{Suhhonenko}}, \citenamefont {{H{\"u}tsi}},
  \citenamefont {{Saar}}, \citenamefont {{Einasto}}, \citenamefont
  {{Liivam{\"a}gi}}, \citenamefont {{M{\"u}ller}}, \citenamefont
  {{Starobinsky}}, \citenamefont {{Tago}},\ and\ \citenamefont
  {{Tempel}}}]{Einasto11}%
  \BibitemOpen
  \bibfield  {author} {\bibinfo {author} {\bibfnamefont {J.}~\bibnamefont
  {{Einasto}}}, \bibinfo {author} {\bibfnamefont {I.}~\bibnamefont
  {{Suhhonenko}}}, \bibinfo {author} {\bibfnamefont {G.}~\bibnamefont
  {{H{\"u}tsi}}}, \bibinfo {author} {\bibfnamefont {E.}~\bibnamefont {{Saar}}},
  \bibinfo {author} {\bibfnamefont {M.}~\bibnamefont {{Einasto}}}, \bibinfo
  {author} {\bibfnamefont {L.~J.}\ \bibnamefont {{Liivam{\"a}gi}}}, \bibinfo
  {author} {\bibfnamefont {V.}~\bibnamefont {{M{\"u}ller}}}, \bibinfo {author}
  {\bibfnamefont {A.~A.}\ \bibnamefont {{Starobinsky}}}, \bibinfo {author}
  {\bibfnamefont {E.}~\bibnamefont {{Tago}}}, \ and\ \bibinfo {author}
  {\bibfnamefont {E.}~\bibnamefont {{Tempel}}},\ }\href {\doibase
  10.1051/0004-6361/201117248} {\bibfield  {journal} {\bibinfo  {journal}
  {\aap}\ }\textbf {\bibinfo {volume} {534}},\ \bibinfo {eid} {A128} (\bibinfo
  {year} {2011})},\ \Eprint {http://arxiv.org/abs/1105.2464} {arXiv:1105.2464
  [astro-ph.CO]} \BibitemShut {NoStop}%
\bibitem [{\citenamefont {{Gottl{\"o}ber}}\ \emph {et~al.}(2003)\citenamefont
  {{Gottl{\"o}ber}}, \citenamefont {{{\L}okas}}, \citenamefont {{Klypin}},\
  and\ \citenamefont {{Hoffman}}}]{Gottlober03}%
  \BibitemOpen
  \bibfield  {author} {\bibinfo {author} {\bibfnamefont {S.}~\bibnamefont
  {{Gottl{\"o}ber}}}, \bibinfo {author} {\bibfnamefont {E.~L.}\ \bibnamefont
  {{{\L}okas}}}, \bibinfo {author} {\bibfnamefont {A.}~\bibnamefont
  {{Klypin}}}, \ and\ \bibinfo {author} {\bibfnamefont {Y.}~\bibnamefont
  {{Hoffman}}},\ }\href {\doibase 10.1046/j.1365-8711.2003.06850.x} {\bibfield
  {journal} {\bibinfo  {journal} {\mnras}\ }\textbf {\bibinfo {volume} {344}},\
  \bibinfo {pages} {715} (\bibinfo {year} {2003})},\ \Eprint
  {http://arxiv.org/abs/astro-ph/0305393} {astro-ph/0305393} \BibitemShut
  {NoStop}%
\bibitem [{\citenamefont {{Sheth}}\ and\ \citenamefont {{van de
  Weygaert}}(2004)}]{Sheth04}%
  \BibitemOpen
  \bibfield  {author} {\bibinfo {author} {\bibfnamefont {R.~K.}\ \bibnamefont
  {{Sheth}}}\ and\ \bibinfo {author} {\bibfnamefont {R.}~\bibnamefont {{van de
  Weygaert}}},\ }\href {\doibase 10.1111/j.1365-2966.2004.07661.x} {\bibfield
  {journal} {\bibinfo  {journal} {\mnras}\ }\textbf {\bibinfo {volume} {350}},\
  \bibinfo {pages} {517} (\bibinfo {year} {2004})},\ \Eprint
  {http://arxiv.org/abs/astro-ph/0311260} {astro-ph/0311260} \BibitemShut
  {NoStop}%
\bibitem [{\citenamefont {{Park}}\ and\ \citenamefont {{Lee}}(2009)}]{Park09}%
  \BibitemOpen
  \bibfield  {author} {\bibinfo {author} {\bibfnamefont {D.}~\bibnamefont
  {{Park}}}\ and\ \bibinfo {author} {\bibfnamefont {J.}~\bibnamefont {{Lee}}},\
  }\href {\doibase 10.1111/j.1365-2966.2009.15117.x} {\bibfield  {journal}
  {\bibinfo  {journal} {\mnras}\ }\textbf {\bibinfo {volume} {397}},\ \bibinfo
  {pages} {2163} (\bibinfo {year} {2009})},\ \Eprint
  {http://arxiv.org/abs/0801.0634} {arXiv:0801.0634} \BibitemShut {NoStop}%
\bibitem [{\citenamefont {{Rieder}}\ \emph {et~al.}(2013)\citenamefont
  {{Rieder}}, \citenamefont {{van de Weygaert}}, \citenamefont {{Cautun}},
  \citenamefont {{Beygu}},\ and\ \citenamefont {{Portegies Zwart}}}]{Rieder13}%
  \BibitemOpen
  \bibfield  {author} {\bibinfo {author} {\bibfnamefont {S.}~\bibnamefont
  {{Rieder}}}, \bibinfo {author} {\bibfnamefont {R.}~\bibnamefont {{van de
  Weygaert}}}, \bibinfo {author} {\bibfnamefont {M.}~\bibnamefont {{Cautun}}},
  \bibinfo {author} {\bibfnamefont {B.}~\bibnamefont {{Beygu}}}, \ and\
  \bibinfo {author} {\bibfnamefont {S.}~\bibnamefont {{Portegies Zwart}}},\
  }\href {\doibase 10.1093/mnras/stt1288} {\bibfield  {journal} {\bibinfo
  {journal} {\mnras}\ }\textbf {\bibinfo {volume} {435}},\ \bibinfo {pages}
  {222} (\bibinfo {year} {2013})},\ \Eprint {http://arxiv.org/abs/1307.7182}
  {arXiv:1307.7182 [astro-ph.CO]} \BibitemShut {NoStop}%
\bibitem [{\citenamefont {{Kreckel}}\ \emph {et~al.}(2012)\citenamefont
  {{Kreckel}}, \citenamefont {{Platen}}, \citenamefont {{Arag{\'o}n-Calvo}},
  \citenamefont {{van Gorkom}}, \citenamefont {{van de Weygaert}},
  \citenamefont {{van der Hulst}},\ and\ \citenamefont {{Beygu}}}]{Kreckel12}%
  \BibitemOpen
  \bibfield  {author} {\bibinfo {author} {\bibfnamefont {K.}~\bibnamefont
  {{Kreckel}}}, \bibinfo {author} {\bibfnamefont {E.}~\bibnamefont {{Platen}}},
  \bibinfo {author} {\bibfnamefont {M.~A.}\ \bibnamefont {{Arag{\'o}n-Calvo}}},
  \bibinfo {author} {\bibfnamefont {J.~H.}\ \bibnamefont {{van Gorkom}}},
  \bibinfo {author} {\bibfnamefont {R.}~\bibnamefont {{van de Weygaert}}},
  \bibinfo {author} {\bibfnamefont {J.~M.}\ \bibnamefont {{van der Hulst}}}, \
  and\ \bibinfo {author} {\bibfnamefont {B.}~\bibnamefont {{Beygu}}},\ }\href
  {\doibase 10.1088/0004-6256/144/1/16} {\bibfield  {journal} {\bibinfo
  {journal} {\aj}\ }\textbf {\bibinfo {volume} {144}},\ \bibinfo {eid} {16}
  (\bibinfo {year} {2012})},\ \Eprint {http://arxiv.org/abs/1204.5185}
  {arXiv:1204.5185} \BibitemShut {NoStop}%
\bibitem [{\citenamefont {{Beygu}}\ \emph {et~al.}(2013)\citenamefont
  {{Beygu}}, \citenamefont {{Kreckel}}, \citenamefont {{van de Weygaert}},
  \citenamefont {{van der Hulst}},\ and\ \citenamefont {{van
  Gorkom}}}]{Beygu13}%
  \BibitemOpen
  \bibfield  {author} {\bibinfo {author} {\bibfnamefont {B.}~\bibnamefont
  {{Beygu}}}, \bibinfo {author} {\bibfnamefont {K.}~\bibnamefont {{Kreckel}}},
  \bibinfo {author} {\bibfnamefont {R.}~\bibnamefont {{van de Weygaert}}},
  \bibinfo {author} {\bibfnamefont {J.~M.}\ \bibnamefont {{van der Hulst}}}, \
  and\ \bibinfo {author} {\bibfnamefont {J.~H.}\ \bibnamefont {{van Gorkom}}},\
  }\href {\doibase 10.1088/0004-6256/145/5/120} {\bibfield  {journal} {\bibinfo
   {journal} {\aj}\ }\textbf {\bibinfo {volume} {145}},\ \bibinfo {eid} {120}
  (\bibinfo {year} {2013})},\ \Eprint {http://arxiv.org/abs/1303.0538}
  {arXiv:1303.0538} \BibitemShut {NoStop}%
\bibitem [{\citenamefont {{Alpaslan}}\ \emph {et~al.}(2014)\citenamefont
  {{Alpaslan}}, \citenamefont {{Robotham}}, \citenamefont {{Obreschkow}},
  \citenamefont {{Penny}}, \citenamefont {{Driver}}, \citenamefont {{Norberg}},
  \citenamefont {{Brough}}, \citenamefont {{Brown}}, \citenamefont {{Cluver}},
  \citenamefont {{Holwerda}}, \citenamefont {{Hopkins}}, \citenamefont {{van
  Kampen}}, \citenamefont {{Kelvin}}, \citenamefont {{Lara-Lopez}},
  \citenamefont {{Liske}}, \citenamefont {{Loveday}}, \citenamefont
  {{Mahajan}},\ and\ \citenamefont {{Pimbblet}}}]{Alpaslan14}%
  \BibitemOpen
  \bibfield  {author} {\bibinfo {author} {\bibfnamefont {M.}~\bibnamefont
  {{Alpaslan}}}, \bibinfo {author} {\bibfnamefont {A.~S.~G.}\ \bibnamefont
  {{Robotham}}}, \bibinfo {author} {\bibfnamefont {D.}~\bibnamefont
  {{Obreschkow}}}, \bibinfo {author} {\bibfnamefont {S.}~\bibnamefont
  {{Penny}}}, \bibinfo {author} {\bibfnamefont {S.}~\bibnamefont {{Driver}}},
  \bibinfo {author} {\bibfnamefont {P.}~\bibnamefont {{Norberg}}}, \bibinfo
  {author} {\bibfnamefont {S.}~\bibnamefont {{Brough}}}, \bibinfo {author}
  {\bibfnamefont {M.}~\bibnamefont {{Brown}}}, \bibinfo {author} {\bibfnamefont
  {M.}~\bibnamefont {{Cluver}}}, \bibinfo {author} {\bibfnamefont
  {B.}~\bibnamefont {{Holwerda}}}, \bibinfo {author} {\bibfnamefont {A.~M.}\
  \bibnamefont {{Hopkins}}}, \bibinfo {author} {\bibfnamefont {E.}~\bibnamefont
  {{van Kampen}}}, \bibinfo {author} {\bibfnamefont {L.~S.}\ \bibnamefont
  {{Kelvin}}}, \bibinfo {author} {\bibfnamefont {M.~A.}\ \bibnamefont
  {{Lara-Lopez}}}, \bibinfo {author} {\bibfnamefont {J.}~\bibnamefont
  {{Liske}}}, \bibinfo {author} {\bibfnamefont {J.}~\bibnamefont {{Loveday}}},
  \bibinfo {author} {\bibfnamefont {S.}~\bibnamefont {{Mahajan}}}, \ and\
  \bibinfo {author} {\bibfnamefont {K.}~\bibnamefont {{Pimbblet}}},\ }\href
  {\doibase 10.1093/mnrasl/slu019} {\bibfield  {journal} {\bibinfo  {journal}
  {\mnras}\ }\textbf {\bibinfo {volume} {440}},\ \bibinfo {pages} {L106}
  (\bibinfo {year} {2014})},\ \Eprint {http://arxiv.org/abs/1401.7331}
  {arXiv:1401.7331} \BibitemShut {NoStop}%
\bibitem [{\citenamefont {{Bauermeister}}\ \emph {et~al.}(2010)\citenamefont
  {{Bauermeister}}, \citenamefont {{Blitz}},\ and\ \citenamefont
  {{Ma}}}]{Bauermeister10}%
  \BibitemOpen
  \bibfield  {author} {\bibinfo {author} {\bibfnamefont {A.}~\bibnamefont
  {{Bauermeister}}}, \bibinfo {author} {\bibfnamefont {L.}~\bibnamefont
  {{Blitz}}}, \ and\ \bibinfo {author} {\bibfnamefont {C.-P.}\ \bibnamefont
  {{Ma}}},\ }\href {\doibase 10.1088/0004-637X/717/1/323} {\bibfield  {journal}
  {\bibinfo  {journal} {\apj}\ }\textbf {\bibinfo {volume} {717}},\ \bibinfo
  {pages} {323} (\bibinfo {year} {2010})},\ \Eprint
  {http://arxiv.org/abs/0909.3840} {arXiv:0909.3840 [astro-ph.CO]} \BibitemShut
  {NoStop}%
\bibitem [{\citenamefont {{Mori}}\ and\ \citenamefont
  {{Burkert}}(2000)}]{Mori00}%
  \BibitemOpen
  \bibfield  {author} {\bibinfo {author} {\bibfnamefont {M.}~\bibnamefont
  {{Mori}}}\ and\ \bibinfo {author} {\bibfnamefont {A.}~\bibnamefont
  {{Burkert}}},\ }\href {\doibase 10.1086/309140} {\bibfield  {journal}
  {\bibinfo  {journal} {\apj}\ }\textbf {\bibinfo {volume} {538}},\ \bibinfo
  {pages} {559} (\bibinfo {year} {2000})},\ \Eprint
  {http://arxiv.org/abs/astro-ph/0001422} {astro-ph/0001422} \BibitemShut
  {NoStop}%
\bibitem [{\citenamefont {{Quilis}}\ \emph {et~al.}(2000)\citenamefont
  {{Quilis}}, \citenamefont {{Moore}},\ and\ \citenamefont
  {{Bower}}}]{Quilis00}%
  \BibitemOpen
  \bibfield  {author} {\bibinfo {author} {\bibfnamefont {V.}~\bibnamefont
  {{Quilis}}}, \bibinfo {author} {\bibfnamefont {B.}~\bibnamefont {{Moore}}}, \
  and\ \bibinfo {author} {\bibfnamefont {R.}~\bibnamefont {{Bower}}},\ }\href
  {\doibase 10.1126/science.288.5471.1617} {\bibfield  {journal} {\bibinfo
  {journal} {Science}\ }\textbf {\bibinfo {volume} {288}},\ \bibinfo {pages}
  {1617} (\bibinfo {year} {2000})},\ \Eprint
  {http://arxiv.org/abs/astro-ph/0006031} {astro-ph/0006031} \BibitemShut
  {NoStop}%
\bibitem [{\citenamefont {{Schaap}}\ and\ \citenamefont {{van de
  Weygaert}}(2000)}]{Schaap00}%
  \BibitemOpen
  \bibfield  {author} {\bibinfo {author} {\bibfnamefont {W.~E.}\ \bibnamefont
  {{Schaap}}}\ and\ \bibinfo {author} {\bibfnamefont {R.}~\bibnamefont {{van de
  Weygaert}}},\ }\href@noop {} {\bibfield  {journal} {\bibinfo  {journal}
  {\aap}\ }\textbf {\bibinfo {volume} {363}},\ \bibinfo {pages} {L29} (\bibinfo
  {year} {2000})},\ \Eprint {http://arxiv.org/abs/astro-ph/0011007}
  {astro-ph/0011007} \BibitemShut {NoStop}%
\bibitem [{\citenamefont {{Faucher-Gigu{\`e}re}}\ \emph
  {et~al.}(2011)\citenamefont {{Faucher-Gigu{\`e}re}}, \citenamefont {{Kere{\v
  s}}},\ and\ \citenamefont {{Ma}}}]{Faucher11b}%
  \BibitemOpen
  \bibfield  {author} {\bibinfo {author} {\bibfnamefont {C.-A.}\ \bibnamefont
  {{Faucher-Gigu{\`e}re}}}, \bibinfo {author} {\bibfnamefont {D.}~\bibnamefont
  {{Kere{\v s}}}}, \ and\ \bibinfo {author} {\bibfnamefont {C.-P.}\
  \bibnamefont {{Ma}}},\ }\href {\doibase 10.1111/j.1365-2966.2011.19457.x}
  {\bibfield  {journal} {\bibinfo  {journal} {\mnras}\ }\textbf {\bibinfo
  {volume} {417}},\ \bibinfo {pages} {2982} (\bibinfo {year} {2011})},\ \Eprint
  {http://arxiv.org/abs/1103.0001} {arXiv:1103.0001} \BibitemShut {NoStop}%
\bibitem [{\citenamefont {{Springel}}\ \emph {et~al.}(2005)\citenamefont
  {{Springel}}, \citenamefont {{White}}, \citenamefont {{Jenkins}},
  \citenamefont {{Frenk}}, \citenamefont {{Yoshida}}, \citenamefont {{Gao}},
  \citenamefont {{Navarro}}, \citenamefont {{Thacker}}, \citenamefont
  {{Croton}}, \citenamefont {{Helly}}, \citenamefont {{Peacock}}, \citenamefont
  {{Cole}}, \citenamefont {{Thomas}}, \citenamefont {{Couchman}}, \citenamefont
  {{Evrard}}, \citenamefont {{Colberg}},\ and\ \citenamefont
  {{Pearce}}}]{Springel05}%
  \BibitemOpen
  \bibfield  {author} {\bibinfo {author} {\bibfnamefont {V.}~\bibnamefont
  {{Springel}}}, \bibinfo {author} {\bibfnamefont {S.~D.~M.}\ \bibnamefont
  {{White}}}, \bibinfo {author} {\bibfnamefont {A.}~\bibnamefont {{Jenkins}}},
  \bibinfo {author} {\bibfnamefont {C.~S.}\ \bibnamefont {{Frenk}}}, \bibinfo
  {author} {\bibfnamefont {N.}~\bibnamefont {{Yoshida}}}, \bibinfo {author}
  {\bibfnamefont {L.}~\bibnamefont {{Gao}}}, \bibinfo {author} {\bibfnamefont
  {J.}~\bibnamefont {{Navarro}}}, \bibinfo {author} {\bibfnamefont
  {R.}~\bibnamefont {{Thacker}}}, \bibinfo {author} {\bibfnamefont
  {D.}~\bibnamefont {{Croton}}}, \bibinfo {author} {\bibfnamefont
  {J.}~\bibnamefont {{Helly}}}, \bibinfo {author} {\bibfnamefont {J.~A.}\
  \bibnamefont {{Peacock}}}, \bibinfo {author} {\bibfnamefont {S.}~\bibnamefont
  {{Cole}}}, \bibinfo {author} {\bibfnamefont {P.}~\bibnamefont {{Thomas}}},
  \bibinfo {author} {\bibfnamefont {H.}~\bibnamefont {{Couchman}}}, \bibinfo
  {author} {\bibfnamefont {A.}~\bibnamefont {{Evrard}}}, \bibinfo {author}
  {\bibfnamefont {J.}~\bibnamefont {{Colberg}}}, \ and\ \bibinfo {author}
  {\bibfnamefont {F.}~\bibnamefont {{Pearce}}},\ }\href {\doibase
  10.1038/nature03597} {\bibfield  {journal} {\bibinfo  {journal} {\nat}\
  }\textbf {\bibinfo {volume} {435}},\ \bibinfo {pages} {629} (\bibinfo {year}
  {2005})},\ \Eprint {http://arxiv.org/abs/arXiv:astro-ph/0504097}
  {arXiv:astro-ph/0504097} \BibitemShut {NoStop}%
\bibitem [{\citenamefont {{Ben{\'{\i}}tez-Llambay}}\ \emph
  {et~al.}(2013)\citenamefont {{Ben{\'{\i}}tez-Llambay}}, \citenamefont
  {{Navarro}}, \citenamefont {{Abadi}}, \citenamefont {{Gottl{\"o}ber}},
  \citenamefont {{Yepes}}, \citenamefont {{Hoffman}},\ and\ \citenamefont
  {{Steinmetz}}}]{Benitez13}%
  \BibitemOpen
  \bibfield  {author} {\bibinfo {author} {\bibfnamefont {A.}~\bibnamefont
  {{Ben{\'{\i}}tez-Llambay}}}, \bibinfo {author} {\bibfnamefont {J.~F.}\
  \bibnamefont {{Navarro}}}, \bibinfo {author} {\bibfnamefont {M.~G.}\
  \bibnamefont {{Abadi}}}, \bibinfo {author} {\bibfnamefont {S.}~\bibnamefont
  {{Gottl{\"o}ber}}}, \bibinfo {author} {\bibfnamefont {G.}~\bibnamefont
  {{Yepes}}}, \bibinfo {author} {\bibfnamefont {Y.}~\bibnamefont {{Hoffman}}},
  \ and\ \bibinfo {author} {\bibfnamefont {M.}~\bibnamefont {{Steinmetz}}},\
  }\href {\doibase 10.1088/2041-8205/763/2/L41} {\bibfield  {journal} {\bibinfo
   {journal} {\apjl}\ }\textbf {\bibinfo {volume} {763}},\ \bibinfo {eid} {L41}
  (\bibinfo {year} {2013})},\ \Eprint {http://arxiv.org/abs/1211.0536}
  {arXiv:1211.0536 [astro-ph.CO]} \BibitemShut {NoStop}%
\bibitem [{\citenamefont {{Aragon-Calvo}}\ \emph {et~al.}(2010)\citenamefont
  {{Aragon-Calvo}}, \citenamefont {{van de Weygaert}}, \citenamefont
  {{Araya-Melo}}, \citenamefont {{Platen}},\ and\ \citenamefont
  {{Szalay}}}]{Aragon10b}%
  \BibitemOpen
  \bibfield  {author} {\bibinfo {author} {\bibfnamefont {M.~A.}\ \bibnamefont
  {{Aragon-Calvo}}}, \bibinfo {author} {\bibfnamefont {R.}~\bibnamefont {{van
  de Weygaert}}}, \bibinfo {author} {\bibfnamefont {P.~A.}\ \bibnamefont
  {{Araya-Melo}}}, \bibinfo {author} {\bibfnamefont {E.}~\bibnamefont
  {{Platen}}}, \ and\ \bibinfo {author} {\bibfnamefont {A.~S.}\ \bibnamefont
  {{Szalay}}},\ }\href {\doibase 10.1111/j.1745-3933.2010.00841.x} {\bibfield
  {journal} {\bibinfo  {journal} {\mnras}\ }\textbf {\bibinfo {volume} {404}},\
  \bibinfo {pages} {L89} (\bibinfo {year} {2010})},\ \Eprint
  {http://arxiv.org/abs/1002.1503} {arXiv:1002.1503 [astro-ph.CO]} \BibitemShut
  {NoStop}%
\bibitem [{\citenamefont {{Suhhonenko}}\ \emph {et~al.}(2011)\citenamefont
  {{Suhhonenko}}, \citenamefont {{Einasto}}, \citenamefont {{Liivam{\"a}gi}},
  \citenamefont {{Saar}}, \citenamefont {{Einasto}}, \citenamefont
  {{H{\"u}tsi}}, \citenamefont {{M{\"u}ller}}, \citenamefont {{Starobinsky}},
  \citenamefont {{Tago}},\ and\ \citenamefont {{Tempel}}}]{Suhhonenko11}%
  \BibitemOpen
  \bibfield  {author} {\bibinfo {author} {\bibfnamefont {I.}~\bibnamefont
  {{Suhhonenko}}}, \bibinfo {author} {\bibfnamefont {J.}~\bibnamefont
  {{Einasto}}}, \bibinfo {author} {\bibfnamefont {L.~J.}\ \bibnamefont
  {{Liivam{\"a}gi}}}, \bibinfo {author} {\bibfnamefont {E.}~\bibnamefont
  {{Saar}}}, \bibinfo {author} {\bibfnamefont {M.}~\bibnamefont {{Einasto}}},
  \bibinfo {author} {\bibfnamefont {G.}~\bibnamefont {{H{\"u}tsi}}}, \bibinfo
  {author} {\bibfnamefont {V.}~\bibnamefont {{M{\"u}ller}}}, \bibinfo {author}
  {\bibfnamefont {A.~A.}\ \bibnamefont {{Starobinsky}}}, \bibinfo {author}
  {\bibfnamefont {E.}~\bibnamefont {{Tago}}}, \ and\ \bibinfo {author}
  {\bibfnamefont {E.}~\bibnamefont {{Tempel}}},\ }\href {\doibase
  10.1051/0004-6361/201016394} {\bibfield  {journal} {\bibinfo  {journal}
  {\aap}\ }\textbf {\bibinfo {volume} {531}},\ \bibinfo {eid} {A149} (\bibinfo
  {year} {2011})},\ \Eprint {http://arxiv.org/abs/1101.0123} {arXiv:1101.0123
  [astro-ph.CO]} \BibitemShut {NoStop}%
\bibitem [{\citenamefont {{Cattaneo}}\ \emph {et~al.}(2006)\citenamefont
  {{Cattaneo}}, \citenamefont {{Dekel}}, \citenamefont {{Devriendt}},
  \citenamefont {{Guiderdoni}},\ and\ \citenamefont {{Blaizot}}}]{Cattaneo06}%
  \BibitemOpen
  \bibfield  {author} {\bibinfo {author} {\bibfnamefont {A.}~\bibnamefont
  {{Cattaneo}}}, \bibinfo {author} {\bibfnamefont {A.}~\bibnamefont {{Dekel}}},
  \bibinfo {author} {\bibfnamefont {J.}~\bibnamefont {{Devriendt}}}, \bibinfo
  {author} {\bibfnamefont {B.}~\bibnamefont {{Guiderdoni}}}, \ and\ \bibinfo
  {author} {\bibfnamefont {J.}~\bibnamefont {{Blaizot}}},\ }\href {\doibase
  10.1111/j.1365-2966.2006.10608.x} {\bibfield  {journal} {\bibinfo  {journal}
  {\mnras}\ }\textbf {\bibinfo {volume} {370}},\ \bibinfo {pages} {1651}
  (\bibinfo {year} {2006})},\ \Eprint {http://arxiv.org/abs/astro-ph/0601295}
  {astro-ph/0601295} \BibitemShut {NoStop}%
\bibitem [{\citenamefont {{Birnboim}}\ \emph {et~al.}(2007)\citenamefont
  {{Birnboim}}, \citenamefont {{Dekel}},\ and\ \citenamefont
  {{Neistein}}}]{Birnboim07}%
  \BibitemOpen
  \bibfield  {author} {\bibinfo {author} {\bibfnamefont {Y.}~\bibnamefont
  {{Birnboim}}}, \bibinfo {author} {\bibfnamefont {A.}~\bibnamefont {{Dekel}}},
  \ and\ \bibinfo {author} {\bibfnamefont {E.}~\bibnamefont {{Neistein}}},\
  }\href {\doibase 10.1111/j.1365-2966.2007.12074.x} {\bibfield  {journal}
  {\bibinfo  {journal} {\mnras}\ }\textbf {\bibinfo {volume} {380}},\ \bibinfo
  {pages} {339} (\bibinfo {year} {2007})},\ \Eprint
  {http://arxiv.org/abs/astro-ph/0703435} {astro-ph/0703435} \BibitemShut
  {NoStop}%
\bibitem [{\citenamefont {{Shandarin}}\ \emph {et~al.}(2012)\citenamefont
  {{Shandarin}}, \citenamefont {{Habib}},\ and\ \citenamefont
  {{Heitmann}}}]{Shandarin12}%
  \BibitemOpen
  \bibfield  {author} {\bibinfo {author} {\bibfnamefont {S.}~\bibnamefont
  {{Shandarin}}}, \bibinfo {author} {\bibfnamefont {S.}~\bibnamefont
  {{Habib}}}, \ and\ \bibinfo {author} {\bibfnamefont {K.}~\bibnamefont
  {{Heitmann}}},\ }\href {\doibase 10.1103/PhysRevD.85.083005} {\bibfield
  {journal} {\bibinfo  {journal} {\prd}\ }\textbf {\bibinfo {volume} {85}},\
  \bibinfo {eid} {083005} (\bibinfo {year} {2012})},\ \Eprint
  {http://arxiv.org/abs/1111.2366} {arXiv:1111.2366 [astro-ph.CO]} \BibitemShut
  {NoStop}%
\bibitem [{\citenamefont {{Abel}}\ \emph {et~al.}(2012)\citenamefont {{Abel}},
  \citenamefont {{Hahn}},\ and\ \citenamefont {{Kaehler}}}]{Abel12}%
  \BibitemOpen
  \bibfield  {author} {\bibinfo {author} {\bibfnamefont {T.}~\bibnamefont
  {{Abel}}}, \bibinfo {author} {\bibfnamefont {O.}~\bibnamefont {{Hahn}}}, \
  and\ \bibinfo {author} {\bibfnamefont {R.}~\bibnamefont {{Kaehler}}},\ }\href
  {\doibase 10.1111/j.1365-2966.2012.21754.x} {\bibfield  {journal} {\bibinfo
  {journal} {\mnras}\ }\textbf {\bibinfo {volume} {427}},\ \bibinfo {pages}
  {61} (\bibinfo {year} {2012})},\ \Eprint {http://arxiv.org/abs/1111.3944}
  {arXiv:1111.3944 [astro-ph.CO]} \BibitemShut {NoStop}%
\bibitem [{\citenamefont {{Shandarin}}(2011)}]{Shandarin11}%
  \BibitemOpen
  \bibfield  {author} {\bibinfo {author} {\bibfnamefont {S.~F.}\ \bibnamefont
  {{Shandarin}}},\ }\href {\doibase 10.1088/1475-7516/2011/05/015} {\bibfield
  {journal} {\bibinfo  {journal} {\jcap}\ }\textbf {\bibinfo {volume} {5}},\
  \bibinfo {eid} {015} (\bibinfo {year} {2011})},\ \Eprint
  {http://arxiv.org/abs/1011.1924} {arXiv:1011.1924} \BibitemShut {NoStop}%
\bibitem [{\citenamefont {{Falck}}\ \emph {et~al.}(2012)\citenamefont
  {{Falck}}, \citenamefont {{Neyrinck}},\ and\ \citenamefont
  {{Szalay}}}]{Falck12}%
  \BibitemOpen
  \bibfield  {author} {\bibinfo {author} {\bibfnamefont {B.~L.}\ \bibnamefont
  {{Falck}}}, \bibinfo {author} {\bibfnamefont {M.~C.}\ \bibnamefont
  {{Neyrinck}}}, \ and\ \bibinfo {author} {\bibfnamefont {A.~S.}\ \bibnamefont
  {{Szalay}}},\ }\href {\doibase 10.1088/0004-637X/754/2/126} {\bibfield
  {journal} {\bibinfo  {journal} {\apj}\ }\textbf {\bibinfo {volume} {754}},\
  \bibinfo {eid} {126} (\bibinfo {year} {2012})},\ \Eprint
  {http://arxiv.org/abs/1201.2353} {arXiv:1201.2353} \BibitemShut {NoStop}%
\bibitem [{\citenamefont {{Neyrinck}}(2012)}]{Neyrinck12}%
  \BibitemOpen
  \bibfield  {author} {\bibinfo {author} {\bibfnamefont {M.~C.}\ \bibnamefont
  {{Neyrinck}}},\ }\href {\doibase 10.1111/j.1365-2966.2012.21956.x} {\bibfield
   {journal} {\bibinfo  {journal} {\mnras}\ }\textbf {\bibinfo {volume}
  {427}},\ \bibinfo {pages} {494} (\bibinfo {year} {2012})},\ \Eprint
  {http://arxiv.org/abs/1202.3364} {arXiv:1202.3364 [astro-ph.CO]} \BibitemShut
  {NoStop}%
\bibitem [{\citenamefont {{Neyrinck}}\ \emph {et~al.}(2015)\citenamefont
  {{Neyrinck}}, \citenamefont {{Falck}},\ and\ \citenamefont
  {{Szalay}}}]{Neyrinck15}%
  \BibitemOpen
  \bibfield  {author} {\bibinfo {author} {\bibfnamefont {M.~C.}\ \bibnamefont
  {{Neyrinck}}}, \bibinfo {author} {\bibfnamefont {B.~L.}\ \bibnamefont
  {{Falck}}}, \ and\ \bibinfo {author} {\bibfnamefont {A.~S.}\ \bibnamefont
  {{Szalay}}},\ }in\ \href@noop {} {\emph {\bibinfo {booktitle} {Thirteenth
  Marcel Grossmann Meeting: On Recent Developments in Theoretical and
  Experimental General Relativity, Astrophysics and Relativistic Field
  Theories}}},\ \bibinfo {editor} {edited by\ \bibinfo {editor} {\bibfnamefont
  {K.}~\bibnamefont {{Rosquist}}}}\ (\bibinfo {year} {2015})\ pp.\ \bibinfo
  {pages} {2136--2138},\ \Eprint {http://arxiv.org/abs/1309.4787}
  {arXiv:1309.4787} \BibitemShut {NoStop}%
\bibitem [{\citenamefont {{Falck}}\ and\ \citenamefont
  {{Neyrinck}}(2015)}]{Falck15}%
  \BibitemOpen
  \bibfield  {author} {\bibinfo {author} {\bibfnamefont {B.}~\bibnamefont
  {{Falck}}}\ and\ \bibinfo {author} {\bibfnamefont {M.~C.}\ \bibnamefont
  {{Neyrinck}}},\ }\href {\doibase 10.1093/mnras/stv879} {\bibfield  {journal}
  {\bibinfo  {journal} {\mnras}\ }\textbf {\bibinfo {volume} {450}},\ \bibinfo
  {pages} {3239} (\bibinfo {year} {2015})},\ \Eprint
  {http://arxiv.org/abs/1410.4751} {arXiv:1410.4751} \BibitemShut {NoStop}%
\bibitem [{\citenamefont {{Behroozi}}\ and\ \citenamefont
  {{Silk}}(2015)}]{Behroozi15}%
  \BibitemOpen
  \bibfield  {author} {\bibinfo {author} {\bibfnamefont {P.~S.}\ \bibnamefont
  {{Behroozi}}}\ and\ \bibinfo {author} {\bibfnamefont {J.}~\bibnamefont
  {{Silk}}},\ }\href {\doibase 10.1088/0004-637X/799/1/32} {\bibfield
  {journal} {\bibinfo  {journal} {\apj}\ }\textbf {\bibinfo {volume} {799}},\
  \bibinfo {eid} {32} (\bibinfo {year} {2015})},\ \Eprint
  {http://arxiv.org/abs/1404.5299} {arXiv:1404.5299} \BibitemShut {NoStop}%
\bibitem [{\citenamefont {{Wechsler}}\ \emph {et~al.}(2002)\citenamefont
  {{Wechsler}}, \citenamefont {{Bullock}}, \citenamefont {{Primack}},
  \citenamefont {{Kravtsov}},\ and\ \citenamefont {{Dekel}}}]{Wechsler02}%
  \BibitemOpen
  \bibfield  {author} {\bibinfo {author} {\bibfnamefont {R.~H.}\ \bibnamefont
  {{Wechsler}}}, \bibinfo {author} {\bibfnamefont {J.~S.}\ \bibnamefont
  {{Bullock}}}, \bibinfo {author} {\bibfnamefont {J.~R.}\ \bibnamefont
  {{Primack}}}, \bibinfo {author} {\bibfnamefont {A.~V.}\ \bibnamefont
  {{Kravtsov}}}, \ and\ \bibinfo {author} {\bibfnamefont {A.}~\bibnamefont
  {{Dekel}}},\ }\href {\doibase 10.1086/338765} {\bibfield  {journal} {\bibinfo
   {journal} {\apj}\ }\textbf {\bibinfo {volume} {568}},\ \bibinfo {pages} {52}
  (\bibinfo {year} {2002})},\ \Eprint {http://arxiv.org/abs/astro-ph/0108151}
  {astro-ph/0108151} \BibitemShut {NoStop}%
\bibitem [{\citenamefont {{Peebles}}(2001)}]{Peebles01}%
  \BibitemOpen
  \bibfield  {author} {\bibinfo {author} {\bibfnamefont {P.~J.~E.}\
  \bibnamefont {{Peebles}}},\ }\href {\doibase 10.1086/322254} {\bibfield
  {journal} {\bibinfo  {journal} {\apj}\ }\textbf {\bibinfo {volume} {557}},\
  \bibinfo {pages} {495} (\bibinfo {year} {2001})},\ \Eprint
  {http://arxiv.org/abs/arXiv:astro-ph/0101127} {arXiv:astro-ph/0101127}
  \BibitemShut {NoStop}%
\bibitem [{\citenamefont {{Cowie}}\ \emph {et~al.}(1996)\citenamefont
  {{Cowie}}, \citenamefont {{Songaila}}, \citenamefont {{Hu}},\ and\
  \citenamefont {{Cohen}}}]{Cowie96}%
  \BibitemOpen
  \bibfield  {author} {\bibinfo {author} {\bibfnamefont {L.~L.}\ \bibnamefont
  {{Cowie}}}, \bibinfo {author} {\bibfnamefont {A.}~\bibnamefont {{Songaila}}},
  \bibinfo {author} {\bibfnamefont {E.~M.}\ \bibnamefont {{Hu}}}, \ and\
  \bibinfo {author} {\bibfnamefont {J.~G.}\ \bibnamefont {{Cohen}}},\ }\href
  {\doibase 10.1086/118058} {\bibfield  {journal} {\bibinfo  {journal} {\aj}\
  }\textbf {\bibinfo {volume} {112}},\ \bibinfo {pages} {839} (\bibinfo {year}
  {1996})},\ \Eprint {http://arxiv.org/abs/astro-ph/9606079} {astro-ph/9606079}
  \BibitemShut {NoStop}%
\bibitem [{\citenamefont {{Heavens}}\ \emph {et~al.}(2004)\citenamefont
  {{Heavens}}, \citenamefont {{Panter}}, \citenamefont {{Jimenez}},\ and\
  \citenamefont {{Dunlop}}}]{Heavens04}%
  \BibitemOpen
  \bibfield  {author} {\bibinfo {author} {\bibfnamefont {A.}~\bibnamefont
  {{Heavens}}}, \bibinfo {author} {\bibfnamefont {B.}~\bibnamefont {{Panter}}},
  \bibinfo {author} {\bibfnamefont {R.}~\bibnamefont {{Jimenez}}}, \ and\
  \bibinfo {author} {\bibfnamefont {J.}~\bibnamefont {{Dunlop}}},\ }\href
  {\doibase 10.1038/nature02474} {\bibfield  {journal} {\bibinfo  {journal}
  {\nat}\ }\textbf {\bibinfo {volume} {428}},\ \bibinfo {pages} {625} (\bibinfo
  {year} {2004})},\ \Eprint {http://arxiv.org/abs/astro-ph/0403293}
  {astro-ph/0403293} \BibitemShut {NoStop}%
\bibitem [{\citenamefont {{Thomas}}\ \emph {et~al.}(2005)\citenamefont
  {{Thomas}}, \citenamefont {{Maraston}}, \citenamefont {{Bender}},\ and\
  \citenamefont {{Mendes de Oliveira}}}]{Thomas05}%
  \BibitemOpen
  \bibfield  {author} {\bibinfo {author} {\bibfnamefont {D.}~\bibnamefont
  {{Thomas}}}, \bibinfo {author} {\bibfnamefont {C.}~\bibnamefont
  {{Maraston}}}, \bibinfo {author} {\bibfnamefont {R.}~\bibnamefont
  {{Bender}}}, \ and\ \bibinfo {author} {\bibfnamefont {C.}~\bibnamefont
  {{Mendes de Oliveira}}},\ }\href {\doibase 10.1086/426932} {\bibfield
  {journal} {\bibinfo  {journal} {\apj}\ }\textbf {\bibinfo {volume} {621}},\
  \bibinfo {pages} {673} (\bibinfo {year} {2005})},\ \Eprint
  {http://arxiv.org/abs/arXiv:astro-ph/0410209} {arXiv:astro-ph/0410209}
  \BibitemShut {NoStop}%
\bibitem [{\citenamefont {{Hopkins}}\ \emph {et~al.}(2006)\citenamefont
  {{Hopkins}}, \citenamefont {{Hernquist}}, \citenamefont {{Cox}},
  \citenamefont {{Di Matteo}}, \citenamefont {{Robertson}},\ and\ \citenamefont
  {{Springel}}}]{Hopkins06}%
  \BibitemOpen
  \bibfield  {author} {\bibinfo {author} {\bibfnamefont {P.~F.}\ \bibnamefont
  {{Hopkins}}}, \bibinfo {author} {\bibfnamefont {L.}~\bibnamefont
  {{Hernquist}}}, \bibinfo {author} {\bibfnamefont {T.~J.}\ \bibnamefont
  {{Cox}}}, \bibinfo {author} {\bibfnamefont {T.}~\bibnamefont {{Di Matteo}}},
  \bibinfo {author} {\bibfnamefont {B.}~\bibnamefont {{Robertson}}}, \ and\
  \bibinfo {author} {\bibfnamefont {V.}~\bibnamefont {{Springel}}},\ }\href
  {\doibase 10.1086/499298} {\bibfield  {journal} {\bibinfo  {journal} {\apjs}\
  }\textbf {\bibinfo {volume} {163}},\ \bibinfo {pages} {1} (\bibinfo {year}
  {2006})},\ \Eprint {http://arxiv.org/abs/astro-ph/0506398} {astro-ph/0506398}
  \BibitemShut {NoStop}%
\bibitem [{\citenamefont {{Neistein}}\ \emph {et~al.}(2006)\citenamefont
  {{Neistein}}, \citenamefont {{van den Bosch}},\ and\ \citenamefont
  {{Dekel}}}]{Neistein06}%
  \BibitemOpen
  \bibfield  {author} {\bibinfo {author} {\bibfnamefont {E.}~\bibnamefont
  {{Neistein}}}, \bibinfo {author} {\bibfnamefont {F.~C.}\ \bibnamefont {{van
  den Bosch}}}, \ and\ \bibinfo {author} {\bibfnamefont {A.}~\bibnamefont
  {{Dekel}}},\ }\href {\doibase 10.1111/j.1365-2966.2006.10918.x} {\bibfield
  {journal} {\bibinfo  {journal} {\mnras}\ }\textbf {\bibinfo {volume} {372}},\
  \bibinfo {pages} {933} (\bibinfo {year} {2006})},\ \Eprint
  {http://arxiv.org/abs/astro-ph/0605045} {astro-ph/0605045} \BibitemShut
  {NoStop}%
\bibitem [{\citenamefont {{Reddy}}\ \emph {et~al.}(2012)\citenamefont
  {{Reddy}}, \citenamefont {{Pettini}}, \citenamefont {{Steidel}},
  \citenamefont {{Shapley}}, \citenamefont {{Erb}},\ and\ \citenamefont
  {{Law}}}]{Reddy12}%
  \BibitemOpen
  \bibfield  {author} {\bibinfo {author} {\bibfnamefont {N.~A.}\ \bibnamefont
  {{Reddy}}}, \bibinfo {author} {\bibfnamefont {M.}~\bibnamefont {{Pettini}}},
  \bibinfo {author} {\bibfnamefont {C.~C.}\ \bibnamefont {{Steidel}}}, \bibinfo
  {author} {\bibfnamefont {A.~E.}\ \bibnamefont {{Shapley}}}, \bibinfo {author}
  {\bibfnamefont {D.~K.}\ \bibnamefont {{Erb}}}, \ and\ \bibinfo {author}
  {\bibfnamefont {D.~R.}\ \bibnamefont {{Law}}},\ }\href {\doibase
  10.1088/0004-637X/754/1/25} {\bibfield  {journal} {\bibinfo  {journal}
  {\apj}\ }\textbf {\bibinfo {volume} {754}},\ \bibinfo {eid} {25} (\bibinfo
  {year} {2012})},\ \Eprint {http://arxiv.org/abs/1205.0555} {arXiv:1205.0555
  [astro-ph.CO]} \BibitemShut {NoStop}%
\bibitem [{\citenamefont {{Geha}}\ \emph {et~al.}(2012)\citenamefont {{Geha}},
  \citenamefont {{Blanton}}, \citenamefont {{Yan}},\ and\ \citenamefont
  {{Tinker}}}]{Geha12}%
  \BibitemOpen
  \bibfield  {author} {\bibinfo {author} {\bibfnamefont {M.}~\bibnamefont
  {{Geha}}}, \bibinfo {author} {\bibfnamefont {M.~R.}\ \bibnamefont
  {{Blanton}}}, \bibinfo {author} {\bibfnamefont {R.}~\bibnamefont {{Yan}}}, \
  and\ \bibinfo {author} {\bibfnamefont {J.~L.}\ \bibnamefont {{Tinker}}},\
  }\href {\doibase 10.1088/0004-637X/757/1/85} {\bibfield  {journal} {\bibinfo
  {journal} {\apj}\ }\textbf {\bibinfo {volume} {757}},\ \bibinfo {eid} {85}
  (\bibinfo {year} {2012})},\ \Eprint {http://arxiv.org/abs/1206.3573}
  {arXiv:1206.3573} \BibitemShut {NoStop}%
\bibitem [{\citenamefont {{Ceccarelli}}\ \emph {et~al.}(2008)\citenamefont
  {{Ceccarelli}}, \citenamefont {{Padilla}},\ and\ \citenamefont
  {{Lambas}}}]{Ceccarelli08}%
  \BibitemOpen
  \bibfield  {author} {\bibinfo {author} {\bibfnamefont {L.}~\bibnamefont
  {{Ceccarelli}}}, \bibinfo {author} {\bibfnamefont {N.}~\bibnamefont
  {{Padilla}}}, \ and\ \bibinfo {author} {\bibfnamefont {D.~G.}\ \bibnamefont
  {{Lambas}}},\ }\href {\doibase 10.1111/j.1745-3933.2008.00520.x} {\bibfield
  {journal} {\bibinfo  {journal} {\mnras}\ }\textbf {\bibinfo {volume} {390}},\
  \bibinfo {pages} {L9} (\bibinfo {year} {2008})},\ \Eprint
  {http://arxiv.org/abs/0805.0790} {arXiv:0805.0790} \BibitemShut {NoStop}%
\bibitem [{\citenamefont {{Liu}}\ \emph {et~al.}(2015)\citenamefont {{Liu}},
  \citenamefont {{Pan}}, \citenamefont {{Hao}}, \citenamefont {{Hoyle}},
  \citenamefont {{Constantin}},\ and\ \citenamefont {{Vogeley}}}]{Liu15}%
  \BibitemOpen
  \bibfield  {author} {\bibinfo {author} {\bibfnamefont {C.-X.}\ \bibnamefont
  {{Liu}}}, \bibinfo {author} {\bibfnamefont {D.~C.}\ \bibnamefont {{Pan}}},
  \bibinfo {author} {\bibfnamefont {L.}~\bibnamefont {{Hao}}}, \bibinfo
  {author} {\bibfnamefont {F.}~\bibnamefont {{Hoyle}}}, \bibinfo {author}
  {\bibfnamefont {A.}~\bibnamefont {{Constantin}}}, \ and\ \bibinfo {author}
  {\bibfnamefont {M.~S.}\ \bibnamefont {{Vogeley}}},\ }\href {\doibase
  10.1088/0004-637X/810/2/165} {\bibfield  {journal} {\bibinfo  {journal}
  {\apj}\ }\textbf {\bibinfo {volume} {810}},\ \bibinfo {eid} {165} (\bibinfo
  {year} {2015})},\ \Eprint {http://arxiv.org/abs/1509.04430}
  {arXiv:1509.04430} \BibitemShut {NoStop}%
\bibitem [{\citenamefont {{Moorman}}\ \emph {et~al.}(2016)\citenamefont
  {{Moorman}}, \citenamefont {{Moreno}}, \citenamefont {{White}}, \citenamefont
  {{Vogeley}}, \citenamefont {{Hoyle}}, \citenamefont {{Giovanelli}},\ and\
  \citenamefont {{Haynes}}}]{Moorman16}%
  \BibitemOpen
  \bibfield  {author} {\bibinfo {author} {\bibfnamefont {C.~M.}\ \bibnamefont
  {{Moorman}}}, \bibinfo {author} {\bibfnamefont {J.}~\bibnamefont {{Moreno}}},
  \bibinfo {author} {\bibfnamefont {A.}~\bibnamefont {{White}}}, \bibinfo
  {author} {\bibfnamefont {M.~S.}\ \bibnamefont {{Vogeley}}}, \bibinfo {author}
  {\bibfnamefont {F.}~\bibnamefont {{Hoyle}}}, \bibinfo {author} {\bibfnamefont
  {R.}~\bibnamefont {{Giovanelli}}}, \ and\ \bibinfo {author} {\bibfnamefont
  {M.~P.}\ \bibnamefont {{Haynes}}},\ }\href@noop {} {\bibfield  {journal}
  {\bibinfo  {journal} {ArXiv e-prints}\ } (\bibinfo {year} {2016})},\ \Eprint
  {http://arxiv.org/abs/1601.04092} {arXiv:1601.04092} \BibitemShut {NoStop}%
\bibitem [{\citenamefont {{Aragon-Calvo}}\ \emph {et~al.}(2011)\citenamefont
  {{Aragon-Calvo}}, \citenamefont {{Silk}},\ and\ \citenamefont
  {{Szalay}}}]{Aragon11}%
  \BibitemOpen
  \bibfield  {author} {\bibinfo {author} {\bibfnamefont {M.~A.}\ \bibnamefont
  {{Aragon-Calvo}}}, \bibinfo {author} {\bibfnamefont {J.}~\bibnamefont
  {{Silk}}}, \ and\ \bibinfo {author} {\bibfnamefont {A.~S.}\ \bibnamefont
  {{Szalay}}},\ }\href {\doibase 10.1111/j.1745-3933.2011.01071.x} {\bibfield
  {journal} {\bibinfo  {journal} {\mnras}\ }\textbf {\bibinfo {volume} {415}},\
  \bibinfo {pages} {L16} (\bibinfo {year} {2011})},\ \Eprint
  {http://arxiv.org/abs/1103.1901} {arXiv:1103.1901 [astro-ph.CO]} \BibitemShut
  {NoStop}%
\bibitem [{\citenamefont {{Ricciardelli}}\ \emph {et~al.}(2014)\citenamefont
  {{Ricciardelli}}, \citenamefont {{Cava}}, \citenamefont {{Varela}},\ and\
  \citenamefont {{Quilis}}}]{Ricciardelli14}%
  \BibitemOpen
  \bibfield  {author} {\bibinfo {author} {\bibfnamefont {E.}~\bibnamefont
  {{Ricciardelli}}}, \bibinfo {author} {\bibfnamefont {A.}~\bibnamefont
  {{Cava}}}, \bibinfo {author} {\bibfnamefont {J.}~\bibnamefont {{Varela}}}, \
  and\ \bibinfo {author} {\bibfnamefont {V.}~\bibnamefont {{Quilis}}},\ }\href
  {\doibase 10.1093/mnras/stu2061} {\bibfield  {journal} {\bibinfo  {journal}
  {\mnras}\ }\textbf {\bibinfo {volume} {445}},\ \bibinfo {pages} {4045}
  (\bibinfo {year} {2014})},\ \Eprint {http://arxiv.org/abs/1410.0023}
  {arXiv:1410.0023} \BibitemShut {NoStop}%
\bibitem [{\citenamefont {{Das}}\ \emph {et~al.}(2015)\citenamefont {{Das}},
  \citenamefont {{Saito}}, \citenamefont {{Iono}}, \citenamefont {{Honey}},\
  and\ \citenamefont {{Ramya}}}]{Das15}%
  \BibitemOpen
  \bibfield  {author} {\bibinfo {author} {\bibfnamefont {M.}~\bibnamefont
  {{Das}}}, \bibinfo {author} {\bibfnamefont {T.}~\bibnamefont {{Saito}}},
  \bibinfo {author} {\bibfnamefont {D.}~\bibnamefont {{Iono}}}, \bibinfo
  {author} {\bibfnamefont {M.}~\bibnamefont {{Honey}}}, \ and\ \bibinfo
  {author} {\bibfnamefont {S.}~\bibnamefont {{Ramya}}},\ }\href {\doibase
  10.1088/0004-637X/815/1/40} {\bibfield  {journal} {\bibinfo  {journal}
  {\apj}\ }\textbf {\bibinfo {volume} {815}},\ \bibinfo {eid} {40} (\bibinfo
  {year} {2015})},\ \Eprint {http://arxiv.org/abs/1510.07411}
  {arXiv:1510.07411} \BibitemShut {NoStop}%
\bibitem [{\citenamefont {{Tully}}\ and\ \citenamefont
  {{Fisher}}(1987)}]{Tully87}%
  \BibitemOpen
  \bibfield  {author} {\bibinfo {author} {\bibfnamefont {R.~B.}\ \bibnamefont
  {{Tully}}}\ and\ \bibinfo {author} {\bibfnamefont {J.~R.}\ \bibnamefont
  {{Fisher}}},\ }\href@noop {} {\bibfield  {journal} {\bibinfo  {journal}
  {Journal of the British Astronomical Association}\ }\textbf {\bibinfo
  {volume} {98}},\ \bibinfo {pages} {48} (\bibinfo {year} {1987})}\BibitemShut
  {NoStop}%
\bibitem [{\citenamefont {{Noeske}}\ \emph {et~al.}(2007)\citenamefont
  {{Noeske}}, \citenamefont {{Faber}}, \citenamefont {{Weiner}}, \citenamefont
  {{Koo}}, \citenamefont {{Primack}}, \citenamefont {{Dekel}}, \citenamefont
  {{Papovich}}, \citenamefont {{Conselice}}, \citenamefont {{Le Floc'h}},
  \citenamefont {{Rieke}}, \citenamefont {{Coil}}, \citenamefont {{Lotz}},
  \citenamefont {{Somerville}},\ and\ \citenamefont {{Bundy}}}]{Noeske07}%
  \BibitemOpen
  \bibfield  {author} {\bibinfo {author} {\bibfnamefont {K.~G.}\ \bibnamefont
  {{Noeske}}}, \bibinfo {author} {\bibfnamefont {S.~M.}\ \bibnamefont
  {{Faber}}}, \bibinfo {author} {\bibfnamefont {B.~J.}\ \bibnamefont
  {{Weiner}}}, \bibinfo {author} {\bibfnamefont {D.~C.}\ \bibnamefont {{Koo}}},
  \bibinfo {author} {\bibfnamefont {J.~R.}\ \bibnamefont {{Primack}}}, \bibinfo
  {author} {\bibfnamefont {A.}~\bibnamefont {{Dekel}}}, \bibinfo {author}
  {\bibfnamefont {C.}~\bibnamefont {{Papovich}}}, \bibinfo {author}
  {\bibfnamefont {C.~J.}\ \bibnamefont {{Conselice}}}, \bibinfo {author}
  {\bibfnamefont {E.}~\bibnamefont {{Le Floc'h}}}, \bibinfo {author}
  {\bibfnamefont {G.~H.}\ \bibnamefont {{Rieke}}}, \bibinfo {author}
  {\bibfnamefont {A.~L.}\ \bibnamefont {{Coil}}}, \bibinfo {author}
  {\bibfnamefont {J.~M.}\ \bibnamefont {{Lotz}}}, \bibinfo {author}
  {\bibfnamefont {R.~S.}\ \bibnamefont {{Somerville}}}, \ and\ \bibinfo
  {author} {\bibfnamefont {K.}~\bibnamefont {{Bundy}}},\ }\href {\doibase
  10.1086/517927} {\bibfield  {journal} {\bibinfo  {journal} {\apjl}\ }\textbf
  {\bibinfo {volume} {660}},\ \bibinfo {pages} {L47} (\bibinfo {year}
  {2007})},\ \Eprint {http://arxiv.org/abs/astro-ph/0703056} {astro-ph/0703056}
  \BibitemShut {NoStop}%
\bibitem [{\citenamefont {{Salim}}\ \emph {et~al.}(2007)\citenamefont
  {{Salim}}, \citenamefont {{Rich}}, \citenamefont {{Charlot}}, \citenamefont
  {{Brinchmann}}, \citenamefont {{Johnson}}, \citenamefont {{Schiminovich}},
  \citenamefont {{Seibert}}, \citenamefont {{Mallery}}, \citenamefont
  {{Heckman}}, \citenamefont {{Forster}}, \citenamefont {{Friedman}},
  \citenamefont {{Martin}}, \citenamefont {{Morrissey}}, \citenamefont
  {{Neff}}, \citenamefont {{Small}}, \citenamefont {{Wyder}}, \citenamefont
  {{Bianchi}}, \citenamefont {{Donas}}, \citenamefont {{Lee}}, \citenamefont
  {{Madore}}, \citenamefont {{Milliard}}, \citenamefont {{Szalay}},
  \citenamefont {{Welsh}},\ and\ \citenamefont {{Yi}}}]{Salim07}%
  \BibitemOpen
  \bibfield  {author} {\bibinfo {author} {\bibfnamefont {S.}~\bibnamefont
  {{Salim}}}, \bibinfo {author} {\bibfnamefont {R.~M.}\ \bibnamefont {{Rich}}},
  \bibinfo {author} {\bibfnamefont {S.}~\bibnamefont {{Charlot}}}, \bibinfo
  {author} {\bibfnamefont {J.}~\bibnamefont {{Brinchmann}}}, \bibinfo {author}
  {\bibfnamefont {B.~D.}\ \bibnamefont {{Johnson}}}, \bibinfo {author}
  {\bibfnamefont {D.}~\bibnamefont {{Schiminovich}}}, \bibinfo {author}
  {\bibfnamefont {M.}~\bibnamefont {{Seibert}}}, \bibinfo {author}
  {\bibfnamefont {R.}~\bibnamefont {{Mallery}}}, \bibinfo {author}
  {\bibfnamefont {T.~M.}\ \bibnamefont {{Heckman}}}, \bibinfo {author}
  {\bibfnamefont {K.}~\bibnamefont {{Forster}}}, \bibinfo {author}
  {\bibfnamefont {P.~G.}\ \bibnamefont {{Friedman}}}, \bibinfo {author}
  {\bibfnamefont {D.~C.}\ \bibnamefont {{Martin}}}, \bibinfo {author}
  {\bibfnamefont {P.}~\bibnamefont {{Morrissey}}}, \bibinfo {author}
  {\bibfnamefont {S.~G.}\ \bibnamefont {{Neff}}}, \bibinfo {author}
  {\bibfnamefont {T.}~\bibnamefont {{Small}}}, \bibinfo {author} {\bibfnamefont
  {T.~K.}\ \bibnamefont {{Wyder}}}, \bibinfo {author} {\bibfnamefont
  {L.}~\bibnamefont {{Bianchi}}}, \bibinfo {author} {\bibfnamefont
  {J.}~\bibnamefont {{Donas}}}, \bibinfo {author} {\bibfnamefont {Y.-W.}\
  \bibnamefont {{Lee}}}, \bibinfo {author} {\bibfnamefont {B.~F.}\ \bibnamefont
  {{Madore}}}, \bibinfo {author} {\bibfnamefont {B.}~\bibnamefont
  {{Milliard}}}, \bibinfo {author} {\bibfnamefont {A.~S.}\ \bibnamefont
  {{Szalay}}}, \bibinfo {author} {\bibfnamefont {B.~Y.}\ \bibnamefont
  {{Welsh}}}, \ and\ \bibinfo {author} {\bibfnamefont {S.~K.}\ \bibnamefont
  {{Yi}}},\ }\href {\doibase 10.1086/519218} {\bibfield  {journal} {\bibinfo
  {journal} {\apjs}\ }\textbf {\bibinfo {volume} {173}},\ \bibinfo {pages}
  {267} (\bibinfo {year} {2007})},\ \Eprint {http://arxiv.org/abs/0704.3611}
  {arXiv:0704.3611} \BibitemShut {NoStop}%
\bibitem [{\citenamefont {{Binney}}\ \emph {et~al.}(2000)\citenamefont
  {{Binney}}, \citenamefont {{Dehnen}},\ and\ \citenamefont
  {{Bertelli}}}]{Binney00}%
  \BibitemOpen
  \bibfield  {author} {\bibinfo {author} {\bibfnamefont {J.}~\bibnamefont
  {{Binney}}}, \bibinfo {author} {\bibfnamefont {W.}~\bibnamefont {{Dehnen}}},
  \ and\ \bibinfo {author} {\bibfnamefont {G.}~\bibnamefont {{Bertelli}}},\
  }\href {\doibase 10.1046/j.1365-8711.2000.03720.x} {\bibfield  {journal}
  {\bibinfo  {journal} {\mnras}\ }\textbf {\bibinfo {volume} {318}},\ \bibinfo
  {pages} {658} (\bibinfo {year} {2000})},\ \Eprint
  {http://arxiv.org/abs/astro-ph/0003479} {astro-ph/0003479} \BibitemShut
  {NoStop}%
\bibitem [{\citenamefont {{Pisano}}(2014)}]{Pisano14}%
  \BibitemOpen
  \bibfield  {author} {\bibinfo {author} {\bibfnamefont {D.~J.}\ \bibnamefont
  {{Pisano}}},\ }\href {\doibase 10.1088/0004-6256/147/3/48} {\bibfield
  {journal} {\bibinfo  {journal} {\aj}\ }\textbf {\bibinfo {volume} {147}},\
  \bibinfo {eid} {48} (\bibinfo {year} {2014})},\ \Eprint
  {http://arxiv.org/abs/1312.3953} {arXiv:1312.3953 [astro-ph.GA]} \BibitemShut
  {NoStop}%
\bibitem [{\citenamefont {{Masters}}\ \emph {et~al.}(2010)\citenamefont
  {{Masters}}, \citenamefont {{Mosleh}}, \citenamefont {{Romer}}, \citenamefont
  {{Nichol}}, \citenamefont {{Bamford}}, \citenamefont {{Schawinski}},
  \citenamefont {{Lintott}}, \citenamefont {{Andreescu}}, \citenamefont
  {{Campbell}}, \citenamefont {{Crowcroft}}, \citenamefont {{Doyle}},
  \citenamefont {{Edmondson}}, \citenamefont {{Murray}}, \citenamefont
  {{Raddick}}, \citenamefont {{Slosar}}, \citenamefont {{Szalay}},\ and\
  \citenamefont {{Vandenberg}}}]{Masters10}%
  \BibitemOpen
  \bibfield  {author} {\bibinfo {author} {\bibfnamefont {K.~L.}\ \bibnamefont
  {{Masters}}}, \bibinfo {author} {\bibfnamefont {M.}~\bibnamefont {{Mosleh}}},
  \bibinfo {author} {\bibfnamefont {A.~K.}\ \bibnamefont {{Romer}}}, \bibinfo
  {author} {\bibfnamefont {R.~C.}\ \bibnamefont {{Nichol}}}, \bibinfo {author}
  {\bibfnamefont {S.~P.}\ \bibnamefont {{Bamford}}}, \bibinfo {author}
  {\bibfnamefont {K.}~\bibnamefont {{Schawinski}}}, \bibinfo {author}
  {\bibfnamefont {C.~J.}\ \bibnamefont {{Lintott}}}, \bibinfo {author}
  {\bibfnamefont {D.}~\bibnamefont {{Andreescu}}}, \bibinfo {author}
  {\bibfnamefont {H.~C.}\ \bibnamefont {{Campbell}}}, \bibinfo {author}
  {\bibfnamefont {B.}~\bibnamefont {{Crowcroft}}}, \bibinfo {author}
  {\bibfnamefont {I.}~\bibnamefont {{Doyle}}}, \bibinfo {author} {\bibfnamefont
  {E.~M.}\ \bibnamefont {{Edmondson}}}, \bibinfo {author} {\bibfnamefont
  {P.}~\bibnamefont {{Murray}}}, \bibinfo {author} {\bibfnamefont {M.~J.}\
  \bibnamefont {{Raddick}}}, \bibinfo {author} {\bibfnamefont {A.}~\bibnamefont
  {{Slosar}}}, \bibinfo {author} {\bibfnamefont {A.~S.}\ \bibnamefont
  {{Szalay}}}, \ and\ \bibinfo {author} {\bibfnamefont {J.}~\bibnamefont
  {{Vandenberg}}},\ }\href {\doibase 10.1111/j.1365-2966.2010.16503.x}
  {\bibfield  {journal} {\bibinfo  {journal} {\mnras}\ }\textbf {\bibinfo
  {volume} {405}},\ \bibinfo {pages} {783} (\bibinfo {year} {2010})},\ \Eprint
  {http://arxiv.org/abs/0910.4113} {arXiv:0910.4113} \BibitemShut {NoStop}%
\bibitem [{\citenamefont {{Larson}}\ \emph {et~al.}(1980)\citenamefont
  {{Larson}}, \citenamefont {{Tinsley}},\ and\ \citenamefont
  {{Caldwell}}}]{Larson80}%
  \BibitemOpen
  \bibfield  {author} {\bibinfo {author} {\bibfnamefont {R.~B.}\ \bibnamefont
  {{Larson}}}, \bibinfo {author} {\bibfnamefont {B.~M.}\ \bibnamefont
  {{Tinsley}}}, \ and\ \bibinfo {author} {\bibfnamefont {C.~N.}\ \bibnamefont
  {{Caldwell}}},\ }\href {\doibase 10.1086/157917} {\bibfield  {journal}
  {\bibinfo  {journal} {\apj}\ }\textbf {\bibinfo {volume} {237}},\ \bibinfo
  {pages} {692} (\bibinfo {year} {1980})}\BibitemShut {NoStop}%
\bibitem [{\citenamefont {{van den Bosch}}\ \emph {et~al.}(2008)\citenamefont
  {{van den Bosch}}, \citenamefont {{Aquino}}, \citenamefont {{Yang}},
  \citenamefont {{Mo}}, \citenamefont {{Pasquali}}, \citenamefont {{McIntosh}},
  \citenamefont {{Weinmann}},\ and\ \citenamefont {{Kang}}}]{Bosch08}%
  \BibitemOpen
  \bibfield  {author} {\bibinfo {author} {\bibfnamefont {F.~C.}\ \bibnamefont
  {{van den Bosch}}}, \bibinfo {author} {\bibfnamefont {D.}~\bibnamefont
  {{Aquino}}}, \bibinfo {author} {\bibfnamefont {X.}~\bibnamefont {{Yang}}},
  \bibinfo {author} {\bibfnamefont {H.~J.}\ \bibnamefont {{Mo}}}, \bibinfo
  {author} {\bibfnamefont {A.}~\bibnamefont {{Pasquali}}}, \bibinfo {author}
  {\bibfnamefont {D.~H.}\ \bibnamefont {{McIntosh}}}, \bibinfo {author}
  {\bibfnamefont {S.~M.}\ \bibnamefont {{Weinmann}}}, \ and\ \bibinfo {author}
  {\bibfnamefont {X.}~\bibnamefont {{Kang}}},\ }\href {\doibase
  10.1111/j.1365-2966.2008.13230.x} {\bibfield  {journal} {\bibinfo  {journal}
  {\mnras}\ }\textbf {\bibinfo {volume} {387}},\ \bibinfo {pages} {79}
  (\bibinfo {year} {2008})},\ \Eprint {http://arxiv.org/abs/0710.3164}
  {arXiv:0710.3164} \BibitemShut {NoStop}%
\bibitem [{\citenamefont {{Shaver}}\ \emph {et~al.}(1996)\citenamefont
  {{Shaver}}, \citenamefont {{Wall}}, \citenamefont {{Kellermann}},
  \citenamefont {{Jackson}},\ and\ \citenamefont {{Hawkins}}}]{Shaver96}%
  \BibitemOpen
  \bibfield  {author} {\bibinfo {author} {\bibfnamefont {P.~A.}\ \bibnamefont
  {{Shaver}}}, \bibinfo {author} {\bibfnamefont {J.~V.}\ \bibnamefont
  {{Wall}}}, \bibinfo {author} {\bibfnamefont {K.~I.}\ \bibnamefont
  {{Kellermann}}}, \bibinfo {author} {\bibfnamefont {C.~A.}\ \bibnamefont
  {{Jackson}}}, \ and\ \bibinfo {author} {\bibfnamefont {M.~R.~S.}\
  \bibnamefont {{Hawkins}}},\ }\href {\doibase 10.1038/384439a0} {\bibfield
  {journal} {\bibinfo  {journal} {\nat}\ }\textbf {\bibinfo {volume} {384}},\
  \bibinfo {pages} {439} (\bibinfo {year} {1996})}\BibitemShut {NoStop}%
\bibitem [{\citenamefont {{Page}}\ \emph {et~al.}(2001)\citenamefont {{Page}},
  \citenamefont {{Stevens}}, \citenamefont {{Mittaz}},\ and\ \citenamefont
  {{Carrera}}}]{Page01}%
  \BibitemOpen
  \bibfield  {author} {\bibinfo {author} {\bibfnamefont {M.~J.}\ \bibnamefont
  {{Page}}}, \bibinfo {author} {\bibfnamefont {J.~A.}\ \bibnamefont
  {{Stevens}}}, \bibinfo {author} {\bibfnamefont {J.~P.~D.}\ \bibnamefont
  {{Mittaz}}}, \ and\ \bibinfo {author} {\bibfnamefont {F.~J.}\ \bibnamefont
  {{Carrera}}},\ }\href {\doibase 10.1126/science.1065880} {\bibfield
  {journal} {\bibinfo  {journal} {Science}\ }\textbf {\bibinfo {volume}
  {294}},\ \bibinfo {pages} {2516} (\bibinfo {year} {2001})},\ \Eprint
  {http://arxiv.org/abs/astro-ph/0202102} {astro-ph/0202102} \BibitemShut
  {NoStop}%
\bibitem [{\citenamefont {{Hasinger}}\ \emph {et~al.}(2005)\citenamefont
  {{Hasinger}}, \citenamefont {{Miyaji}},\ and\ \citenamefont
  {{Schmidt}}}]{Hasinger05}%
  \BibitemOpen
  \bibfield  {author} {\bibinfo {author} {\bibfnamefont {G.}~\bibnamefont
  {{Hasinger}}}, \bibinfo {author} {\bibfnamefont {T.}~\bibnamefont
  {{Miyaji}}}, \ and\ \bibinfo {author} {\bibfnamefont {M.}~\bibnamefont
  {{Schmidt}}},\ }\href {\doibase 10.1051/0004-6361:20042134} {\bibfield
  {journal} {\bibinfo  {journal} {\aap}\ }\textbf {\bibinfo {volume} {441}},\
  \bibinfo {pages} {417} (\bibinfo {year} {2005})},\ \Eprint
  {http://arxiv.org/abs/astro-ph/0506118} {astro-ph/0506118} \BibitemShut
  {NoStop}%
\bibitem [{\citenamefont {{Babi{\'c}}}\ \emph {et~al.}(2007)\citenamefont
  {{Babi{\'c}}}, \citenamefont {{Miller}}, \citenamefont {{Jarvis}},
  \citenamefont {{Turner}}, \citenamefont {{Alexander}},\ and\ \citenamefont
  {{Croom}}}]{Babi07}%
  \BibitemOpen
  \bibfield  {author} {\bibinfo {author} {\bibfnamefont {A.}~\bibnamefont
  {{Babi{\'c}}}}, \bibinfo {author} {\bibfnamefont {L.}~\bibnamefont
  {{Miller}}}, \bibinfo {author} {\bibfnamefont {M.~J.}\ \bibnamefont
  {{Jarvis}}}, \bibinfo {author} {\bibfnamefont {T.~J.}\ \bibnamefont
  {{Turner}}}, \bibinfo {author} {\bibfnamefont {D.~M.}\ \bibnamefont
  {{Alexander}}}, \ and\ \bibinfo {author} {\bibfnamefont {S.~M.}\ \bibnamefont
  {{Croom}}},\ }\href {\doibase 10.1051/0004-6361:20078286} {\bibfield
  {journal} {\bibinfo  {journal} {\aap}\ }\textbf {\bibinfo {volume} {474}},\
  \bibinfo {pages} {755} (\bibinfo {year} {2007})},\ \Eprint
  {http://arxiv.org/abs/0709.0786} {arXiv:0709.0786} \BibitemShut {NoStop}%
\bibitem [{\citenamefont {{Enoki}}\ \emph {et~al.}(2014)\citenamefont
  {{Enoki}}, \citenamefont {{Ishiyama}}, \citenamefont {{Kobayashi}},\ and\
  \citenamefont {{Nagashima}}}]{Enoki14}%
  \BibitemOpen
  \bibfield  {author} {\bibinfo {author} {\bibfnamefont {M.}~\bibnamefont
  {{Enoki}}}, \bibinfo {author} {\bibfnamefont {T.}~\bibnamefont {{Ishiyama}}},
  \bibinfo {author} {\bibfnamefont {M.~A.~R.}\ \bibnamefont {{Kobayashi}}}, \
  and\ \bibinfo {author} {\bibfnamefont {M.}~\bibnamefont {{Nagashima}}},\
  }\href {\doibase 10.1088/0004-637X/794/1/69} {\bibfield  {journal} {\bibinfo
  {journal} {\apj}\ }\textbf {\bibinfo {volume} {794}},\ \bibinfo {eid} {69}
  (\bibinfo {year} {2014})},\ \Eprint {http://arxiv.org/abs/1408.3726}
  {arXiv:1408.3726} \BibitemShut {NoStop}%
\bibitem [{\citenamefont {{Suh}}\ \emph {et~al.}(2017)\citenamefont {{Suh}},
  \citenamefont {{Civano}}, \citenamefont {{Hasinger}}, \citenamefont
  {{Lusso}}, \citenamefont {{Lanzuisi}}, \citenamefont {{Marchesi}},
  \citenamefont {{Trakhtenbrot}}, \citenamefont {{Allevato}}, \citenamefont
  {{Cappelluti}}, \citenamefont {{Capak}}, \citenamefont {{Elvis}},
  \citenamefont {{Griffiths}}, \citenamefont {{Laigle}}, \citenamefont
  {{Lira}}, \citenamefont {{Riguccini}}, \citenamefont {{Rosario}},
  \citenamefont {{Salvato}}, \citenamefont {{Schawinski}},\ and\ \citenamefont
  {{Vignali}}}]{Suh17}%
  \BibitemOpen
  \bibfield  {author} {\bibinfo {author} {\bibfnamefont {H.}~\bibnamefont
  {{Suh}}}, \bibinfo {author} {\bibfnamefont {F.}~\bibnamefont {{Civano}}},
  \bibinfo {author} {\bibfnamefont {G.}~\bibnamefont {{Hasinger}}}, \bibinfo
  {author} {\bibfnamefont {E.}~\bibnamefont {{Lusso}}}, \bibinfo {author}
  {\bibfnamefont {G.}~\bibnamefont {{Lanzuisi}}}, \bibinfo {author}
  {\bibfnamefont {S.}~\bibnamefont {{Marchesi}}}, \bibinfo {author}
  {\bibfnamefont {B.}~\bibnamefont {{Trakhtenbrot}}}, \bibinfo {author}
  {\bibfnamefont {V.}~\bibnamefont {{Allevato}}}, \bibinfo {author}
  {\bibfnamefont {N.}~\bibnamefont {{Cappelluti}}}, \bibinfo {author}
  {\bibfnamefont {P.~L.}\ \bibnamefont {{Capak}}}, \bibinfo {author}
  {\bibfnamefont {M.}~\bibnamefont {{Elvis}}}, \bibinfo {author} {\bibfnamefont
  {R.~E.}\ \bibnamefont {{Griffiths}}}, \bibinfo {author} {\bibfnamefont
  {C.}~\bibnamefont {{Laigle}}}, \bibinfo {author} {\bibfnamefont
  {P.}~\bibnamefont {{Lira}}}, \bibinfo {author} {\bibfnamefont
  {L.}~\bibnamefont {{Riguccini}}}, \bibinfo {author} {\bibfnamefont {D.~J.}\
  \bibnamefont {{Rosario}}}, \bibinfo {author} {\bibfnamefont {M.}~\bibnamefont
  {{Salvato}}}, \bibinfo {author} {\bibfnamefont {K.}~\bibnamefont
  {{Schawinski}}}, \ and\ \bibinfo {author} {\bibfnamefont {C.}~\bibnamefont
  {{Vignali}}},\ }\href {\doibase 10.3847/1538-4357/aa725c} {\bibfield
  {journal} {\bibinfo  {journal} {\apj}\ }\textbf {\bibinfo {volume} {841}},\
  \bibinfo {eid} {102} (\bibinfo {year} {2017})},\ \Eprint
  {http://arxiv.org/abs/1705.03890} {arXiv:1705.03890} \BibitemShut {NoStop}%
\bibitem [{\citenamefont {{Planck Collaboration}}\ \emph
  {et~al.}(2015)\citenamefont {{Planck Collaboration}}, \citenamefont {{Ade}},
  \citenamefont {{Aghanim}}, \citenamefont {{Arnaud}}, \citenamefont
  {{Ashdown}}, \citenamefont {{Aumont}}, \citenamefont {{Baccigalupi}},
  \citenamefont {{Banday}}, \citenamefont {{Barreiro}}, \citenamefont
  {{Bartlett}},\ and\ \citenamefont {et~al.}}]{Planck15}%
  \BibitemOpen
  \bibfield  {author} {\bibinfo {author} {\bibnamefont {{Planck
  Collaboration}}}, \bibinfo {author} {\bibfnamefont {P.~A.~R.}\ \bibnamefont
  {{Ade}}}, \bibinfo {author} {\bibfnamefont {N.}~\bibnamefont {{Aghanim}}},
  \bibinfo {author} {\bibfnamefont {M.}~\bibnamefont {{Arnaud}}}, \bibinfo
  {author} {\bibfnamefont {M.}~\bibnamefont {{Ashdown}}}, \bibinfo {author}
  {\bibfnamefont {J.}~\bibnamefont {{Aumont}}}, \bibinfo {author}
  {\bibfnamefont {C.}~\bibnamefont {{Baccigalupi}}}, \bibinfo {author}
  {\bibfnamefont {A.~J.}\ \bibnamefont {{Banday}}}, \bibinfo {author}
  {\bibfnamefont {R.~B.}\ \bibnamefont {{Barreiro}}}, \bibinfo {author}
  {\bibfnamefont {J.~G.}\ \bibnamefont {{Bartlett}}}, \ and\ \bibinfo {author}
  {\bibnamefont {et~al.}},\ }\href@noop {} {\bibfield  {journal} {\bibinfo
  {journal} {ArXiv e-prints}\ } (\bibinfo {year} {2015})},\ \Eprint
  {http://arxiv.org/abs/1502.01589} {arXiv:1502.01589} \BibitemShut {NoStop}%
\bibitem [{\citenamefont {{Aragon-Calvo}}(2012)}]{Aragon12}%
  \BibitemOpen
  \bibfield  {author} {\bibinfo {author} {\bibfnamefont {M.~A.}\ \bibnamefont
  {{Aragon-Calvo}}},\ }\href@noop {} {\bibfield  {journal} {\bibinfo  {journal}
  {ArXiv e-prints}\ } (\bibinfo {year} {2012})},\ \Eprint
  {http://arxiv.org/abs/1210.7871} {arXiv:1210.7871 [astro-ph.CO]} \BibitemShut
  {NoStop}%
\bibitem [{\citenamefont {{Aragon-Calvo}}\ and\ \citenamefont
  {{Yang}}(2014)}]{Aragon14}%
  \BibitemOpen
  \bibfield  {author} {\bibinfo {author} {\bibfnamefont {M.~A.}\ \bibnamefont
  {{Aragon-Calvo}}}\ and\ \bibinfo {author} {\bibfnamefont {L.~F.}\
  \bibnamefont {{Yang}}},\ }\href {\doibase 10.1093/mnrasl/slu009} {\bibfield
  {journal} {\bibinfo  {journal} {\mnras}\ }\textbf {\bibinfo {volume} {440}},\
  \bibinfo {pages} {L46} (\bibinfo {year} {2014})},\ \Eprint
  {http://arxiv.org/abs/1303.1590} {arXiv:1303.1590 [astro-ph.CO]} \BibitemShut
  {NoStop}%
\bibitem [{\citenamefont {{Hockney}}\ and\ \citenamefont
  {{Eastwood}}(1988)}]{Hockney88}%
  \BibitemOpen
  \bibfield  {author} {\bibinfo {author} {\bibfnamefont {R.~W.}\ \bibnamefont
  {{Hockney}}}\ and\ \bibinfo {author} {\bibfnamefont {J.~W.}\ \bibnamefont
  {{Eastwood}}},\ }\href@noop {} {\emph {\bibinfo {title} {Bristol: Hilger,
  1988}}}\ (\bibinfo {year} {1988})\BibitemShut {NoStop}%
\bibitem [{\citenamefont {{Barnes}}\ and\ \citenamefont
  {{Hut}}(1986)}]{Barnes86}%
  \BibitemOpen
  \bibfield  {author} {\bibinfo {author} {\bibfnamefont {J.}~\bibnamefont
  {{Barnes}}}\ and\ \bibinfo {author} {\bibfnamefont {P.}~\bibnamefont
  {{Hut}}},\ }\href {\doibase 10.1038/324446a0} {\bibfield  {journal} {\bibinfo
   {journal} {\nat}\ }\textbf {\bibinfo {volume} {324}},\ \bibinfo {pages}
  {446} (\bibinfo {year} {1986})}\BibitemShut {NoStop}%
\bibitem [{\citenamefont {{Lucy}}(1977)}]{Lucy77}%
  \BibitemOpen
  \bibfield  {author} {\bibinfo {author} {\bibfnamefont {L.~B.}\ \bibnamefont
  {{Lucy}}},\ }\href {\doibase 10.1086/112164} {\bibfield  {journal} {\bibinfo
  {journal} {\aj}\ }\textbf {\bibinfo {volume} {82}},\ \bibinfo {pages} {1013}
  (\bibinfo {year} {1977})}\BibitemShut {NoStop}%
\bibitem [{\citenamefont {{Springel}}\ and\ \citenamefont
  {{Hernquist}}(2002)}]{Springel02}%
  \BibitemOpen
  \bibfield  {author} {\bibinfo {author} {\bibfnamefont {V.}~\bibnamefont
  {{Springel}}}\ and\ \bibinfo {author} {\bibfnamefont {L.}~\bibnamefont
  {{Hernquist}}},\ }\href {\doibase 10.1046/j.1365-8711.2002.05445.x}
  {\bibfield  {journal} {\bibinfo  {journal} {\mnras}\ }\textbf {\bibinfo
  {volume} {333}},\ \bibinfo {pages} {649} (\bibinfo {year} {2002})},\ \Eprint
  {http://arxiv.org/abs/astro-ph/0111016} {astro-ph/0111016} \BibitemShut
  {NoStop}%
\bibitem [{\citenamefont {{Springel}}\ and\ \citenamefont
  {{Hernquist}}(2003)}]{Springel03}%
  \BibitemOpen
  \bibfield  {author} {\bibinfo {author} {\bibfnamefont {V.}~\bibnamefont
  {{Springel}}}\ and\ \bibinfo {author} {\bibfnamefont {L.}~\bibnamefont
  {{Hernquist}}},\ }\href {\doibase 10.1046/j.1365-8711.2003.06207.x}
  {\bibfield  {journal} {\bibinfo  {journal} {\mnras}\ }\textbf {\bibinfo
  {volume} {339}},\ \bibinfo {pages} {312} (\bibinfo {year} {2003})},\ \Eprint
  {http://arxiv.org/abs/astro-ph/0206395} {astro-ph/0206395} \BibitemShut
  {NoStop}%
\bibitem [{\citenamefont {{Davis}}\ \emph {et~al.}(1985)\citenamefont
  {{Davis}}, \citenamefont {{Efstathiou}}, \citenamefont {{Frenk}},\ and\
  \citenamefont {{White}}}]{Davis85}%
  \BibitemOpen
  \bibfield  {author} {\bibinfo {author} {\bibfnamefont {M.}~\bibnamefont
  {{Davis}}}, \bibinfo {author} {\bibfnamefont {G.}~\bibnamefont
  {{Efstathiou}}}, \bibinfo {author} {\bibfnamefont {C.~S.}\ \bibnamefont
  {{Frenk}}}, \ and\ \bibinfo {author} {\bibfnamefont {S.~D.~M.}\ \bibnamefont
  {{White}}},\ }\href {\doibase 10.1086/163168} {\bibfield  {journal} {\bibinfo
   {journal} {\apj}\ }\textbf {\bibinfo {volume} {292}},\ \bibinfo {pages}
  {371} (\bibinfo {year} {1985})}\BibitemShut {NoStop}%
\bibitem [{\citenamefont {{Onions}}\ \emph {et~al.}(2012)\citenamefont
  {{Onions}}, \citenamefont {{Knebe}}, \citenamefont {{Pearce}}, \citenamefont
  {{Muldrew}}, \citenamefont {{Lux}}, \citenamefont {{Knollmann}},
  \citenamefont {{Ascasibar}}, \citenamefont {{Behroozi}}, \citenamefont
  {{Elahi}}, \citenamefont {{Han}}, \citenamefont {{Maciejewski}},
  \citenamefont {{Merch{\'a}n}}, \citenamefont {{Neyrinck}}, \citenamefont
  {{Ruiz}}, \citenamefont {{Sgr{\'o}}}, \citenamefont {{Springel}},\ and\
  \citenamefont {{Tweed}}}]{Onions12}%
  \BibitemOpen
  \bibfield  {author} {\bibinfo {author} {\bibfnamefont {J.}~\bibnamefont
  {{Onions}}}, \bibinfo {author} {\bibfnamefont {A.}~\bibnamefont {{Knebe}}},
  \bibinfo {author} {\bibfnamefont {F.~R.}\ \bibnamefont {{Pearce}}}, \bibinfo
  {author} {\bibfnamefont {S.~I.}\ \bibnamefont {{Muldrew}}}, \bibinfo {author}
  {\bibfnamefont {H.}~\bibnamefont {{Lux}}}, \bibinfo {author} {\bibfnamefont
  {S.~R.}\ \bibnamefont {{Knollmann}}}, \bibinfo {author} {\bibfnamefont
  {Y.}~\bibnamefont {{Ascasibar}}}, \bibinfo {author} {\bibfnamefont
  {P.}~\bibnamefont {{Behroozi}}}, \bibinfo {author} {\bibfnamefont
  {P.}~\bibnamefont {{Elahi}}}, \bibinfo {author} {\bibfnamefont
  {J.}~\bibnamefont {{Han}}}, \bibinfo {author} {\bibfnamefont
  {M.}~\bibnamefont {{Maciejewski}}}, \bibinfo {author} {\bibfnamefont {M.~E.}\
  \bibnamefont {{Merch{\'a}n}}}, \bibinfo {author} {\bibfnamefont
  {M.}~\bibnamefont {{Neyrinck}}}, \bibinfo {author} {\bibfnamefont {A.~N.}\
  \bibnamefont {{Ruiz}}}, \bibinfo {author} {\bibfnamefont {M.~A.}\
  \bibnamefont {{Sgr{\'o}}}}, \bibinfo {author} {\bibfnamefont
  {V.}~\bibnamefont {{Springel}}}, \ and\ \bibinfo {author} {\bibfnamefont
  {D.}~\bibnamefont {{Tweed}}},\ }\href {\doibase
  10.1111/j.1365-2966.2012.20947.x} {\bibfield  {journal} {\bibinfo  {journal}
  {\mnras}\ }\textbf {\bibinfo {volume} {423}},\ \bibinfo {pages} {1200}
  (\bibinfo {year} {2012})},\ \Eprint {http://arxiv.org/abs/1203.3695}
  {arXiv:1203.3695 [astro-ph.CO]} \BibitemShut {NoStop}%
\bibitem [{\citenamefont {{Conroy}}\ \emph {et~al.}(2006)\citenamefont
  {{Conroy}}, \citenamefont {{Wechsler}},\ and\ \citenamefont
  {{Kravtsov}}}]{Conroy06}%
  \BibitemOpen
  \bibfield  {author} {\bibinfo {author} {\bibfnamefont {C.}~\bibnamefont
  {{Conroy}}}, \bibinfo {author} {\bibfnamefont {R.~H.}\ \bibnamefont
  {{Wechsler}}}, \ and\ \bibinfo {author} {\bibfnamefont {A.~V.}\ \bibnamefont
  {{Kravtsov}}},\ }\href {\doibase 10.1086/503602} {\bibfield  {journal}
  {\bibinfo  {journal} {\apj}\ }\textbf {\bibinfo {volume} {647}},\ \bibinfo
  {pages} {201} (\bibinfo {year} {2006})},\ \Eprint
  {http://arxiv.org/abs/astro-ph/0512234} {astro-ph/0512234} \BibitemShut
  {NoStop}%
\bibitem [{\citenamefont {{Vale}}\ and\ \citenamefont
  {{Ostriker}}(2006)}]{Vale06}%
  \BibitemOpen
  \bibfield  {author} {\bibinfo {author} {\bibfnamefont {A.}~\bibnamefont
  {{Vale}}}\ and\ \bibinfo {author} {\bibfnamefont {J.~P.}\ \bibnamefont
  {{Ostriker}}},\ }\href {\doibase 10.1111/j.1365-2966.2006.10605.x} {\bibfield
   {journal} {\bibinfo  {journal} {\mnras}\ }\textbf {\bibinfo {volume}
  {371}},\ \bibinfo {pages} {1173} (\bibinfo {year} {2006})},\ \Eprint
  {http://arxiv.org/abs/astro-ph/0511816} {astro-ph/0511816} \BibitemShut
  {NoStop}%
\bibitem [{\citenamefont {{York}}\ \emph {et~al.}(2000)\citenamefont {{York}},
  \citenamefont {{Adelman}}, \citenamefont {{Anderson}}, \citenamefont
  {{Anderson}}, \citenamefont {{Annis}}, \citenamefont {{Bahcall}},
  \citenamefont {{Bakken}}, \citenamefont {{Barkhouser}}, \citenamefont
  {{Bastian}}, \citenamefont {{Berman}}, \citenamefont {{Boroski}},
  \citenamefont {{Bracker}}, \citenamefont {{Briegel}}, \citenamefont
  {{Briggs}}, \citenamefont {{Brinkmann}}, \citenamefont {{Brunner}},
  \citenamefont {{Burles}}, \citenamefont {{Carey}}, \citenamefont {{Carr}},
  \citenamefont {{Castander}}, \citenamefont {{Chen}}, \citenamefont
  {{Colestock}}, \citenamefont {{Connolly}}, \citenamefont {{Crocker}},
  \citenamefont {{Csabai}}, \citenamefont {{Czarapata}}, \citenamefont
  {{Davis}}, \citenamefont {{Doi}}, \citenamefont {{Dombeck}}, \citenamefont
  {{Eisenstein}}, \citenamefont {{Ellman}}, \citenamefont {{Elms}},
  \citenamefont {{Evans}}, \citenamefont {{Fan}}, \citenamefont {{Federwitz}},
  \citenamefont {{Fiscelli}}, \citenamefont {{Friedman}}, \citenamefont
  {{Frieman}}, \citenamefont {{Fukugita}}, \citenamefont {{Gillespie}},
  \citenamefont {{Gunn}}, \citenamefont {{Gurbani}}, \citenamefont {{de Haas}},
  \citenamefont {{Haldeman}}, \citenamefont {{Harris}}, \citenamefont
  {{Hayes}}, \citenamefont {{Heckman}}, \citenamefont {{Hennessy}},
  \citenamefont {{Hindsley}}, \citenamefont {{Holm}}, \citenamefont
  {{Holmgren}}, \citenamefont {{Huang}}, \citenamefont {{Hull}}, \citenamefont
  {{Husby}}, \citenamefont {{Ichikawa}}, \citenamefont {{Ichikawa}},
  \citenamefont {{Ivezi{\'c}}}, \citenamefont {{Kent}}, \citenamefont {{Kim}},
  \citenamefont {{Kinney}}, \citenamefont {{Klaene}}, \citenamefont
  {{Kleinman}}, \citenamefont {{Kleinman}}, \citenamefont {{Knapp}},
  \citenamefont {{Korienek}}, \citenamefont {{Kron}}, \citenamefont {{Kunszt}},
  \citenamefont {{Lamb}}, \citenamefont {{Lee}}, \citenamefont {{Leger}},
  \citenamefont {{Limmongkol}}, \citenamefont {{Lindenmeyer}}, \citenamefont
  {{Long}}, \citenamefont {{Loomis}}, \citenamefont {{Loveday}}, \citenamefont
  {{Lucinio}}, \citenamefont {{Lupton}}, \citenamefont {{MacKinnon}},
  \citenamefont {{Mannery}}, \citenamefont {{Mantsch}}, \citenamefont
  {{Margon}}, \citenamefont {{McGehee}}, \citenamefont {{McKay}}, \citenamefont
  {{Meiksin}}, \citenamefont {{Merelli}}, \citenamefont {{Monet}},
  \citenamefont {{Munn}}, \citenamefont {{Narayanan}}, \citenamefont {{Nash}},
  \citenamefont {{Neilsen}}, \citenamefont {{Neswold}}, \citenamefont
  {{Newberg}}, \citenamefont {{Nichol}}, \citenamefont {{Nicinski}},
  \citenamefont {{Nonino}}, \citenamefont {{Okada}}, \citenamefont {{Okamura}},
  \citenamefont {{Ostriker}}, \citenamefont {{Owen}}, \citenamefont {{Pauls}},
  \citenamefont {{Peoples}}, \citenamefont {{Peterson}}, \citenamefont
  {{Petravick}}, \citenamefont {{Pier}}, \citenamefont {{Pope}}, \citenamefont
  {{Pordes}}, \citenamefont {{Prosapio}}, \citenamefont {{Rechenmacher}},
  \citenamefont {{Quinn}}, \citenamefont {{Richards}}, \citenamefont
  {{Richmond}}, \citenamefont {{Rivetta}}, \citenamefont {{Rockosi}},
  \citenamefont {{Ruthmansdorfer}}, \citenamefont {{Sandford}}, \citenamefont
  {{Schlegel}}, \citenamefont {{Schneider}}, \citenamefont {{Sekiguchi}},
  \citenamefont {{Sergey}}, \citenamefont {{Shimasaku}}, \citenamefont
  {{Siegmund}}, \citenamefont {{Smee}}, \citenamefont {{Smith}}, \citenamefont
  {{Snedden}}, \citenamefont {{Stone}}, \citenamefont {{Stoughton}},
  \citenamefont {{Strauss}}, \citenamefont {{Stubbs}}, \citenamefont
  {{SubbaRao}}, \citenamefont {{Szalay}}, \citenamefont {{Szapudi}},
  \citenamefont {{Szokoly}}, \citenamefont {{Thakar}}, \citenamefont
  {{Tremonti}}, \citenamefont {{Tucker}}, \citenamefont {{Uomoto}},
  \citenamefont {{Vanden Berk}}, \citenamefont {{Vogeley}}, \citenamefont
  {{Waddell}}, \citenamefont {{Wang}}, \citenamefont {{Watanabe}},
  \citenamefont {{Weinberg}}, \citenamefont {{Yanny}}, \citenamefont
  {{Yasuda}},\ and\ \citenamefont {{SDSS Collaboration}}}]{York00}%
  \BibitemOpen
  \bibfield  {author} {\bibinfo {author} {\bibfnamefont {D.~G.}\ \bibnamefont
  {{York}}}, \bibinfo {author} {\bibfnamefont {J.}~\bibnamefont {{Adelman}}},
  \bibinfo {author} {\bibfnamefont {J.~E.}\ \bibnamefont {{Anderson}},
  \bibfnamefont {Jr.}}, \bibinfo {author} {\bibfnamefont {S.~F.}\ \bibnamefont
  {{Anderson}}}, \bibinfo {author} {\bibfnamefont {J.}~\bibnamefont {{Annis}}},
  \bibinfo {author} {\bibfnamefont {N.~A.}\ \bibnamefont {{Bahcall}}}, \bibinfo
  {author} {\bibfnamefont {J.~A.}\ \bibnamefont {{Bakken}}}, \bibinfo {author}
  {\bibfnamefont {R.}~\bibnamefont {{Barkhouser}}}, \bibinfo {author}
  {\bibfnamefont {S.}~\bibnamefont {{Bastian}}}, \bibinfo {author}
  {\bibfnamefont {E.}~\bibnamefont {{Berman}}}, \bibinfo {author}
  {\bibfnamefont {W.~N.}\ \bibnamefont {{Boroski}}}, \bibinfo {author}
  {\bibfnamefont {S.}~\bibnamefont {{Bracker}}}, \bibinfo {author}
  {\bibfnamefont {C.}~\bibnamefont {{Briegel}}}, \bibinfo {author}
  {\bibfnamefont {J.~W.}\ \bibnamefont {{Briggs}}}, \bibinfo {author}
  {\bibfnamefont {J.}~\bibnamefont {{Brinkmann}}}, \bibinfo {author}
  {\bibfnamefont {R.}~\bibnamefont {{Brunner}}}, \bibinfo {author}
  {\bibfnamefont {S.}~\bibnamefont {{Burles}}}, \bibinfo {author}
  {\bibfnamefont {L.}~\bibnamefont {{Carey}}}, \bibinfo {author} {\bibfnamefont
  {M.~A.}\ \bibnamefont {{Carr}}}, \bibinfo {author} {\bibfnamefont {F.~J.}\
  \bibnamefont {{Castander}}}, \bibinfo {author} {\bibfnamefont
  {B.}~\bibnamefont {{Chen}}}, \bibinfo {author} {\bibfnamefont {P.~L.}\
  \bibnamefont {{Colestock}}}, \bibinfo {author} {\bibfnamefont {A.~J.}\
  \bibnamefont {{Connolly}}}, \bibinfo {author} {\bibfnamefont {J.~H.}\
  \bibnamefont {{Crocker}}}, \bibinfo {author} {\bibfnamefont {I.}~\bibnamefont
  {{Csabai}}}, \bibinfo {author} {\bibfnamefont {P.~C.}\ \bibnamefont
  {{Czarapata}}}, \bibinfo {author} {\bibfnamefont {J.~E.}\ \bibnamefont
  {{Davis}}}, \bibinfo {author} {\bibfnamefont {M.}~\bibnamefont {{Doi}}},
  \bibinfo {author} {\bibfnamefont {T.}~\bibnamefont {{Dombeck}}}, \bibinfo
  {author} {\bibfnamefont {D.}~\bibnamefont {{Eisenstein}}}, \bibinfo {author}
  {\bibfnamefont {N.}~\bibnamefont {{Ellman}}}, \bibinfo {author}
  {\bibfnamefont {B.~R.}\ \bibnamefont {{Elms}}}, \bibinfo {author}
  {\bibfnamefont {M.~L.}\ \bibnamefont {{Evans}}}, \bibinfo {author}
  {\bibfnamefont {X.}~\bibnamefont {{Fan}}}, \bibinfo {author} {\bibfnamefont
  {G.~R.}\ \bibnamefont {{Federwitz}}}, \bibinfo {author} {\bibfnamefont
  {L.}~\bibnamefont {{Fiscelli}}}, \bibinfo {author} {\bibfnamefont
  {S.}~\bibnamefont {{Friedman}}}, \bibinfo {author} {\bibfnamefont {J.~A.}\
  \bibnamefont {{Frieman}}}, \bibinfo {author} {\bibfnamefont {M.}~\bibnamefont
  {{Fukugita}}}, \bibinfo {author} {\bibfnamefont {B.}~\bibnamefont
  {{Gillespie}}}, \bibinfo {author} {\bibfnamefont {J.~E.}\ \bibnamefont
  {{Gunn}}}, \bibinfo {author} {\bibfnamefont {V.~K.}\ \bibnamefont
  {{Gurbani}}}, \bibinfo {author} {\bibfnamefont {E.}~\bibnamefont {{de
  Haas}}}, \bibinfo {author} {\bibfnamefont {M.}~\bibnamefont {{Haldeman}}},
  \bibinfo {author} {\bibfnamefont {F.~H.}\ \bibnamefont {{Harris}}}, \bibinfo
  {author} {\bibfnamefont {J.}~\bibnamefont {{Hayes}}}, \bibinfo {author}
  {\bibfnamefont {T.~M.}\ \bibnamefont {{Heckman}}}, \bibinfo {author}
  {\bibfnamefont {G.~S.}\ \bibnamefont {{Hennessy}}}, \bibinfo {author}
  {\bibfnamefont {R.~B.}\ \bibnamefont {{Hindsley}}}, \bibinfo {author}
  {\bibfnamefont {S.}~\bibnamefont {{Holm}}}, \bibinfo {author} {\bibfnamefont
  {D.~J.}\ \bibnamefont {{Holmgren}}}, \bibinfo {author} {\bibfnamefont
  {C.-h.}\ \bibnamefont {{Huang}}}, \bibinfo {author} {\bibfnamefont
  {C.}~\bibnamefont {{Hull}}}, \bibinfo {author} {\bibfnamefont
  {D.}~\bibnamefont {{Husby}}}, \bibinfo {author} {\bibfnamefont {S.-I.}\
  \bibnamefont {{Ichikawa}}}, \bibinfo {author} {\bibfnamefont
  {T.}~\bibnamefont {{Ichikawa}}}, \bibinfo {author} {\bibfnamefont {{\v
  Z}.}~\bibnamefont {{Ivezi{\'c}}}}, \bibinfo {author} {\bibfnamefont
  {S.}~\bibnamefont {{Kent}}}, \bibinfo {author} {\bibfnamefont {R.~S.~J.}\
  \bibnamefont {{Kim}}}, \bibinfo {author} {\bibfnamefont {E.}~\bibnamefont
  {{Kinney}}}, \bibinfo {author} {\bibfnamefont {M.}~\bibnamefont {{Klaene}}},
  \bibinfo {author} {\bibfnamefont {A.~N.}\ \bibnamefont {{Kleinman}}},
  \bibinfo {author} {\bibfnamefont {S.}~\bibnamefont {{Kleinman}}}, \bibinfo
  {author} {\bibfnamefont {G.~R.}\ \bibnamefont {{Knapp}}}, \bibinfo {author}
  {\bibfnamefont {J.}~\bibnamefont {{Korienek}}}, \bibinfo {author}
  {\bibfnamefont {R.~G.}\ \bibnamefont {{Kron}}}, \bibinfo {author}
  {\bibfnamefont {P.~Z.}\ \bibnamefont {{Kunszt}}}, \bibinfo {author}
  {\bibfnamefont {D.~Q.}\ \bibnamefont {{Lamb}}}, \bibinfo {author}
  {\bibfnamefont {B.}~\bibnamefont {{Lee}}}, \bibinfo {author} {\bibfnamefont
  {R.~F.}\ \bibnamefont {{Leger}}}, \bibinfo {author} {\bibfnamefont
  {S.}~\bibnamefont {{Limmongkol}}}, \bibinfo {author} {\bibfnamefont
  {C.}~\bibnamefont {{Lindenmeyer}}}, \bibinfo {author} {\bibfnamefont {D.~C.}\
  \bibnamefont {{Long}}}, \bibinfo {author} {\bibfnamefont {C.}~\bibnamefont
  {{Loomis}}}, \bibinfo {author} {\bibfnamefont {J.}~\bibnamefont {{Loveday}}},
  \bibinfo {author} {\bibfnamefont {R.}~\bibnamefont {{Lucinio}}}, \bibinfo
  {author} {\bibfnamefont {R.~H.}\ \bibnamefont {{Lupton}}}, \bibinfo {author}
  {\bibfnamefont {B.}~\bibnamefont {{MacKinnon}}}, \bibinfo {author}
  {\bibfnamefont {E.~J.}\ \bibnamefont {{Mannery}}}, \bibinfo {author}
  {\bibfnamefont {P.~M.}\ \bibnamefont {{Mantsch}}}, \bibinfo {author}
  {\bibfnamefont {B.}~\bibnamefont {{Margon}}}, \bibinfo {author}
  {\bibfnamefont {P.}~\bibnamefont {{McGehee}}}, \bibinfo {author}
  {\bibfnamefont {T.~A.}\ \bibnamefont {{McKay}}}, \bibinfo {author}
  {\bibfnamefont {A.}~\bibnamefont {{Meiksin}}}, \bibinfo {author}
  {\bibfnamefont {A.}~\bibnamefont {{Merelli}}}, \bibinfo {author}
  {\bibfnamefont {D.~G.}\ \bibnamefont {{Monet}}}, \bibinfo {author}
  {\bibfnamefont {J.~A.}\ \bibnamefont {{Munn}}}, \bibinfo {author}
  {\bibfnamefont {V.~K.}\ \bibnamefont {{Narayanan}}}, \bibinfo {author}
  {\bibfnamefont {T.}~\bibnamefont {{Nash}}}, \bibinfo {author} {\bibfnamefont
  {E.}~\bibnamefont {{Neilsen}}}, \bibinfo {author} {\bibfnamefont
  {R.}~\bibnamefont {{Neswold}}}, \bibinfo {author} {\bibfnamefont {H.~J.}\
  \bibnamefont {{Newberg}}}, \bibinfo {author} {\bibfnamefont {R.~C.}\
  \bibnamefont {{Nichol}}}, \bibinfo {author} {\bibfnamefont {T.}~\bibnamefont
  {{Nicinski}}}, \bibinfo {author} {\bibfnamefont {M.}~\bibnamefont
  {{Nonino}}}, \bibinfo {author} {\bibfnamefont {N.}~\bibnamefont {{Okada}}},
  \bibinfo {author} {\bibfnamefont {S.}~\bibnamefont {{Okamura}}}, \bibinfo
  {author} {\bibfnamefont {J.~P.}\ \bibnamefont {{Ostriker}}}, \bibinfo
  {author} {\bibfnamefont {R.}~\bibnamefont {{Owen}}}, \bibinfo {author}
  {\bibfnamefont {A.~G.}\ \bibnamefont {{Pauls}}}, \bibinfo {author}
  {\bibfnamefont {J.}~\bibnamefont {{Peoples}}}, \bibinfo {author}
  {\bibfnamefont {R.~L.}\ \bibnamefont {{Peterson}}}, \bibinfo {author}
  {\bibfnamefont {D.}~\bibnamefont {{Petravick}}}, \bibinfo {author}
  {\bibfnamefont {J.~R.}\ \bibnamefont {{Pier}}}, \bibinfo {author}
  {\bibfnamefont {A.}~\bibnamefont {{Pope}}}, \bibinfo {author} {\bibfnamefont
  {R.}~\bibnamefont {{Pordes}}}, \bibinfo {author} {\bibfnamefont
  {A.}~\bibnamefont {{Prosapio}}}, \bibinfo {author} {\bibfnamefont
  {R.}~\bibnamefont {{Rechenmacher}}}, \bibinfo {author} {\bibfnamefont
  {T.~R.}\ \bibnamefont {{Quinn}}}, \bibinfo {author} {\bibfnamefont {G.~T.}\
  \bibnamefont {{Richards}}}, \bibinfo {author} {\bibfnamefont {M.~W.}\
  \bibnamefont {{Richmond}}}, \bibinfo {author} {\bibfnamefont {C.~H.}\
  \bibnamefont {{Rivetta}}}, \bibinfo {author} {\bibfnamefont {C.~M.}\
  \bibnamefont {{Rockosi}}}, \bibinfo {author} {\bibfnamefont {K.}~\bibnamefont
  {{Ruthmansdorfer}}}, \bibinfo {author} {\bibfnamefont {D.}~\bibnamefont
  {{Sandford}}}, \bibinfo {author} {\bibfnamefont {D.~J.}\ \bibnamefont
  {{Schlegel}}}, \bibinfo {author} {\bibfnamefont {D.~P.}\ \bibnamefont
  {{Schneider}}}, \bibinfo {author} {\bibfnamefont {M.}~\bibnamefont
  {{Sekiguchi}}}, \bibinfo {author} {\bibfnamefont {G.}~\bibnamefont
  {{Sergey}}}, \bibinfo {author} {\bibfnamefont {K.}~\bibnamefont
  {{Shimasaku}}}, \bibinfo {author} {\bibfnamefont {W.~A.}\ \bibnamefont
  {{Siegmund}}}, \bibinfo {author} {\bibfnamefont {S.}~\bibnamefont {{Smee}}},
  \bibinfo {author} {\bibfnamefont {J.~A.}\ \bibnamefont {{Smith}}}, \bibinfo
  {author} {\bibfnamefont {S.}~\bibnamefont {{Snedden}}}, \bibinfo {author}
  {\bibfnamefont {R.}~\bibnamefont {{Stone}}}, \bibinfo {author} {\bibfnamefont
  {C.}~\bibnamefont {{Stoughton}}}, \bibinfo {author} {\bibfnamefont {M.~A.}\
  \bibnamefont {{Strauss}}}, \bibinfo {author} {\bibfnamefont {C.}~\bibnamefont
  {{Stubbs}}}, \bibinfo {author} {\bibfnamefont {M.}~\bibnamefont
  {{SubbaRao}}}, \bibinfo {author} {\bibfnamefont {A.~S.}\ \bibnamefont
  {{Szalay}}}, \bibinfo {author} {\bibfnamefont {I.}~\bibnamefont {{Szapudi}}},
  \bibinfo {author} {\bibfnamefont {G.~P.}\ \bibnamefont {{Szokoly}}}, \bibinfo
  {author} {\bibfnamefont {A.~R.}\ \bibnamefont {{Thakar}}}, \bibinfo {author}
  {\bibfnamefont {C.}~\bibnamefont {{Tremonti}}}, \bibinfo {author}
  {\bibfnamefont {D.~L.}\ \bibnamefont {{Tucker}}}, \bibinfo {author}
  {\bibfnamefont {A.}~\bibnamefont {{Uomoto}}}, \bibinfo {author}
  {\bibfnamefont {D.}~\bibnamefont {{Vanden Berk}}}, \bibinfo {author}
  {\bibfnamefont {M.~S.}\ \bibnamefont {{Vogeley}}}, \bibinfo {author}
  {\bibfnamefont {P.}~\bibnamefont {{Waddell}}}, \bibinfo {author}
  {\bibfnamefont {S.-i.}\ \bibnamefont {{Wang}}}, \bibinfo {author}
  {\bibfnamefont {M.}~\bibnamefont {{Watanabe}}}, \bibinfo {author}
  {\bibfnamefont {D.~H.}\ \bibnamefont {{Weinberg}}}, \bibinfo {author}
  {\bibfnamefont {B.}~\bibnamefont {{Yanny}}}, \bibinfo {author} {\bibfnamefont
  {N.}~\bibnamefont {{Yasuda}}}, \ and\ \bibinfo {author} {\bibnamefont {{SDSS
  Collaboration}}},\ }\href {\doibase 10.1086/301513} {\bibfield  {journal}
  {\bibinfo  {journal} {\aj}\ }\textbf {\bibinfo {volume} {120}},\ \bibinfo
  {pages} {1579} (\bibinfo {year} {2000})},\ \Eprint
  {http://arxiv.org/abs/astro-ph/0006396} {astro-ph/0006396} \BibitemShut
  {NoStop}%
\bibitem [{\citenamefont {{Abazajian}}\ \emph {et~al.}(2009)\citenamefont
  {{Abazajian}}, \citenamefont {{Adelman-McCarthy}}, \citenamefont
  {{Ag{\"u}eros}}, \citenamefont {{Allam}}, \citenamefont {{Allende Prieto}},
  \citenamefont {{An}}, \citenamefont {{Anderson}}, \citenamefont {{Anderson}},
  \citenamefont {{Annis}}, \citenamefont {{Bahcall}},\ and\ \citenamefont
  {et~al.}}]{Abazajian09}%
  \BibitemOpen
  \bibfield  {author} {\bibinfo {author} {\bibfnamefont {K.~N.}\ \bibnamefont
  {{Abazajian}}}, \bibinfo {author} {\bibfnamefont {J.~K.}\ \bibnamefont
  {{Adelman-McCarthy}}}, \bibinfo {author} {\bibfnamefont {M.~A.}\ \bibnamefont
  {{Ag{\"u}eros}}}, \bibinfo {author} {\bibfnamefont {S.~S.}\ \bibnamefont
  {{Allam}}}, \bibinfo {author} {\bibfnamefont {C.}~\bibnamefont {{Allende
  Prieto}}}, \bibinfo {author} {\bibfnamefont {D.}~\bibnamefont {{An}}},
  \bibinfo {author} {\bibfnamefont {K.~S.~J.}\ \bibnamefont {{Anderson}}},
  \bibinfo {author} {\bibfnamefont {S.~F.}\ \bibnamefont {{Anderson}}},
  \bibinfo {author} {\bibfnamefont {J.}~\bibnamefont {{Annis}}}, \bibinfo
  {author} {\bibfnamefont {N.~A.}\ \bibnamefont {{Bahcall}}}, \ and\ \bibinfo
  {author} {\bibnamefont {et~al.}},\ }\href {\doibase
  10.1088/0067-0049/182/2/543} {\bibfield  {journal} {\bibinfo  {journal}
  {\apjs}\ }\textbf {\bibinfo {volume} {182}},\ \bibinfo {eid} {543} (\bibinfo
  {year} {2009})},\ \Eprint {http://arxiv.org/abs/0812.0649} {arXiv:0812.0649}
  \BibitemShut {NoStop}%
\end{thebibliography}%


\appendix

\section{Non-linear mass as a function of redshift}\label{sec:M_star}

\begin{figure}
  \centering
  \includegraphics[width=0.45\textwidth,angle=0.0]{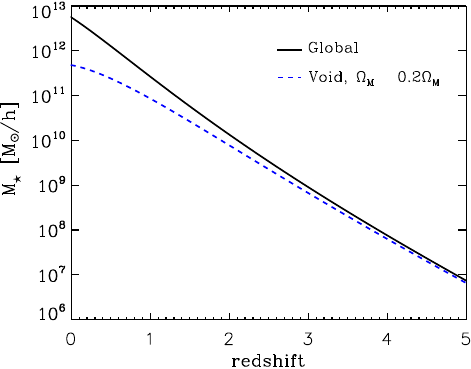}
  \caption{$M_{\star}$ as a function of redshift, both in our fiducial $\Lambda$CDM model, and in an approximation of a void region (see text for details).}\label{fig:M_star}
\end{figure}

Figure \ref{fig:M_star} shows $M_{\star}(z)$, that is, the mass within a Lagrangian sphere of size such that the variance of the linear-theory density equals $\delta_c=1.68$. For this, we integrated a CAMB power spectrum with the parameters of the simulation. We also show an estimate of $M_{\star}(z)$ in a region that is a void at time $z=0$. For the void calculation, we kept fixed the shape of the linear power spectrum, but multiplied $\Omega_m$ by 0.2 (the fiducial density in a void region) in obtaining the growth factor; $\sigma_8$ is effectively lower in a void at $z=0$.

\section{Simulations and galaxy catalogs}\label{sec:simulations}
The results presented here are based on two cosmological boxes of 64 \mpc and 32 \mpc  side. Each box was run at several resolutions and including dark matter only, dark matter plus gas and with and without feedback.
We assumed a $\Lambda$CDM cosmology with values of $\Omega_m=0.3$, $\Omega_{\Lambda}=0.7$, $h=0.73$ and $\sigma_8=0.8$, in good agreement with the latest values obtained from the Planck mission \citep{Planck15}. The results presented here are insensitive to small variations in the cosmological parameters. The 64 \mpc and 32 \mpc boxes were run at a mass resolution of $1.62 \times 10^8$ \msun per particle corresponding to $512^3$ and $256^3$ particles respectively. Several zoom resimulations were run at higher resolution for selected individual halos at $2.02 \times 10^{7}$ and $2.53 \times 10^6$ \msun per  dark matter particle.

The 64 \mpc box was run using the sph Gadget-2 code with gas and no cooling. The 32 \mpc simulation box is the same box described in \citep{Aragon12} and \citep{Aragon14}. Gadget-2 computes long-range gravitational forces using a particle-mesh algorithm \citep{Hockney88} and a tree algorithm for small separations \citep{Barnes86}. Gas dynamics are computed using a smoothed hydrodynamics approach \citep{Lucy77,Springel02}.
The Gadget-3 sph code used to run the simulations with star formation implements recipes for hydrodynamics and metal enrichment including stochastic star formation, SN feedback and winds \citep{Springel03}. The critical density for star formation is set to $n_h = 0.1$ cm$^{-3}$. While more detailed star formation recipes including feedback are possible, in the present work they are not critical, as the CWD mechanisms described here are mechanical/gravitational in nature and take place outside the star-forming regions of the galaxy.

%
\subsection{Halo catalogues}

From the two simulations we computed halo/sub-halo catalogs using Friends-of-Friends (FoF) and SubFind \citep{Davis85,Springel05}. For both halos and sub-halos we generated merger trees by identifying common particles between halos/sub-halos in adjacent snapshots. This produced large and complex trees that were then pruned to identify their ``most massive progenitor'' lines. The mass at each point of the most massive progenitor line forms the mass accretion history (MAH) of the halo/sub-halo. The MAH of FoF halos is in most cases an increasing function of time. Sub-halos, on the other hand, tend to have more erratic MAH as a result of non-linear interactions  and the particle unbinding procedure \citep{Onions12}. One particular problem of sub-halo MAH is the decrease in mass after accretion into a larger halo as a result of mass stripping. Following a similar approach as \citet{Conroy06} and  \citet{Vale06} we allow the mass of a sub-halo to either remain constant or increase with time. In the case of satellite halos with severe mass stripping this means that the satellite is implicitly assumed to retain its mass right before accretion. For consistency we applied the same procedure to the FoF halos although they have a near monotonic increase in mass.

\begin{figure*}
  \centering
  \includegraphics[width=0.32\textwidth,angle=0.0]{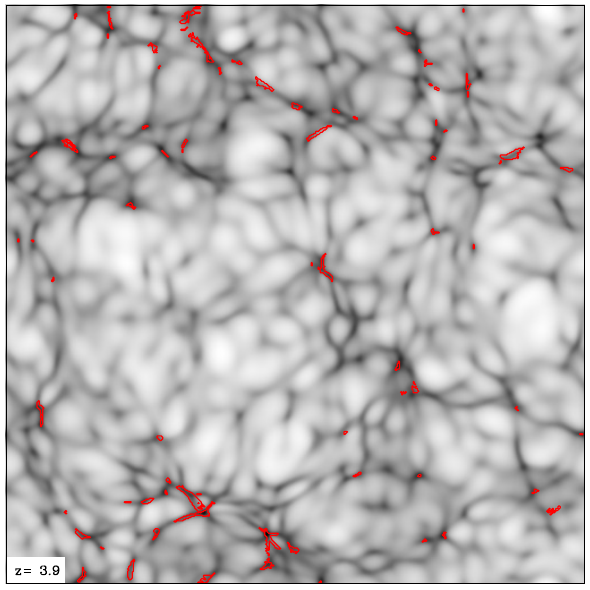}
  \includegraphics[width=0.32\textwidth,angle=0.0]{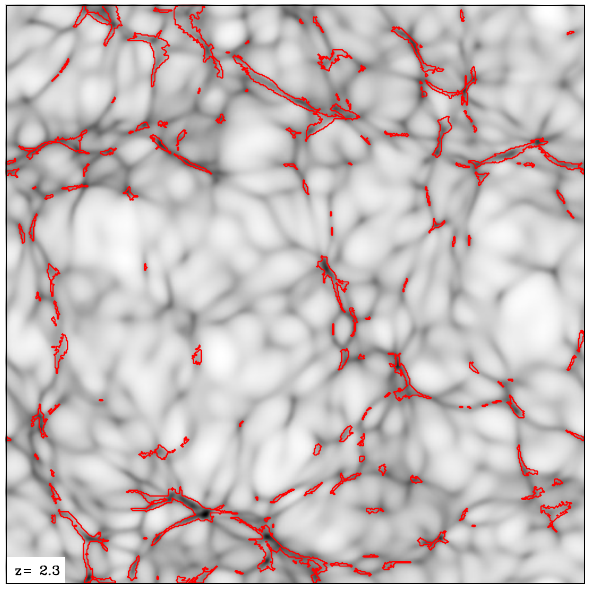}
  \includegraphics[width=0.32\textwidth,angle=0.0]{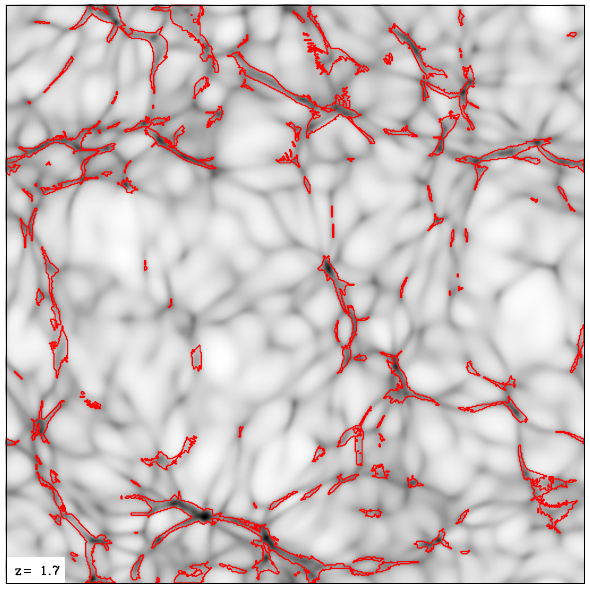}
  \includegraphics[width=0.32\textwidth,angle=0.0]{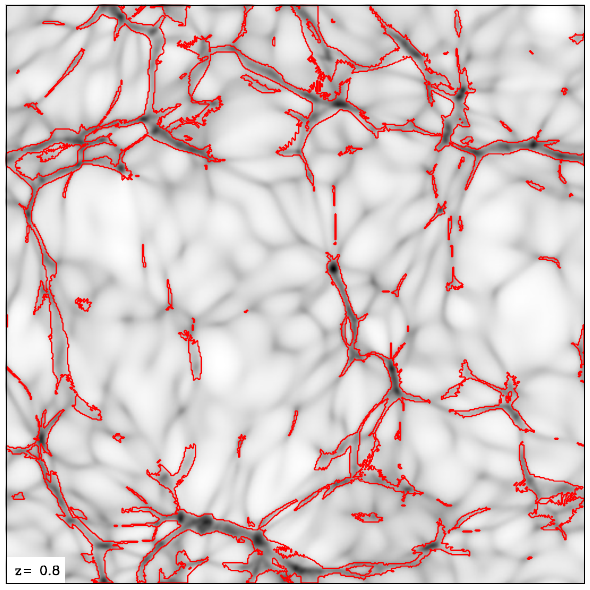}
  \includegraphics[width=0.32\textwidth,angle=0.0]{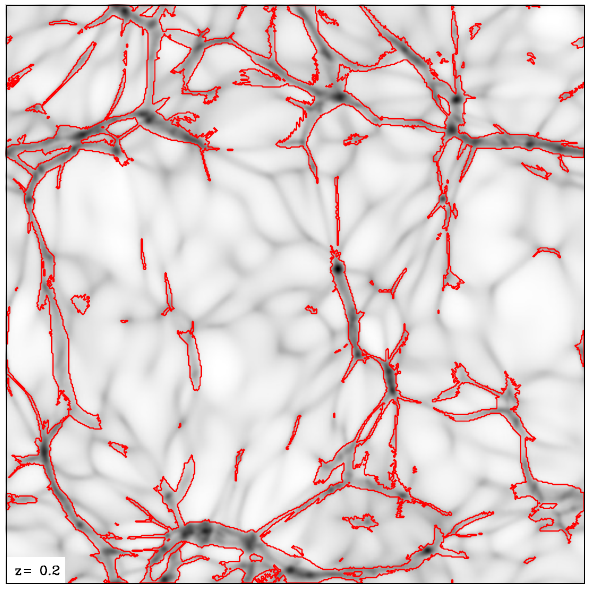}
  \includegraphics[width=0.32\textwidth,angle=0.0]{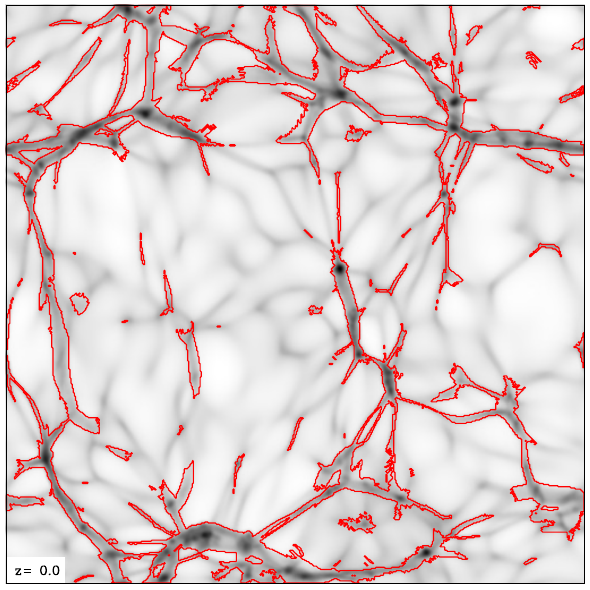}
  \caption{Evolution of shell-crossing regions in a 2\mpc thick slice across the simulation box. The gray-scale background represents the projected smoothed density field.  The red contours enclose the regions where shell-crossing has occurred which correspond to Web Detachment regions. This is an extended version of Fig. \ref{fig:multistreams_grid} in the main text.}\label{fig:multistreams_2}
\end{figure*}

%
\subsection{Galaxy catalogue (SDSS)}\label{app:sdss}

For the study of the galaxy mass-color distribution (Sec. \ref{sec:sdss}) we used a sample of galaxies selected from the Sloan Digital Sky Survey data release 9 \citep{York00,Abazajian09}. We selected galaxies from the spectroscopic sample in the redshift range $z < 0.1$ and no constraints on magnitude. Given the target selection criteria this corresponds to galaxies with a Petrosian magnitude in the red filter of $m_r <17.77$. The total sample contains almost half a million galaxies which is enough to perform the matching with halos in our 64 \mpc simulation at $z=0$.  The SDSS galaxy sample used in this work was queried from the CASJOBS service\footnote{https://skyserver.sdss.org/CasJobs}. We selected galaxies with spectroscopic measurements in the redshift range $z=0-0.1$ using the following (abridged) query:

\begin{verbatim}
SELECT  p.ObjID, p.ra, p.dec, s.z, s.zErr, s.zConf,
     s.cx, s.cy, s.cz,
     p.dered_u, p.dered_g, p.dered_r, ...
     p.petroMag_u,p.petroMag_g,p.petroMag_r,...
     s.specclass,
     s.eClass
INTO myDB.galaxies_dr9
FROM SpecObj  AS s,
         PhotoObj AS p
WHERE s.SpecObjID = p.SpecObjID AND
           s.z BETWEEN 0 and 0.1
\end{verbatim}

\section{CWD examples}\label{app:cwd_examples}

In this section we present several examples of halos/subhalos undergoing CWD events and others that remain attached to their network for primordial filaments. The descriptions are based on observations on several projections as well as mass accretion rates and CWD times (not shown here for clarity). Note that the different times and scales are not the same for all figures.


\begin{figure*}
  \centering
  \includegraphics[width=0.95\textwidth,angle=0.0]{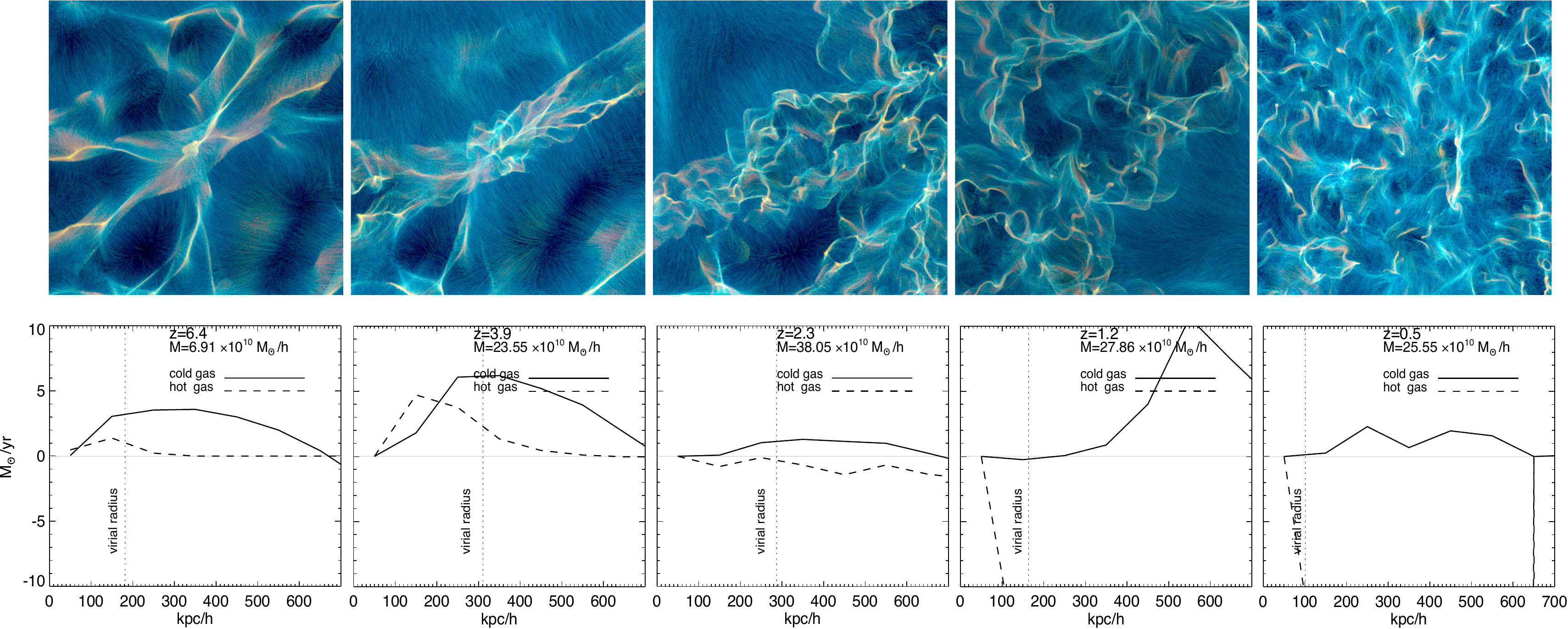}
\caption{Subhalo 002087. Top panels correspond to ZY projections of the  dark matter velocity field (see main paper for details). Subhalo entering a small group. Steady accretion of gas at early times before CWD event and afterwards a large gravitational interaction with a group resulting in mass loss.}
\end{figure*}

\begin{figure*}
  \centering
  \includegraphics[width=0.95\textwidth,angle=0.0]{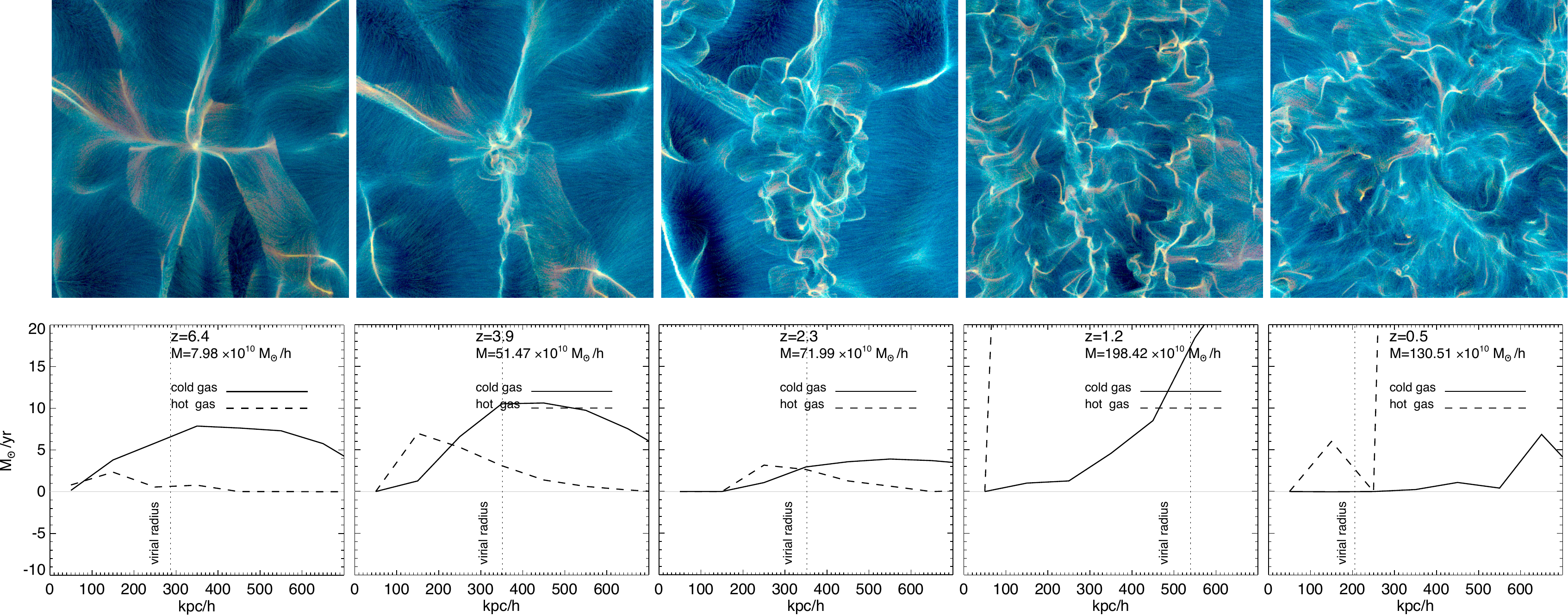}

\caption{Subhalo 005123.  Top three rows correspond to XY XZ and ZY projections of the  dark matter velocity field (see main paper for details). Steady early gas accretion, then quiet CWD event and later mergers that disrupt the surrounding matter configuration.}
\end{figure*}

\begin{figure*}
  \centering
  \includegraphics[width=0.95\textwidth,angle=0.0]{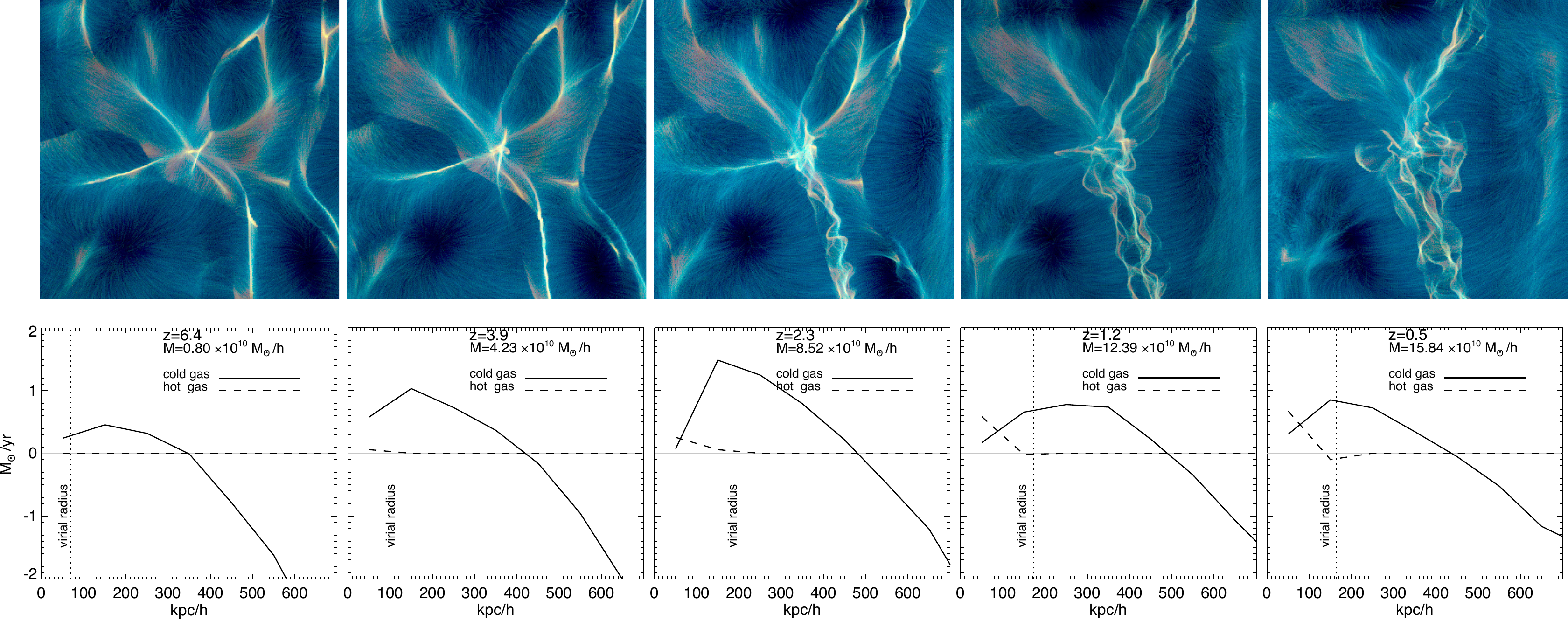}
\caption{Halo 004307.  Top panels correspond to ZY projections of the  dark matter velocity field (see main paper for details). This subhalo presents steady cold gas accretion from early times. Its matter and velocity field configuration indicates that it is still attached to its filaments and this can be seen by the steady cold gas accretion/star formation at the present time. }
\end{figure*}

\begin{figure*}
  \centering

 \includegraphics[width=0.95\textwidth,angle=0.0]{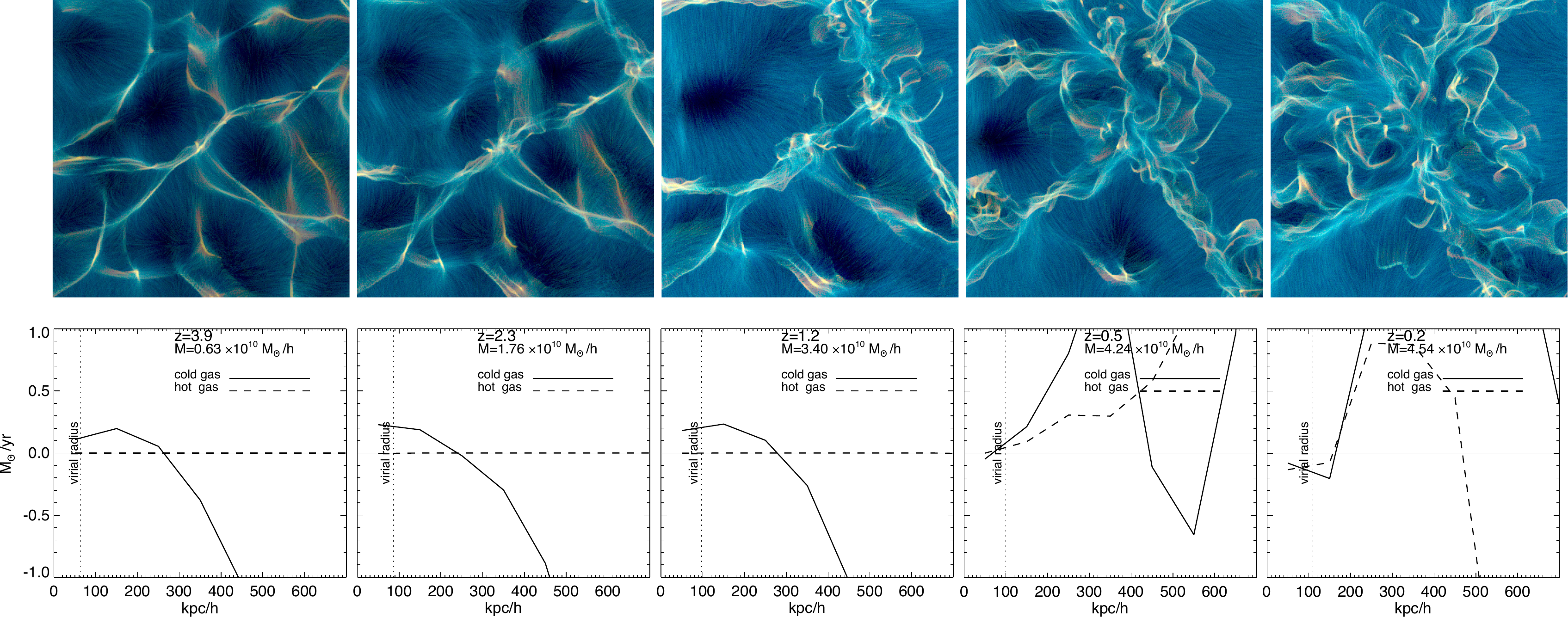}
\caption{Halo 014648.  Top panels correspond to ZY projections of the  dark matter velocity field  of the  dark matter velocity field (see main paper for details). This halo shows cold gas accretion right until its CWD event triggered by accretion into a larger filament, after which point the accretion of cold gas stops.}
\end{figure*}

\begin{figure*}
  \centering

  \includegraphics[width=0.95\textwidth,angle=0.0]{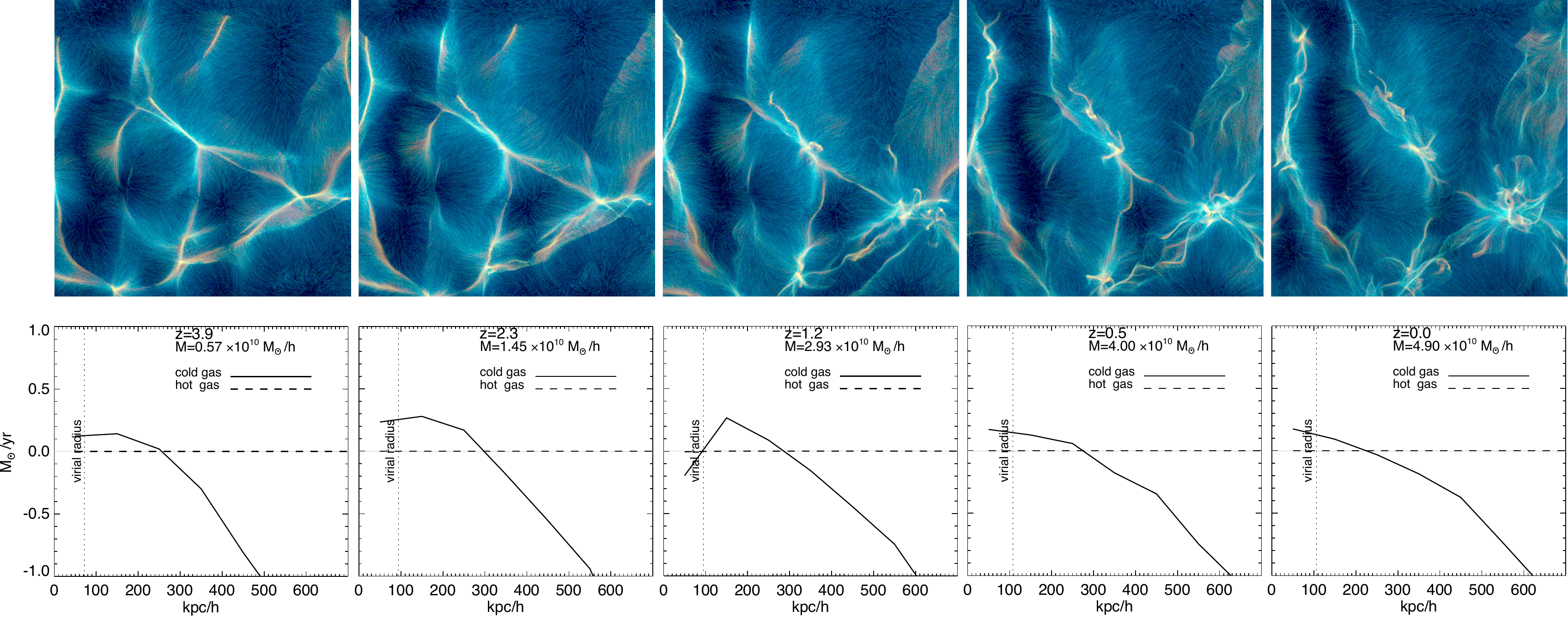}
\caption{Halo 014862. Halo located in a low-density region corresponding to a present-time cosmological void. The halo never enters a non-linear region and therefore, according to our multi-streaming prescription, it is still attached to its primordial filaments. Note that the accretion the cold gas, although very small, remains relatively constant until the present time.
Non-detached, Web stretching.}
\end{figure*}

\end{document}